\documentclass[longauth]{aa}

\usepackage[export]{adjustbox}
\usepackage{graphicx}
\usepackage{natbib}
\usepackage{scalerel}
\usepackage{booktabs}
\usepackage[table]{xcolor}
\usepackage{amsmath}
\usepackage{txfonts}
\usepackage{placeins}
\usepackage{rotating}
\usepackage{lastpage}
\usepackage{bm}
\usepackage{siunitx}
\usepackage{CJKutf8}
\usepackage{tabularx}
\usepackage{pdflscape}
\usepackage{longtable}
\usepackage{glossaries}
\newacronym{dr1}{DR1}{Data Release 1}
\newacronym{q1}{Q1}{Quick Data Release 1}
\newacronym{SED}{SED}{spectral energy distribution}
\newacronym{NIR}{NIR}{Near-Infrared}
\newacronym{EWS}{EWS}{Euclid Wide Survey}
\newacronym{EDFs}{EDFs}{Euclid Deep Fields}
\newacronym{sfr}{SFR}{star-formation rate}
\newacronym{ssfr}{sSFR}{specific \gls{sfr}}
\newcommand{\kpc}{\kilo\parsec}

\newcommand{\Msun}{\si{\solarmass}}
\newcommand{\Mstarwun}{\logten(\Mstellar/\Msun)}
\newcommand*{\JWST}{\textit{James Webb} Space Telescope\xspace}

\defcitealias{Pandya_2024}{P24}
\defcitealias{Zhang_2019}{Z19}
\defcitealias{VanDerWel_2014a}{vdW14a}

\usepackage[pdfencoding=auto,psdextra]{hyperref}
\hypersetup{
    colorlinks=true,
    linkcolor=blue,
    filecolor=magenta,
    urlcolor=blue,
    citecolor=blue
}
\makeatletter
\renewcommand*\aa@pageof{, page \thepage{} of \pageref*{LastPage}}
\makeatother

\usepackage[utf8]{inputenc}
\usepackage[switch, modulo]{lineno}
\usepackage{euclid}
\makeatletter
\newcommand{\capfig}{\def\@captype{figure}}
\newcommand{\captab}{\def\@captype{table}}
\makeatother

\begin{document}

\title{\Euclid: Data Release 1 (DR1) -- Bayesian 3D shape inference of star-forming galaxy populations\thanks{This paper is published on behalf of the Euclid Consortium.}}    

\newcommand{\orcid}[1]{}		   
\author{M.~Dunn\orcid{0000-0001-5374-1644}\thanks{\email{marina.dunn@iac.es}}\inst{\ref{aff1},\ref{aff2},\ref{aff3}}
\and M.~Huertas-Company\orcid{0000-0002-1416-8483}\inst{\ref{aff1},\ref{aff4},\ref{aff5}}
\and J.~H.~Knapen\orcid{0000-0003-1643-0024}\inst{\ref{aff1},\ref{aff2}}
\and V.~Pandya\orcid{0000-0002-2499-9205}\inst{\ref{aff6}}
\and H.~Zhang\orcid{0000-0002-4321-3538}\inst{\ref{aff7}}
\and L.~Quilley\orcid{0009-0008-8375-8605}\inst{\ref{aff8}}
\and A.~van~der~Wel\orcid{0000-0002-5027-0135}\inst{\ref{aff9}}
\and E.~Duran-Camacho\orcid{0000-0002-3153-0536}\inst{\ref{aff1},\ref{aff2}}
\and D.~Scott\orcid{0000-0002-6878-9840}\inst{\ref{aff10}}
\and G.~Helou\orcid{0000-0003-3367-3415}\inst{\ref{aff11}}
\and K.~Bunte\orcid{0000-0002-2930-6172}\inst{\ref{aff3}}
\and B.~Altieri\orcid{0000-0003-3936-0284}\inst{\ref{aff12}}
\and S.~Andreon\orcid{0000-0002-2041-8784}\inst{\ref{aff13}}
\and N.~Auricchio\orcid{0000-0003-4444-8651}\inst{\ref{aff14}}
\and C.~Baccigalupi\orcid{0000-0002-8211-1630}\inst{\ref{aff15},\ref{aff16},\ref{aff17},\ref{aff18}}
\and M.~Baldi\orcid{0000-0003-4145-1943}\inst{\ref{aff19},\ref{aff14},\ref{aff20}}
\and S.~Bardelli\orcid{0000-0002-8900-0298}\inst{\ref{aff14}}
\and P.~Battaglia\orcid{0000-0002-7337-5909}\inst{\ref{aff14}}
\and A.~Biviano\orcid{0000-0002-0857-0732}\inst{\ref{aff16},\ref{aff15}}
\and E.~Branchini\orcid{0000-0002-0808-6908}\inst{\ref{aff21},\ref{aff22},\ref{aff13}}
\and M.~Brescia\orcid{0000-0001-9506-5680}\inst{\ref{aff23},\ref{aff24}}
\and S.~Camera\orcid{0000-0003-3399-3574}\inst{\ref{aff25},\ref{aff26},\ref{aff27}}
\and V.~Capobianco\orcid{0000-0002-3309-7692}\inst{\ref{aff27}}
\and C.~Carbone\orcid{0000-0003-0125-3563}\inst{\ref{aff28}}
\and J.~Carretero\orcid{0000-0002-3130-0204}\inst{\ref{aff29},\ref{aff30}}
\and M.~Castellano\orcid{0000-0001-9875-8263}\inst{\ref{aff31}}
\and G.~Castignani\orcid{0000-0001-6831-0687}\inst{\ref{aff14}}
\and S.~Cavuoti\orcid{0000-0002-3787-4196}\inst{\ref{aff24},\ref{aff32}}
\and K.~C.~Chambers\orcid{0000-0001-6965-7789}\inst{\ref{aff33}}
\and A.~Cimatti\inst{\ref{aff34}}
\and C.~Colodro-Conde\inst{\ref{aff1}}
\and G.~Congedo\orcid{0000-0003-2508-0046}\inst{\ref{aff35}}
\and C.~J.~Conselice\orcid{0000-0003-1949-7638}\inst{\ref{aff8}}
\and L.~Conversi\orcid{0000-0002-6710-8476}\inst{\ref{aff36},\ref{aff12}}
\and Y.~Copin\orcid{0000-0002-5317-7518}\inst{\ref{aff37}}
\and F.~Courbin\orcid{0000-0003-0758-6510}\inst{\ref{aff38},\ref{aff39},\ref{aff40}}
\and H.~M.~Courtois\orcid{0000-0003-0509-1776}\inst{\ref{aff41}}
\and M.~Cropper\orcid{0000-0003-4571-9468}\inst{\ref{aff42}}
\and H.~Degaudenzi\orcid{0000-0002-5887-6799}\inst{\ref{aff43}}
\and G.~De~Lucia\orcid{0000-0002-6220-9104}\inst{\ref{aff16}}
\and C.~Dolding\orcid{0009-0003-7199-6108}\inst{\ref{aff42}}
\and H.~Dole\orcid{0000-0002-9767-3839}\inst{\ref{aff44}}
\and F.~Dubath\orcid{0000-0002-6533-2810}\inst{\ref{aff43}}
\and X.~Dupac\inst{\ref{aff12}}
\and S.~Dusini\orcid{0000-0002-1128-0664}\inst{\ref{aff45}}
\and A.~Ealet\orcid{0000-0003-3070-014X}\inst{\ref{aff37}}
\and S.~Escoffier\orcid{0000-0002-2847-7498}\inst{\ref{aff46}}
\and M.~Farina\orcid{0000-0002-3089-7846}\inst{\ref{aff47}}
\and R.~Farinelli\inst{\ref{aff14}}
\and S.~Ferriol\inst{\ref{aff37}}
\and P.~Fosalba\orcid{0000-0002-1510-5214}\inst{\ref{aff48},\ref{aff49}}
\and S.~Fotopoulou\orcid{0000-0002-9686-254X}\inst{\ref{aff50}}
\and M.~Frailis\orcid{0000-0002-7400-2135}\inst{\ref{aff16}}
\and E.~Franceschi\orcid{0000-0002-0585-6591}\inst{\ref{aff14}}
\and M.~Fumana\orcid{0000-0001-6787-5950}\inst{\ref{aff28}}
\and S.~Galeotta\orcid{0000-0002-3748-5115}\inst{\ref{aff16}}
\and K.~George\orcid{0000-0002-1734-8455}\inst{\ref{aff51}}
\and B.~Gillis\orcid{0000-0002-4478-1270}\inst{\ref{aff35}}
\and C.~Giocoli\orcid{0000-0002-9590-7961}\inst{\ref{aff14},\ref{aff20}}
\and J.~Gracia-Carpio\orcid{0000-0003-4689-3134}\inst{\ref{aff52}}
\and A.~Grazian\orcid{0000-0002-5688-0663}\inst{\ref{aff53}}
\and F.~Grupp\inst{\ref{aff52},\ref{aff54}}
\and S.~Gwyn\orcid{0000-0001-8221-8406}\inst{\ref{aff55}}
\and S.~V.~H.~Haugan\orcid{0000-0001-9648-7260}\inst{\ref{aff56}}
\and W.~Holmes\orcid{0009-0007-8554-4646}\inst{\ref{aff57}}
\and F.~Hormuth\inst{\ref{aff58}}
\and A.~Hornstrup\orcid{0000-0002-3363-0936}\inst{\ref{aff59},\ref{aff60}}
\and K.~Jahnke\orcid{0000-0003-3804-2137}\inst{\ref{aff61}}
\and M.~Jhabvala\inst{\ref{aff62}}
\and B.~Joachimi\orcid{0000-0001-7494-1303}\inst{\ref{aff63}}
\and S.~Kermiche\orcid{0000-0002-0302-5735}\inst{\ref{aff46}}
\and A.~Kiessling\orcid{0000-0002-2590-1273}\inst{\ref{aff57}}
\and B.~Kubik\orcid{0009-0006-5823-4880}\inst{\ref{aff37}}
\and M.~K\"ummel\orcid{0000-0003-2791-2117}\inst{\ref{aff54}}
\and H.~Kurki-Suonio\orcid{0000-0002-4618-3063}\inst{\ref{aff64},\ref{aff65}}
\and A.~M.~C.~Le~Brun\orcid{0000-0002-0936-4594}\inst{\ref{aff66}}
\and S.~Ligori\orcid{0000-0003-4172-4606}\inst{\ref{aff27}}
\and P.~B.~Lilje\orcid{0000-0003-4324-7794}\inst{\ref{aff56}}
\and V.~Lindholm\orcid{0000-0003-2317-5471}\inst{\ref{aff64},\ref{aff65}}
\and I.~Lloro\orcid{0000-0001-5966-1434}\inst{\ref{aff67}}
\and M.~Magliocchetti\orcid{0000-0001-9158-4838}\inst{\ref{aff47}}
\and G.~Mainetti\orcid{0000-0003-2384-2377}\inst{\ref{aff68}}
\and O.~Mansutti\orcid{0000-0001-5758-4658}\inst{\ref{aff16}}
\and O.~Marggraf\orcid{0000-0001-7242-3852}\inst{\ref{aff69}}
\and M.~Martinelli\orcid{0000-0002-6943-7732}\inst{\ref{aff31},\ref{aff70}}
\and N.~Martinet\orcid{0000-0003-2786-7790}\inst{\ref{aff71}}
\and F.~Marulli\orcid{0000-0002-8850-0303}\inst{\ref{aff72},\ref{aff14},\ref{aff20}}
\and R.~J.~Massey\orcid{0000-0002-6085-3780}\inst{\ref{aff73}}
\and E.~Medinaceli\orcid{0000-0002-4040-7783}\inst{\ref{aff14}}
\and S.~Mei\orcid{0000-0002-2849-559X}\inst{\ref{aff74},\ref{aff75}}
\and M.~Melchior\inst{\ref{aff76}}
\and M.~Meneghetti\orcid{0000-0003-1225-7084}\inst{\ref{aff14},\ref{aff20}}
\and E.~Merlin\orcid{0000-0001-6870-8900}\inst{\ref{aff53}}
\and G.~Meylan\orcid{0000-0001-6503-0209}\inst{\ref{aff77}}
\and A.~Mora\orcid{0000-0002-1922-8529}\inst{\ref{aff78}}
\and M.~Moresco\orcid{0000-0002-7616-7136}\inst{\ref{aff72},\ref{aff14}}
\and C.~Moretti\orcid{0000-0003-3314-8936}\inst{\ref{aff16},\ref{aff15},\ref{aff17}}
\and L.~Moscardini\orcid{0000-0002-3473-6716}\inst{\ref{aff72},\ref{aff14},\ref{aff20}}
\and R.~Nakajima\orcid{0009-0009-1213-7040}\inst{\ref{aff69}}
\and C.~Neissner\orcid{0000-0001-8524-4968}\inst{\ref{aff79},\ref{aff30}}
\and S.-M.~Niemi\orcid{0009-0005-0247-0086}\inst{\ref{aff80}}
\and J.~W.~Nightingale\orcid{0000-0002-8987-7401}\inst{\ref{aff81}}
\and C.~Padilla\orcid{0000-0001-7951-0166}\inst{\ref{aff79}}
\and S.~Paltani\orcid{0000-0002-8108-9179}\inst{\ref{aff43}}
\and F.~Pasian\orcid{0000-0002-4869-3227}\inst{\ref{aff16}}
\and K.~Pedersen\inst{\ref{aff82}}
\and W.~J.~Percival\orcid{0000-0002-0644-5727}\inst{\ref{aff83},\ref{aff84},\ref{aff85}}
\and V.~Pettorino\orcid{0000-0002-4203-9320}\inst{\ref{aff80}}
\and A.~Pezzotta\orcid{0000-0003-0726-2268}\inst{\ref{aff13}}
\and G.~Polenta\orcid{0000-0003-4067-9196}\inst{\ref{aff86}}
\and M.~Poncet\inst{\ref{aff87}}
\and L.~A.~Popa\inst{\ref{aff88}}
\and L.~Pozzetti\orcid{0000-0001-7085-0412}\inst{\ref{aff14}}
\and F.~Raison\orcid{0000-0002-7819-6918}\inst{\ref{aff52}}
\and A.~Renzi\orcid{0000-0001-9856-1970}\inst{\ref{aff89},\ref{aff45},\ref{aff14}}
\and J.~Rhodes\orcid{0000-0002-4485-8549}\inst{\ref{aff57}}
\and G.~Riccio\inst{\ref{aff24}}
\and E.~Romelli\orcid{0000-0003-3069-9222}\inst{\ref{aff16}}
\and M.~Roncarelli\orcid{0000-0001-9587-7822}\inst{\ref{aff14}}
\and B.~Rusholme\orcid{0000-0001-7648-4142}\inst{\ref{aff90}}
\and R.~Saglia\orcid{0000-0003-0378-7032}\inst{\ref{aff54},\ref{aff52}}
\and Z.~Sakr\orcid{0000-0002-4823-3757}\inst{\ref{aff91},\ref{aff92},\ref{aff93}}
\and D.~Sapone\orcid{0000-0001-7089-4503}\inst{\ref{aff94}}
\and B.~Sartoris\orcid{0000-0003-1337-5269}\inst{\ref{aff54},\ref{aff16}}
\and P.~Schneider\orcid{0000-0001-8561-2679}\inst{\ref{aff69}}
\and T.~Schrabback\orcid{0000-0002-6987-7834}\inst{\ref{aff95}}
\and M.~Scodeggio\inst{\ref{aff28}}
\and A.~Secroun\orcid{0000-0003-0505-3710}\inst{\ref{aff46}}
\and E.~Sihvola\orcid{0000-0003-1804-7715}\inst{\ref{aff96}}
\and P.~Simon\inst{\ref{aff69}}
\and C.~Sirignano\orcid{0000-0002-0995-7146}\inst{\ref{aff89},\ref{aff45}}
\and G.~Sirri\orcid{0000-0003-2626-2853}\inst{\ref{aff20}}
\and L.~Stanco\orcid{0000-0002-9706-5104}\inst{\ref{aff45}}
\and P.~Tallada-Cresp\'{i}\orcid{0000-0002-1336-8328}\inst{\ref{aff29},\ref{aff30}}
\and A.~N.~Taylor\inst{\ref{aff35}}
\and H.~I.~Teplitz\orcid{0000-0002-7064-5424}\inst{\ref{aff97}}
\and I.~Tereno\orcid{0000-0002-4537-6218}\inst{\ref{aff98},\ref{aff99}}
\and N.~Tessore\orcid{0000-0002-9696-7931}\inst{\ref{aff42}}
\and S.~Toft\orcid{0000-0003-3631-7176}\inst{\ref{aff100},\ref{aff101}}
\and R.~Toledo-Moreo\orcid{0000-0002-2997-4859}\inst{\ref{aff102},\ref{aff103}}
\and F.~Torradeflot\orcid{0000-0003-1160-1517}\inst{\ref{aff30},\ref{aff29}}
\and I.~Tutusaus\orcid{0000-0002-3199-0399}\inst{\ref{aff49},\ref{aff48},\ref{aff92}}
\and J.~Valiviita\orcid{0000-0001-6225-3693}\inst{\ref{aff64},\ref{aff65}}
\and T.~Vassallo\orcid{0000-0001-6512-6358}\inst{\ref{aff16},\ref{aff51}}
\and A.~Veropalumbo\orcid{0000-0003-2387-1194}\inst{\ref{aff13},\ref{aff22},\ref{aff21}}
\and Y.~Wang\orcid{0000-0002-4749-2984}\inst{\ref{aff90}}
\and J.~Weller\orcid{0000-0002-8282-2010}\inst{\ref{aff54},\ref{aff52}}
\and A.~Zacchei\orcid{0000-0003-0396-1192}\inst{\ref{aff16},\ref{aff15}}
\and G.~Zamorani\orcid{0000-0002-2318-301X}\inst{\ref{aff14}}
\and F.~M.~Zerbi\orcid{0000-0002-9996-973X}\inst{\ref{aff13}}
\and I.~A.~Zinchenko\orcid{0000-0002-2944-2449}\inst{\ref{aff104}}
\and E.~Zucca\orcid{0000-0002-5845-8132}\inst{\ref{aff14}}
\and M.~Sereno\orcid{0000-0003-0302-0325}\inst{\ref{aff14},\ref{aff20}}}

\institute{Instituto de Astrof\'{\i}sica de Canarias, E-38205 La Laguna, Tenerife, Spain\label{aff1}
\and
Universidad de La Laguna, Dpto. Astrof\'\i sica, E-38206 La Laguna, Tenerife, Spain\label{aff2}
\and
Bernoulli Institute for Mathematics, Computer Science and Artificial Intelligence, University of Groningen, PO Box 407, 9700 AK Groningen, The Netherlands\label{aff3}
\and
Universit\'e PSL, Observatoire de Paris, Sorbonne Universit\'e, CNRS, LERMA, 75014, Paris, France\label{aff4}
\and
Universit\'e Paris-Cit\'e, 5 Rue Thomas Mann, 75013, Paris, France\label{aff5}
\and
Columbia Astrophysics Laboratory, Columbia University, 550 West 120th Street, New York, NY 10027, USA\label{aff6}
\and
Canadian Institute for Theoretical Astrophysics, Toronto, ON M5S 3H8, Canada\label{aff7}
\and
Jodrell Bank Centre for Astrophysics, Department of Physics and Astronomy, University of Manchester, Oxford Road, Manchester M13 9PL, UK\label{aff8}
\and
Universiteit Gent, Department of Physics and Astronomy, Proeftuinstraat 86 N3, 9000 Ghent, Belgium
\label{aff9}
\and
Department of Physics and Astronomy, University of British Columbia, Vancouver, BC V6T 1Z1, Canada\label{aff10}
\and
California Institute of Technology, 1200 E California Blvd, Pasadena, CA 91125, USA\label{aff11}
\and
ESAC/ESA, Camino Bajo del Castillo, s/n., Urb. Villafranca del Castillo, 28692 Villanueva de la Ca\~nada, Madrid, Spain\label{aff12}
\and
INAF-Osservatorio Astronomico di Brera, Via Brera 28, 20122 Milano, Italy\label{aff13}
\and
INAF-Osservatorio di Astrofisica e Scienza dello Spazio di Bologna, Via Piero Gobetti 93/3, 40129 Bologna, Italy\label{aff14}
\and
IFPU, Institute for Fundamental Physics of the Universe, via Beirut 2, 34151 Trieste, Italy\label{aff15}
\and
INAF-Osservatorio Astronomico di Trieste, Via G. B. Tiepolo 11, 34143 Trieste, Italy\label{aff16}
\and
INFN, Sezione di Trieste, Via Valerio 2, 34127 Trieste TS, Italy\label{aff17}
\and
SISSA, International School for Advanced Studies, Via Bonomea 265, 34136 Trieste TS, Italy\label{aff18}
\and
Dipartimento di Fisica e Astronomia, Universit\`a di Bologna, Via Gobetti 93/2, 40129 Bologna, Italy\label{aff19}
\and
INFN-Sezione di Bologna, Viale Berti Pichat 6/2, 40127 Bologna, Italy\label{aff20}
\and
Dipartimento di Fisica, Universit\`a di Genova, Via Dodecaneso 33, 16146, Genova, Italy\label{aff21}
\and
INFN-Sezione di Genova, Via Dodecaneso 33, 16146, Genova, Italy\label{aff22}
\and
Department of Physics "E. Pancini", University Federico II, Via Cinthia 6, 80126, Napoli, Italy\label{aff23}
\and
INAF-Osservatorio Astronomico di Capodimonte, Via Moiariello 16, 80131 Napoli, Italy\label{aff24}
\and
Dipartimento di Fisica, Universit\`a degli Studi di Torino, Via P. Giuria 1, 10125 Torino, Italy\label{aff25}
\and
INFN-Sezione di Torino, Via P. Giuria 1, 10125 Torino, Italy\label{aff26}
\and
INAF-Osservatorio Astrofisico di Torino, Via Osservatorio 20, 10025 Pino Torinese (TO), Italy\label{aff27}
\and
INAF-IASF Milano, Via Alfonso Corti 12, 20133 Milano, Italy\label{aff28}
\and
Centro de Investigaciones Energ\'eticas, Medioambientales y Tecnol\'ogicas (CIEMAT), Avenida Complutense 40, 28040 Madrid, Spain\label{aff29}
\and
Port d'Informaci\'{o} Cient\'{i}fica, Campus UAB, C. Albareda s/n, 08193 Bellaterra (Barcelona), Spain\label{aff30}
\and
INAF-Osservatorio Astronomico di Roma, Via Frascati 33, 00078 Monteporzio Catone, Italy\label{aff31}
\and
INFN -- Sezione di Napoli, Via Cinthia 6, 80126, Napoli, Italy\label{aff32}
\and
Institute for Astronomy, University of Hawaii, 2680 Woodlawn Drive, Honolulu, HI 96822, USA\label{aff33}
\and
Dipartimento di Fisica e Astronomia "Augusto Righi" - Alma Mater Studiorum Universit\`a di Bologna, Viale Berti Pichat 6/2, 40127 Bologna, Italy\label{aff34}
\and
Institute for Astronomy, University of Edinburgh, Royal Observatory, Blackford Hill, Edinburgh EH9 3HJ, UK\label{aff35}
\and
European Space Agency/ESRIN, Largo Galileo Galilei 1, 00044 Frascati, Roma, Italy\label{aff36}
\and
Universit\'e Claude Bernard Lyon 1, CNRS/IN2P3, IP2I Lyon, UMR 5822, Villeurbanne, F-69100, France\label{aff37}
\and
Institut de Ci\`{e}ncies del Cosmos (ICCUB), Universitat de Barcelona (IEEC-UB), Mart\'{i} i Franqu\`{e}s 1, 08028 Barcelona, Spain\label{aff38}
\and
Instituci\'o Catalana de Recerca i Estudis Avan\c{c}ats (ICREA), Passeig de Llu\'{\i}s Companys 23, 08010 Barcelona, Spain\label{aff39}
\and
Institut de Ciencies de l'Espai (IEEC-CSIC), Campus UAB, Carrer de Can Magrans, s/n Cerdanyola del Vall\'es, 08193 Barcelona, Spain\label{aff40}
\and
UCB Lyon 1, CNRS/IN2P3, IUF, IP2I Lyon, 4 rue Enrico Fermi, 69622 Villeurbanne, France\label{aff41}
\and
Mullard Space Science Laboratory, University College London, Holmbury St Mary, Dorking, Surrey RH5 6NT, UK\label{aff42}
\and
Department of Astronomy, University of Geneva, ch. d'Ecogia 16, 1290 Versoix, Switzerland\label{aff43}
\and
Universit\'e Paris-Saclay, CNRS, Institut d'astrophysique spatiale, 91405, Orsay, France\label{aff44}
\and
INFN-Padova, Via Marzolo 8, 35131 Padova, Italy\label{aff45}
\and
Aix-Marseille Universit\'e, CNRS/IN2P3, CPPM, Marseille, France\label{aff46}
\and
INAF-Istituto di Astrofisica e Planetologia Spaziali, via del Fosso del Cavaliere, 100, 00100 Roma, Italy\label{aff47}
\and
Institut d'Estudis Espacials de Catalunya (IEEC),  Edifici RDIT, Campus UPC, 08860 Castelldefels, Barcelona, Spain\label{aff48}
\and
Institute of Space Sciences (ICE, CSIC), Campus UAB, Carrer de Can Magrans, s/n, 08193 Barcelona, Spain\label{aff49}
\and
School of Physics, HH Wills Physics Laboratory, University of Bristol, Tyndall Avenue, Bristol, BS8 1TL, UK\label{aff50}
\and
University Observatory, LMU Faculty of Physics, Scheinerstr.~1, 81679 Munich, Germany\label{aff51}
\and
Max Planck Institute for Extraterrestrial Physics, Giessenbachstr. 1, 85748 Garching, Germany\label{aff52}
\and
INAF-Osservatorio Astronomico di Padova, Via dell'Osservatorio 5, 35122 Padova, Italy\label{aff53}
\and
Universit\"ats-Sternwarte M\"unchen, Fakult\"at f\"ur Physik, Ludwig-Maximilians-Universit\"at M\"unchen, Scheinerstr.~1, 81679 M\"unchen, Germany\label{aff54}
\and
National Research Council, Herzberg Astronomy and Astrophysics Research Centre, 5071 W. Saanich Rd. Victoria, BC, V9E 2E7, Canada\label{aff55}
\and
Institute of Theoretical Astrophysics, University of Oslo, P.O. Box 1029 Blindern, 0315 Oslo, Norway\label{aff56}
\and
Jet Propulsion Laboratory, California Institute of Technology, 4800 Oak Grove Drive, Pasadena, CA, 91109, USA\label{aff57}
\and
Felix Hormuth Engineering, Goethestr. 17, 69181 Leimen, Germany\label{aff58}
\and
Technical University of Denmark, Elektrovej 327, 2800 Kgs. Lyngby, Denmark\label{aff59}
\and
Cosmic Dawn Center (DAWN), Denmark\label{aff60}
\and
Max-Planck-Institut f\"ur Astronomie, K\"onigstuhl 17, 69117 Heidelberg, Germany\label{aff61}
\and
NASA Goddard Space Flight Center, Greenbelt, MD 20771, USA\label{aff62}
\and
Department of Physics and Astronomy, University College London, Gower Street, London WC1E 6BT, UK\label{aff63}
\and
Department of Physics, P.O. Box 64, University of Helsinki, 00014 Helsinki, Finland\label{aff64}
\and
Helsinki Institute of Physics, Gustaf H{\"a}llstr{\"o}min katu 2, University of Helsinki, 00014 Helsinki, Finland\label{aff65}
\and
Laboratoire d'etude de l'Univers et des phenomenes eXtremes, Observatoire de Paris, Universit\'e PSL, Sorbonne Universit\'e, CNRS, 92190 Meudon, France\label{aff66}
\and
SKAO, Jodrell Bank, Lower Withington, Macclesfield SK11 9FT, UK\label{aff67}
\and
Centre de Calcul de l'IN2P3/CNRS, 21 avenue Pierre de Coubertin 69627 Villeurbanne Cedex, France\label{aff68}
\and
Universit\"at Bonn, Argelander-Institut f\"ur Astronomie, Auf dem H\"ugel 71, 53121 Bonn, Germany\label{aff69}
\and
INFN-Sezione di Roma, Piazzale Aldo Moro, 2 - c/o Dipartimento di Fisica, Edificio G. Marconi, 00185 Roma, Italy\label{aff70}
\and
Aix-Marseille Universit\'e, CNRS, CNES, LAM, Marseille, France\label{aff71}
\and
Dipartimento di Fisica e Astronomia "Augusto Righi" - Alma Mater Studiorum Universit\`a di Bologna, via Piero Gobetti 93/2, 40129 Bologna, Italy\label{aff72}
\and
Department of Physics, Institute for Computational Cosmology, Durham University, South Road, Durham, DH1 3LE, UK\label{aff73}
\and
Universit\'e Paris Cit\'e, CNRS, Astroparticule et Cosmologie, 75013 Paris, France\label{aff74}
\and
CNRS-UCB International Research Laboratory, Centre Pierre Bin\'etruy, IRL2007, CPB-IN2P3, Berkeley, USA\label{aff75}
\and
University of Applied Sciences and Arts of Northwestern Switzerland, School of Engineering, 5210 Windisch, Switzerland\label{aff76}
\and
Institute of Physics, Laboratory of Astrophysics, Ecole Polytechnique F\'ed\'erale de Lausanne (EPFL), Observatoire de Sauverny, 1290 Versoix, Switzerland\label{aff77}
\and
Telespazio UK S.L. for European Space Agency (ESA), Camino bajo del Castillo, s/n, Urbanizacion Villafranca del Castillo, Villanueva de la Ca\~nada, 28692 Madrid, Spain\label{aff78}
\and
Institut de F\'{i}sica d'Altes Energies (IFAE), The Barcelona Institute of Science and Technology, Campus UAB, 08193 Bellaterra (Barcelona), Spain\label{aff79}
\and
European Space Agency/ESTEC, Keplerlaan 1, 2201 AZ Noordwijk, The Netherlands\label{aff80}
\and
School of Mathematics, Statistics and Physics, Newcastle University, Herschel Building, Newcastle-upon-Tyne, NE1 7RU, UK\label{aff81}
\and
DARK, Niels Bohr Institute, University of Copenhagen, Jagtvej 155, 2200 Copenhagen, Denmark\label{aff82}
\and
Waterloo Centre for Astrophysics, University of Waterloo, Waterloo, Ontario N2L 3G1, Canada\label{aff83}
\and
Department of Physics and Astronomy, University of Waterloo, Waterloo, Ontario N2L 3G1, Canada\label{aff84}
\and
Perimeter Institute for Theoretical Physics, Waterloo, Ontario N2L 2Y5, Canada\label{aff85}
\and
Space Science Data Center, Italian Space Agency, via del Politecnico snc, 00133 Roma, Italy\label{aff86}
\and
Centre National d'Etudes Spatiales -- Centre spatial de Toulouse, 18 avenue Edouard Belin, 31401 Toulouse Cedex 9, France\label{aff87}
\and
Institute of Space Science, Str. Atomistilor, nr. 409 M\u{a}gurele, Ilfov, 077125, Romania\label{aff88}
\and
Dipartimento di Fisica e Astronomia "G. Galilei", Universit\`a di Padova, Via Marzolo 8, 35131 Padova, Italy\label{aff89}
\and
Caltech/IPAC, 1200 E. California Blvd., Pasadena, CA 91125, USA\label{aff90}
\and
Instituto de F\'isica Te\'orica UAM-CSIC, Campus de Cantoblanco, 28049 Madrid, Spain\label{aff91}
\and
Institut de Recherche en Astrophysique et Plan\'etologie (IRAP), Universit\'e de Toulouse, CNRS, UPS, CNES, 14 Av. Edouard Belin, 31400 Toulouse, France\label{aff92}
\and
Universit\'e St Joseph; Faculty of Sciences, Beirut, Lebanon\label{aff93}
\and
Departamento de F\'isica, FCFM, Universidad de Chile, Blanco Encalada 2008, Santiago, Chile\label{aff94}
\and
Universit\"at Innsbruck, Institut f\"ur Astro- und Teilchenphysik, Technikerstr. 25/8, 6020 Innsbruck, Austria\label{aff95}
\and
Department of Physics and Helsinki Institute of Physics, Gustaf H\"allstr\"omin katu 2, University of Helsinki, 00014 Helsinki, Finland\label{aff96}
\and
Infrared Processing and Analysis Center, California Institute of Technology, Pasadena, CA 91125, USA\label{aff97}
\and
Departamento de F\'isica, Faculdade de Ci\^encias, Universidade de Lisboa, Edif\'icio C8, Campo Grande, PT1749-016 Lisboa, Portugal\label{aff98}
\and
Instituto de Astrof\'isica e Ci\^encias do Espa\c{c}o, Faculdade de Ci\^encias, Universidade de Lisboa, Tapada da Ajuda, 1349-018 Lisboa, Portugal\label{aff99}
\and
Cosmic Dawn Center (DAWN)\label{aff100}
\and
Niels Bohr Institute, University of Copenhagen, Jagtvej 128, 2200 Copenhagen, Denmark\label{aff101}
\and
Universidad Polit\'ecnica de Cartagena, Departamento de Electr\'onica y Tecnolog\'ia de Computadoras,  Plaza del Hospital 1, 30202 Cartagena, Spain\label{aff102}
\and
European University of Technology EUt+, European Union\label{aff103}
\and
Astronomisches Rechen-Institut, Zentrum f\"ur Astronomie der Universit\"at Heidelberg, M\"onchhofstr. 12-14, 69120 Heidelberg, Germany\label{aff104}}    

\date{\bf \today}

\abstract{
Understanding the intrinsic three-dimensional (3D) shapes of galaxies provides crucial insights into their formation and evolution, but distinguishing these from two-dimensional (2D) projected observations remains challenging. We applied a Bayesian inference framework to the optical imaging of galaxies in the Euclid Data Release 1, leveraging its larger sample sizes to test the scalability and robustness of this methodology. Applying Hamiltonian Monte Carlo (HMC) sampling to Sérsic axis ratios and effective radii, we constrained the population-level mean intrinsic 3D ellipticity, triaxiality, and size, together with their covariances. This analysis was performed for 40\,864\,304 star-forming galaxies with stellar masses $\Mstellar=10^{9}$--$10^{10.5}\Msun$ at redshifts $0<z<1.2$. Our sample size exceeds previous shape inference studies by factors of 440--52\,800 relative to early observations by JWST, and factors of 440--17\,600 relative to legacy observations from HST in overlapping stellar mass-redshift bins, enabling substantially finer binning alongside improved statistical constraints on shape evolution. We additionally included over 9 million galaxies at $z<0.5$, a regime historically limited by the sample size in previous shape inference studies. We recovered key evolutionary trends identified in previous smaller-sample studies. We find high mean ellipticities at the population level ($\mu_E \approx 0.600$--$0.821$) across all bins, indicating that the underlying 3D shape distributions are statistically dominated by flattened, non-spherical geometries. Most notably, we recovered the decline of the prolate fraction towards the present day at low masses ($\Mstellar=10^{9}$--$10^{9.5}\Msun$) reported by previous studies. This trend is seen consistently in comparable analyses of independent datasets spanning different depths and resolution regimes, showing that it is not an artefact of small-sample statistics, suggesting it may be a genuine feature of galaxy evolution. We also identified future methodological improvements for subsequent \Euclid data releases.
}

\titlerunning{3D shape inference of galaxy populations}
\authorrunning{M. Dunn et al.}

\maketitle
\nolinenumbers

\section{Introduction \label{sc:Introduction}}
The study of galaxy three-dimensional (3D) geometries through projected shapes spans nearly a century, and it remains crucial to understanding how the diverse morphologies seen today formed and evolved. Such studies assume that galaxies can be approximated as 3D ellipsoids classified by their degree of flattening in certain directions. \citet{Hubble_1926} demonstrated that nearby galaxies have projected ellipticities too round to be explained by randomly orientated disc galaxies alone, leading to his well-known classification system, and the introduction of the flattened `lenticular' galaxy class \citep{De_Vaucouleurs_1959, Sandage_1961}. Subsequent analyses incorporated projected axis ratios for late- and early-type galaxies \citep[e.g.][]{Sandage_1970, Binney_1978, Noerdlinger_1979}. This eventually led to more `triaxial' models to explain the kinematics of nearby elliptical galaxies \citep[e.g.][]{Contopoulos_1956, Stark_1977, Binney_1978}. 

Modern imaging has enabled statistical inference of intrinsic 3D galaxy shapes from projected measurements, revealing several robust trends in the local Universe \citep[e.g.][]{Lambas_1992, Alam_Ryden_2002, Ryden_2004, Vincent_Ryden_2005, Mendez-Abreu_2010, Costantin_2018}. In particular, intrinsic roundness increases with luminosity and stellar mass \citep{Padilla_Strauss_2008, RP_2013}; spiral discs are thicker and less circular than previously assumed, with the thickness increasing with luminosity \citep{RP_2013}; massive elliptical galaxies are more spherical than their low-mass counterparts, as confirmed by stellar kinematics \citep[e.g.][]{Binney_1985, Franx_1991, Ryden_1992, Tremblay_Merritt_1995, Emsellem_2011, Cappellari_2016}; and local dwarf galaxies span a wide range of morphologies, including both discs and spheroids \citep[e.g.][]{Caldwell_1983, Ichikawa_1986, Ferguson_Sandage_1989, Ferguson_Binggeli_1994, Sanchez-Janssen_2010, Kado-Fong_2021}.

\HST (HST) observations allowed the extension of intrinsic shape studies to massive galaxies at higher redshifts, showing that discs and spheroids are typically more compact and thick than at a low redshift \citep[e.g.][]{Elmegreen_2004A, Elmegreen_2004B, VanDerWel_2014b, Zhang_2022, Hamilton-Campos_2023}. Using projected axis ratios, \citet{Chang_2013} find that massive quiescent galaxies, $\Mstarwun=9.0$--$11.0$, at $1.0<z<2.5$ from the Cosmic Assembly Near-infrared Deep Extragalactic Legacy Survey \citep[CANDELS;][]{CANDELS1, CANDELS2} are more disc-like at a high redshift (before mergers destroyed those discs), whereas lower-mass galaxies become increasingly disc-dominated towards $z=0$. Similarly, \citet[hereafter \citetalias{VanDerWel_2014a}]{VanDerWel_2014a} show that star-forming galaxies with $\Mstellar>10^{10}\Msun$ remain disc-dominated up to $z\approx2$, whereas lower-mass galaxies exhibit rising fractions of elongated (prolate) shapes at higher redshifts, consistent with other observational studies \citep[e.g.][]{Cowie_1995, Ravindranath_2006, Law_2012, Yuma_2012}. Incorporating both projected sizes and axis ratios, \citet[hereafter \citetalias{Zhang_2019}]{Zhang_2019} also confirm a transition from predominantly elongated (prolate) populations at a high redshift to disc-dominated systems at later times. \citet{Hoffmann_2022} also find minimal evolution in disc properties in Cosmological Evolution Survey \citep[COSMOS;][]{COSMOS} galaxies, suggesting that most present-day discs formed early and evolved largely undisturbed.

Recently, \citet[hereafter \citetalias{Pandya_2024}]{Pandya_2024} demonstrate the power of differentiable Bayesian modelling with Hamiltonian Monte Carlo (HMC) sampling to constrain intrinsic 3D shapes from \JWST (JWST) Cosmic Evolution Early Release Science \citep[CEERS;][]{CEERS} observations of star-forming galaxies with $\Mstarwun=9.0$--$10.5$ at $0.5<z<8$, successfully reproducing earlier HST-CANDELS results at a dramatically reduced computational cost. They confirm the evolutionary trend identified in previous work in which elongated geometries dominate low-mass galaxies ($\Mstellar=10^{9}$--$10^{9.5}\Msun$) at a high redshift, declining from around 50--80\% at $3<z<8$ to 25\% at $0.5<z<1.0$. Their analysis further reveals a transition from triaxial (oval) to increasingly axisymmetric (circular) disc configurations over time. 

The HST-CANDELS and JWST-CEERS studies firmly establish the prolate-to-disc evolutionary trend, but they are limited by relatively small samples -- often only a few hundred galaxies per mass-redshift bin. Large-scale surveys such as \Euclid \citep{EuclidSkyOverview} provide an opportunity to test the robustness and scalability of the \citetalias{Pandya_2024} methodology on a far larger sample and in a different observational regime. Aimed at probing the evolution of the dark Universe, \Euclid is delivering high-quality imaging and spectroscopic data for millions of galaxies across a wide range of redshifts. Both its wide-field and deep-field surveys provide unprecedented statistical power to test whether 3D shape evolution trends persist across the broader galaxy population and explore dependencies on the environment and other physical properties.

For this study, we applied the Bayesian inference framework of \citetalias{Pandya_2024} to optical imaging from Data Release 1 \citep[DR1;][]{DR1cite, DR1-TP001}. We leveraged substantially larger samples to statistically infer the 3D shape distributions of star-forming galaxies with stellar masses $\Mstarwun=9.0$--$10.5$ at photometric redshifts $0<z<1.2$. In Sect.\,\ref{sc:Methodology}, we detail our methods. Section \ref{sc:Data} describes the \Euclid data sample used for this work. In Sect.\,\ref{sc:Results}, we present the inferred 3D shape distributions and their evolution with mass and redshift, and we visually inspected and examined the properties of galaxies in the sample. We discuss the implications of our findings in Sect.\,\ref{sc:Discussion} and summarise our conclusions in Sect.\,\ref{sc:Conclusions}. We emphasise that the primary goal of this work is to test the scalability and robustness of the \citetalias{Pandya_2024} inference framework on a vastly larger and observationally distinct sample, by applying the same methodology to DR1. A detailed astrophysical interpretation of the inferred shape distributions, together with efforts to address the methodological and observational limitations identified here -- in particular, the move beyond single-band optical imaging and single-Sérsic morphologies -- is deferred to a forthcoming paper. 

Throughout this paper, we use capital letters ($A$, $B$, and $C$) for intrinsic 3D axes, and lowercase letters ($a$ and $b$) for projected two-dimensional (2D) axes. We use $\logten(a)$ and $\logten(A)$ as shorthand for $\logten(a/\si{\kilo\parsec})$ and $\logten(A/{\rm \si{\kilo\parsec}})$, following the convention of \citetalias{Zhang_2019} and \citetalias{Pandya_2024}. We assumed a flat Lambda cold dark matter ($\Lambda$CDM) cosmology with $H_0 = \SI{67.74}{\kmsMpc}$, $\Omm = 0.3089$, and $\OmLa = 0.6911$ \citep{2016A&A...594A..13P}. Magnitudes are given in the AB system \citep{Oke_1974}.

\section{Methodology \label{sc:Methodology}}
In this section, we describe how galaxies are modelled as 3D ellipsoids, how populations of intrinsic shapes are projected onto the observed plane, and the population-level Bayesian inference framework of \citetalias{Pandya_2024} that we adopt. As in previous studies \citepalias[e.g.][]{VanDerWel_2014a,Zhang_2019,Pandya_2024}, we approximated galaxies as triaxial ellipsoids with intrinsic semi-axes $A \geq B \geq C$, parameterised by the length of the semi-major axis $A$, the ellipticity $E$, defined as
\begin{equation}\label{eq:ellipticity}
    E = 1 - C/A\;,
\end{equation}
and triaxiality $T$, defined as
\begin{equation}\label{eq:triaxiality}
    T = \frac{A^{\rm 2} - B^{\rm 2}}{A^{\rm 2} - C^{\rm 2}}\;.
\end{equation}
Using these definitions, ellipsoids can be classified into three general geometries: (1) `spheroidal' ($A\sim B\sim C$ and $E\sim0$), appearing round from any viewing angle; (2) `oblate' ($B\sim A$, $E\sim1$, and $T\rightarrow0$), flattened in one direction, appearing as flat discs when viewed `edge-on' and round when viewed face-on; or (3) `prolate' ($B\sim C$, $E\sim1$, and $T\rightarrow1$), flattened in two directions, appearing as a `cigar-like' shape when viewed edge-on. There are also intermediate cases where all three axes differ greatly ($0<T<1$) known as `triaxial ellipsoids' or non-axisymmetric discs, appearing as flattened oval discs \citepalias{Pandya_2024}. Given this, we expect triaxiality $T$ (Eq.\,\ref{eq:triaxiality}) to become increasingly ill-constrained as ellipticity $E$ (Eq.\,\ref{eq:ellipticity}) approaches zero, since both the numerator and denominator of $T$ vanish for nearly spherical ellipsoids. Figure 5 of \citetalias{Pandya_2024} illustrates the different projections and cross-sections of these 3D ellipsoids, and how their populations appear projected in 2D space.

We adopted the classification boundaries of figure 1 in \citetalias{Zhang_2019} and figure 2 in \citetalias{VanDerWel_2014a}, classifying galaxies by their location in intrinsic 3D axis-ratio space ($C/A$--$B/A$), where $C/A$ and $B/A$ are obtained by rearranging Eqs.\,(\ref{eq:ellipticity}) and (\ref{eq:triaxiality}). The triaxiality parameter $T$ is essential for distinguishing oblate from prolate populations, since both have large ellipticities and similar edge-on appearance. We focus on identifying the dominant intrinsic shape distributions (specifically spheroidal, oblate, or prolate) for galaxy populations in each bin of stellar mass and redshift. Imaging data alone enables us to constrain the intrinsic shape distribution of a population, but it is not enough to recover the intrinsic shapes of individual galaxies in bins where ellipsoid types are ambiguously mixed -- such cases would require additional data (e.g. kinematic measurements).

We used the empirical library of \citetalias{Zhang_2019} and \citetalias{Pandya_2024}, comprising 1 million 3D ellipsoids spanning a uniform grid in $E$--$T$--$\logten(A)$ space, each characterised by an intrinsic shape vector $\vec{\theta} = (E, T, \logten A)$ and observed from \num{100000} simulated viewing angles. Projecting these ellipsoids and marginalising them over all viewing angles \citep{Binney_1985, VDV_2021} produces population-level probability distribution maps in projected $b/a$ versus $\logten(a)$, binned on a $20\times20$ grid. The library further incorporates realistic observational uncertainties: a Rice distribution \citep[][appendix C]{Rix_Zaritsky_1995} scatters each ellipsoid's projected axis ratio assuming $\Delta(b/a) = 0.04$, the resulting perturbation is propagated into the projected size, and $\logten(a)$ is additionally smeared by a Gaussian of width $0.03\,{\rm dex}$ \citepalias[][section 3.1]{Pandya_2024}.

Following \citetalias{Pandya_2024}, we assumed that each ellipsoid's intrinsic shape vector is drawn from a trivariate Gaussian distribution with an unknown mean shape vector $\vec{\mu}=(\mu_E,\mu_T,\mu_{\logten A})$ and an unknown covariance matrix \citepalias[][equation 4]{Pandya_2024}, characterised by standard deviations $\sigma_E$, $\sigma_T$, and $\sigma_{\logten(A)}$, and a correlation $\rho_{E,\logten(A)}$ between 3D ellipticity and size. Central to this approach is the assumption that, within a given bin of stellar mass and redshift, all galaxies are 2D projections of a single parent 3D shape distribution, described by this one trivariate Gaussian. The inference thus constrains the parameters of that parent distribution rather than classify individual galaxies, whose projected shapes are intrinsically degenerate. We discuss the validity of the single-distribution assumption in Sect.\,\ref{ssc:Discussion-Limitations}. These seven population-level parameters constitute the shape parameters we infer for each mass–redshift bin in our sample.

We performed Bayesian parameter estimation following \citetalias{Pandya_2024} using the \texttt{PyMC} probabilistic programming framework \citep{PyMC} with the No U-Turn Sampler \citep[NUTS;][]{NUTS}. We used weakly informative uniform priors for all seven shape parameters, only imposing physical constraints while allowing the posterior distributions to be driven primarily by the observed data. A visualisation of the full Bayesian model used can be found in figure 6 of \citetalias{Pandya_2024}. We evaluated the Poisson likelihood between observed and model 2D $b/a$--$\logten(a)$ histograms; only cells populated by the observed galaxies above the minimum-size cut (Sect.\,\ref{ssc:MER_selection}) and within $\logten(a)\in[-1,1]$ were considered, and the model histogram is normalised over these cells. We ran four parallel sampling chains, with 5000 tuning and 2000 sampling iterations, and a target acceptance rate of 0.95, for all bins. We assessed convergence using standard NUTS diagnostic statistics, such as $\hat{R}$ \citep{Rhat}, effective sample size \citep[ESS;][]{Gelman_2013, Gong_Flegal_2016, Vats_2019}, Monte Carlo standard error (MCSE) in the Markov chain Monte Carlo \citep[MCMC;][]{Flegal_Jones_2010, Flegal_Jones_2011}, and Bayesian fraction of missing information \citep[BFMI;][]{BFMI1}. Together, these diagnostics assessed sampler performance, probing complementary aspects of convergence, sampling efficiency, and goodness-of-fit, providing confidence that the posteriors are both well-sampled and consistent with the observed distributions.

\section{Data sample \label{sc:Data}}

\subsection{DR1 catalogue selection \label{ssc:MER_selection}}
We used DR1 optical Visible Instrument (VIS) imaging \citep[\IE;][]{EuclidSkyVIS, DR1-TP002} from the Euclid Wide Survey \citep[EWS;][]{Scaramella-EP1, DR1-TP027}, with approximately $1900\,\mathrm{deg}^2$ of sky coverage. More detailed information on DR1, the \Euclid observing strategy, and VIS processing and data products can be found in \citet{DR1-TP001}, \citet{Scaramella-EP1, DR1-TP027}, and \citet{DR1-TP002}. We constructed our parent sample from catalogues built with the MER \citep{DR1-TP009} and PHZ \citep{DR1-TP011} processing functions (PFs). Both projected sizes and axis ratios for galaxies come from elliptical 2D Sérsic profile fits \citep{Sersic_1963} of \Euclid VIS images using \texttt{SourceXtractor++} \citep[\texttt{SE++};][]{SE++1, SE++2} from MER, as described in \citet{Q1-SP040}. These parameters are point spread function (PSF)-corrected through forward-modelling during Sérsic fitting. We defined a galaxy's projected size $a$ as its angular Sérsic effective (half-light) semi-major axis (in arcseconds), following the convention of \citet{VanDerWel_2014b}, and the projected axis ratio $b/a$ as the Sérsic axis ratio. We used photometric redshifts, stellar masses, star-formation rates (SFRs), and rest-frame magnitudes from the PHZ PF \citep{DR1-TP012}, adopting the medians of their posterior probability distributions as point estimates. We did not directly propagate the photometric-redshift uncertainties into the inference. Compared with spectroscopic redshifts, \citet{DR1-TP012} report an overall normalised median absolute deviation $\sigma_{\rm NMAD}$ of approximately 0.021 and an outlier fraction $f_{\rm out}$ of 3.7\% for EWS-like processing of Euclid Deep Field North (EDF-N) for the Northern hemisphere (EWS-N), and $\sigma_{\rm NMAD}\approx0.027$ and $f_{\rm out}\approx7.6\%$ for Euclid Deep Field Fornax (EDF-F) for the Southern hemisphere (EWS-S). These degrade to $\sigma_{\rm NMAD}\approx0.034$--$0.048$ and $f_{\rm out}\approx9.4$--$16.2\%$ near the faint end of our magnitude range at $23<\IE\leq25$. This effect on physical sizes is quantified in Sect.\,\ref{ssc:Final_Sample}.

After building a magnitude-limited sample from these catalogues, we applied several quality selection cuts recommended in \citet{Q1-SP040} and \citet{DR1-TP012} to ensure reliable measurements and physical parameters. We also included several of our own selection cuts, such as excluding sources with fractional uncertainties $>\,10\%$ in either the Sérsic effective radius or the axis ratio values (see Appendix \ref{app:sersic_param_comparison} for a comparison of these uncertainties against independent HST measurements). We also excluded sources with projected sizes smaller than the PSF full width at half maximum (FWHM) for the VIS filter \citep[\ang{;;0.16};][]{Scaramella-EP1}. Our model can in principle recover smaller intrinsic sizes, but this limit ensures that resolved galaxies dominate the analysis. As in \citetalias{Zhang_2019} and \citetalias{Pandya_2024}, this cut primarily affects more massive bins and those at higher redshifts. Lastly, we restricted the analysis to galaxies with stellar masses $\Mstellar=10^{9}$--$10^{10.5}\Msun$ at photometric redshifts $0<z<1.2$, and excluded sources with missing or non-finite Sérsic, stellar mass, or redshift values. The full list of these selection cuts is detailed in Appendix \ref{app:DR1_Selection_Cuts}.

\subsection{sSFR selection \label{ssc:sSFR_selection}}
Although galaxy populations within a given bin of stellar mass and redshift span a range of intrinsic geometries, our methodology models each population as a mixture of predominantly spheroidal, oblate, and prolate shapes. To maximise population homogeneity, we restricted our analysis to star-forming galaxies, since quiescent galaxies not only differ in their 3D shape distributions \citep{Chang_2013}, but are harder to detect and grow increasingly rare at higher redshifts \citep[e.g.][]{Pandya_2017}. In principle, the same analysis could be applied to purely quiescent populations instead. Following \citet{Pandya_2017}, we computed the median specific star formation rate (sSFR) of the zero-slope star-forming main sequence (SFMS) in each redshift bin, adopting a large SFMS scatter of $0.45\,{\rm dex}$ ($1\,\sigma$) for all redshift slices \citepalias{Pandya_2024}. We then classified galaxies in each redshift bin by their offset below the median SFMS: (1) `star-forming', offset by less than $0.45\,{\rm dex}$ ($1\,\sigma$) below the SFMS; (2) `transition', offset by $0.675$--$1.575\,{\rm dex}$ ($1.5$--$3.5\,\sigma$) below the SFMS; and (3) `quiescent', offset by more than $1.575\,{\rm dex}$ ($3.5\,\sigma$) below the SFMS. This selection is shown in the sSFR--$\Mstellar$ diagrams for each redshift slice in Fig.\,\ref{fig:SFMS_diagram_app}.

\subsection{Colour-based selection \label{ssc:Colour_selection}}
To further mitigate contamination by quiescent galaxies, we also applied a colour-based galaxy classification. Following \citet[][appendix A]{Pandya_2017}, we constructed a ($U-V$) versus ($V-J$) colour diagram \citep{Van_Dokkum_2015} for the sSFR-classified sample in each redshift slice, using rest-frame magnitudes and the optical dust attenuation $A_V$ derived from spectral energy distribution (SED) fitting by \citet{DR1-TP012}. As shown in Fig.\,\ref{fig:UVJ_Av_diagram_app}, points are colour-coded by $A_V$ and organised into three columns by the sSFR-based classification described above, with each column containing a `quiescent wedge' that is defined in \citet[][equation 3]{Van_Dokkum_2015} as galaxies that satisfy the relations
\begin{equation} \label{eq:UVJ_quiescent}
    \begin{aligned}
        (V - J) &< 1.5\;, \\
        (U - V) &> 1.3\;, \\
        (U - V) &> 0.8\,(V - J) + 0.7\;.
    \end{aligned}
\end{equation}
Excluding transition and quiescent galaxies and star-forming galaxies in the quiescent wedges, the sample size became 104\,540\,419.

\subsection{Final sample\label{ssc:Final_Sample}}
\begin{figure*}[htp!]
\centering
\includegraphics[width=0.8\hsize]{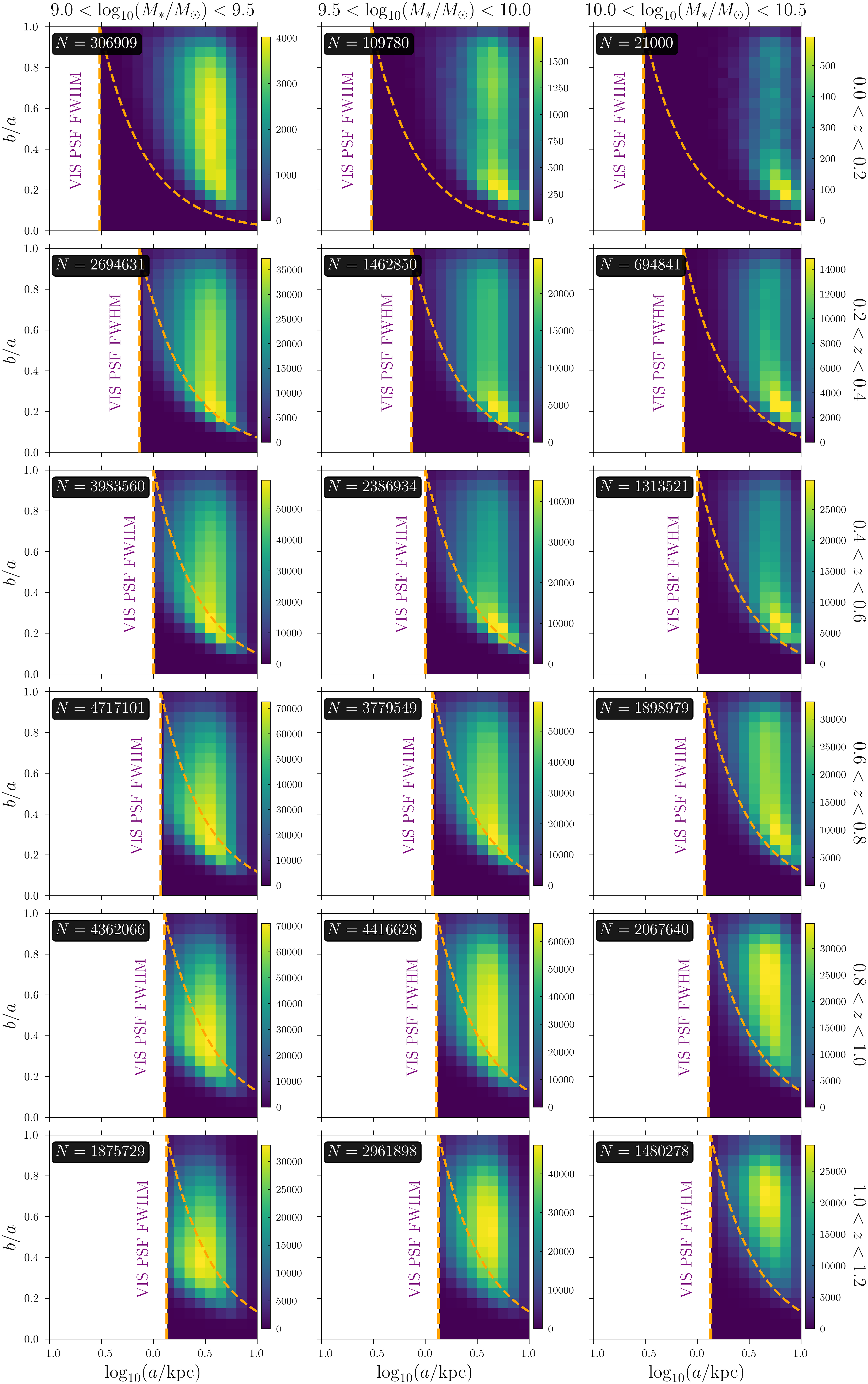}
\caption{Projected $b/a$ versus $\logten(a)$ histograms for each mass-redshift bin. The dashed orange line indicates the VIS PSF FWHM converted to units of kiloparsec at the redshift midpoint; the curved portion assumes $b=$\,PSF FWHM. Galaxy counts are shown in the top left of each panel.}
\label{fig:euclid_2Dhists_PSF_cut}
\end{figure*}

After applying the quality cuts described in Sect.\,\ref{ssc:MER_selection} to our star-forming sample, the resulting sample size is \num{70498316}. Our final sample consists of \num{40864304} star-forming galaxies within the stellar mass range $\Mstellar=10^{9}$--$10^{10.5}\Msun$ at photometric redshifts $0<z<1.2$, with a limiting magnitude of $\IE\leq24.5$. We binned galaxies by $\Delta \Mstarwun=0.5$ and $\Delta z=0.2$. We converted the projected sizes $a$ from arcseconds to physical units of kiloparsec in each redshift bin. Photometric-redshift uncertainties propagate into the physical sizes through the angular-diameter distance. We estimated this contribution by redrawing each galaxy's redshift from its 68\% credible interval estimated by the PHZ PF and recomputing $\log_{10}(a/{\rm kpc})$. The resulting median scatter at all redshift intervals was at or below the $0.03\,{\rm dex}$ size smearing already included in the model library (Sect.\,\ref{sc:Methodology}), so photometric-redshift uncertainties did not significantly affect the inferred shape parameters. The 2D projected $b/a$--$\logten(a)$ histograms for $b/a\in[0,1]$ and $\logten(a)\in[-1,1]$ (corresponding to $a=0.1$--$10\,\si{\kilo\parsec}$) for each \Euclid mass-redshift bin are shown in Fig.\,\ref{fig:euclid_2Dhists_PSF_cut}, with the galaxy counts indicated in each panel. These distributions serve as input to our shape inference model.

We adopted a limiting magnitude of $\IE\leq24.5$ to balance the sample size against reliable Sérsic measurements and completeness. In Appendix \ref{app:sersic_param_comparison}, we assessed the reliability of the Sérsic measurements themselves by comparing single-visit EDF-F DR1 measurements at EWS depth against independent HST-based measurements in the Galaxy Evolution from Morphology and SEDs \citep[GEMS;][]{GEMS} field (Figs.\,\ref{fig:sersic_params_comparison} and \ref{fig:sersic_error_comparison}), finding behaviour consistent with the equivalent Quick Data Release 1 (Q1) comparison of \citet{Q1-SP040}. In Appendix \ref{app:Completeness}, we assessed the detection fraction using single-Sérsic fits with \texttt{SourceXtractor++} on mock sources from the \Euclid Morphology Challenge \citep{Merlin-EP25, Bretonniere-EP26}. Morphological completeness remains high across the $b/a$--$\mathrm{size}$ plane for our magnitude-limited sample, including for large face-on discs that are most vulnerable to surface-brightness biases (Fig.\,\ref{fig:completeness-EMC}). We also assessed the stellar mass completeness of our magnitude-limited sample. Future \Euclid data releases are expected to improve mass completeness through a larger sample size, allow wider mass and redshift ranges, and allow finer binning.

\section{Results \label{sc:Results}}
In this section, we present constraints on the 3D shapes of DR1 star-forming galaxies with $\Mstellar=10^{9}$--$10^{10.5}\Msun$ at $0<z<1.2$. Table \ref{tab:3D_shape_results} summarises the inference results for each mass-redshift bin based on random draws from the posterior. The convergence diagnostic statistics for each bin can be found in Table \ref{tab:bin_conv_diag}. Across all mass-redshift bins, we find $\hat{R}<1.01$ for all sampling chains, no divergent transitions, no low BFMI values, and sufficiently large effective sample sizes, together indicating good convergence and fit \citep{Rhat, BFMI1}. The observed and posterior-mean model histograms, along with their residuals, are provided for each bin in Appendix \ref{app:Residuals}.

We examine the evolution of inferred 3D ellipticity (Sect.\,\ref{ssc:Results-Ellipticity_Evolution}), triaxiality (Sect.\,\ref{ssc:Results-Triaxiality_Evolution}), and size (Sect.\,\ref{ssc:Results-Size_Evolution}), as well as axis ratios (Sect.\,\ref{ssc:Results-Pred_3D_Axis_Ratios}) and class fractions (Sect.\,\ref{ssc:Results-Pred_3D_Fractions}). Figures \ref{fig:3DE_evolution}--\ref{fig:3DlogArho_evolution} and \ref{fig:3Dfrac_evolution_random_draws} specifically show how the inferred shape parameters and class fractions evolve with mass and redshift in this work compared to the JWST-CEERS results from table 2 in \citetalias{Pandya_2024}, and the HST-CANDELS results from table 1 in \citetalias{Zhang_2019} (accounting for variance and covariance conversions). Because our redshift bins are finer than those of \citetalias{Pandya_2024} and \citetalias{Zhang_2019}, and our inference is limited to $0<z<1.2$, our bins do not align exactly with theirs. We therefore restricted the comparisons to our $0.6<z<1.2$ range, which overlaps with their $0.5<z<1.5$ range. There are no uncertainties available for the $\Mstellar=10^{10}$--$10^{10.5}\Msun$ bins for \citetalias{Zhang_2019} due to their bootstrap algorithm giving unrealistic uncertainties given the small sample size \citepalias[][section 4.3]{Zhang_2019}. In Appendix \ref{app:Visualisation}, we computed 3D shape class probabilities of individual galaxies from observed $b/a$ and $\logten(a)$ values, and show several example cutouts for each class only for illustrative purposes (Fig.\,\ref{fig:cutouts}).

\subsection{Evolution of 3D ellipticity \label{ssc:Results-Ellipticity_Evolution}}
\begin{figure*}[!t]
\sidecaption
\includegraphics[width=12cm]{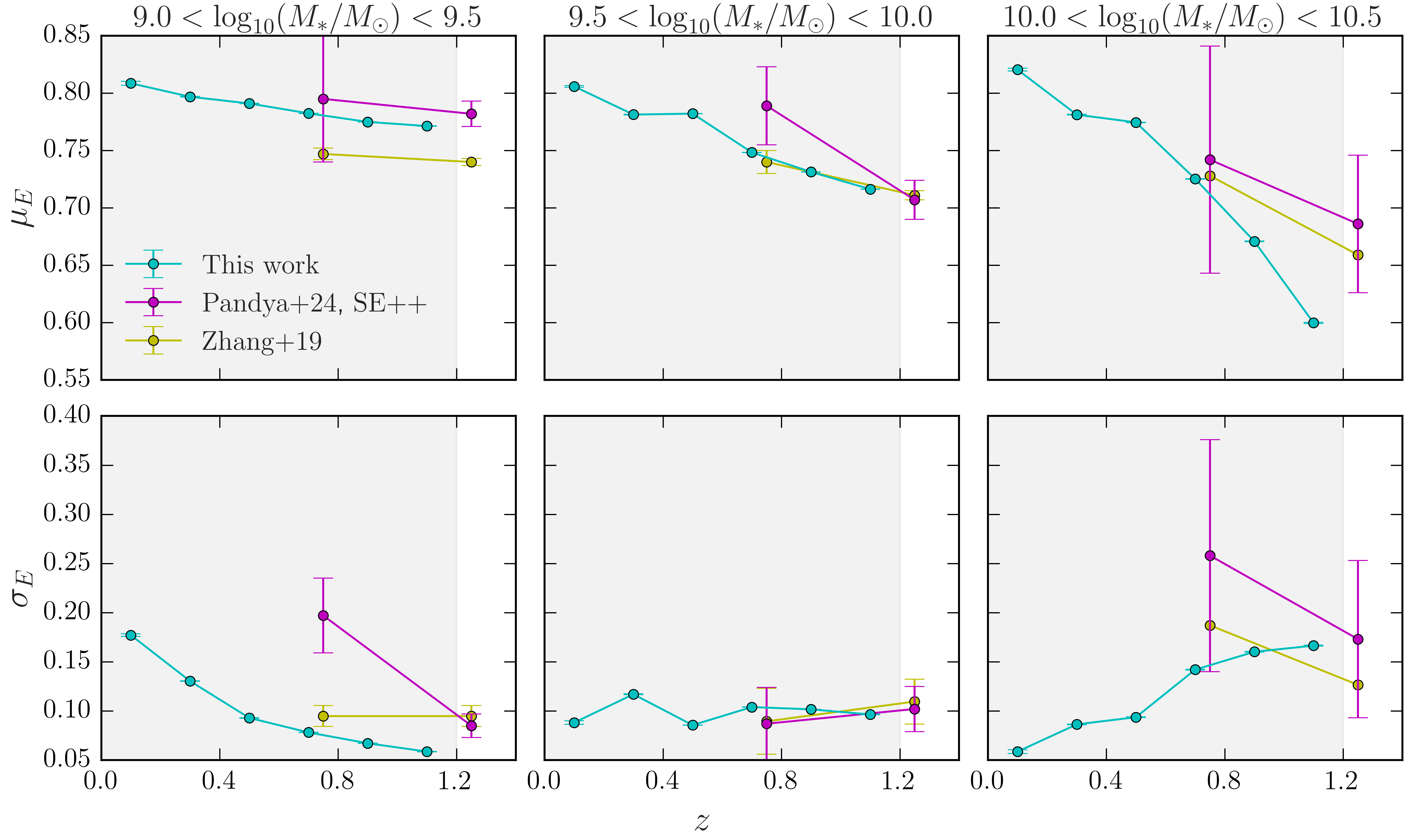}
\caption{Comparison of inferred 3D ellipticity evolution with redshift (left to right) and stellar mass: mean ellipticity ($\mu_E$; top row) and ellipticity dispersion ($\sigma_E$; bottom row) in this work (cyan), JWST-CEERS \citetalias{Pandya_2024} (magenta), and HST-CANDELS \citetalias{Zhang_2019} (gold). The shaded region marks this work's redshift coverage ($z \le 1.2$); points outside it are \citetalias{Pandya_2024} and \citetalias{Zhang_2019}'s overlapping $1 < z < 1.5$ bin, which overlaps our $0.8 < z < 1.2$ bin. Points show posterior means, with error bars from random draws; \citetalias{Zhang_2019} uncertainties are unavailable for the high-mass bins due to small sample size.}
\label{fig:3DE_evolution}
\end{figure*}

In Fig.\,\ref{fig:3DE_evolution}, we show the redshift evolution of mean ellipticity ($\mu_E$; top row) and ellipticity dispersion ($\sigma_E$; bottom row) across the three stellar mass bins for $0<z<1.2$, compared to \citetalias{Zhang_2019} and \citetalias{Pandya_2024}, evaluated at the midpoint of each redshift bin. Similarly to \citetalias{Zhang_2019} and \citetalias{Pandya_2024}, we find high mean ellipticities ($\mu_E=0.600\pm0.001$--$0.821\pm0.001$) across all mass bins, indicating that most galaxies in the sample are intrinsically non-spherical. Our intrinsic scatter $\sigma_E$ values are also consistent with both studies, although we find slightly lower values than \citetalias{Pandya_2024} at $z\sim0.75$ for low-mass ($\Mstellar=10^{9}$--$10^{9.5}\Msun$) and high-mass ($\Mstellar=10^{10}$--$10^{10.5}\Msun$) bins. Across all comparisons, our uncertainties are substantially smaller, reflecting our larger sample: those for $\mu_E$ are lower than \citetalias{Pandya_2024} by a factor of approximately 58--100 and \citetalias{Zhang_2019} by 6--27, while those for $\sigma_E$ are lower by 60--96 and 17--84, respectively. Our additional bins at $0.0 < z<0.2$ and $0.2<z<0.4$, which extend below the redshift range probed by these earlier studies, show that towards the present day, $\mu_E$ continues to rise just above $0.8$ across all mass bins, whereas the dispersion increases at lower masses.

\subsection{Evolution of 3D triaxiality \label{ssc:Results-Triaxiality_Evolution}}
\begin{figure*}[!t]
\sidecaption
\includegraphics[width=12cm]{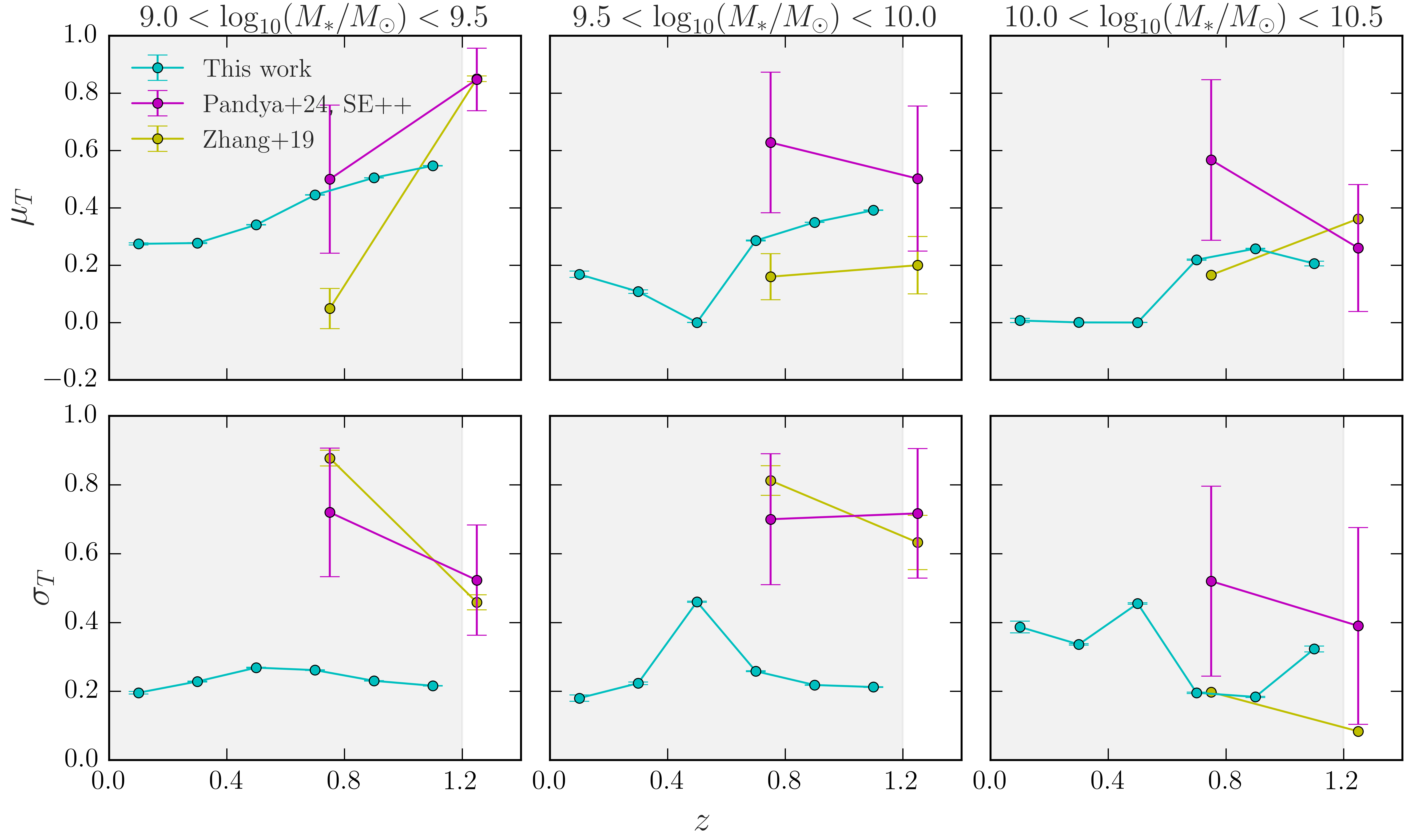}
\caption{Same as Fig.\,\ref{fig:3DE_evolution} but for the mean triaxiality ($\mu_T$; top row) and triaxiality dispersion ($\sigma_T$; bottom row).}
\label{fig:3DT_evolution}
\end{figure*}

The top and bottom rows of Fig.\,\ref{fig:3DT_evolution} show the redshift evolution of the mean triaxiality $\mu_T$ and its standard deviation $\sigma_T$. We find that in all three mass bins, $\mu_T$ decreases towards the present day, and at fixed redshift is higher in low-mass galaxies. The results of \citetalias{Pandya_2024} show systematically higher $\mu_T$ across all mass bins, decreasing over time for low-mass galaxies but rising over time for intermediate- and high-mass galaxies. The values of $\mu_T$ reported by \citetalias{Zhang_2019} are comparable to those of \citetalias{Pandya_2024} at $1.0<z<1.5$ for the low- and high-mass bins, but lower than both our work and \citetalias{Pandya_2024} at $0.5<z<1.0$ for all bins. Comparisons are more difficult to make quantitatively for triaxiality than for ellipticity, since the uncertainties in $\mu_T$ are much larger in previous studies (by a factor of approximately 25--332 in \citetalias{Pandya_2024} and 9--30 in \citetalias{Zhang_2019}), as are the uncertainties in $\sigma_T$. Both studies find substantially higher $\sigma_T$ values than our work in the low- and intermediate-mass bins. At high mass, \citetalias{Pandya_2024} again show slightly higher dispersion than our work at $z\sim0.75$, whereas \citetalias{Zhang_2019} is more comparable to ours. Both \citetalias{Zhang_2019} and \citetalias{Pandya_2024} show that $\sigma_T$ increases from $1.0<z<1.5$ to $0.5<z<1.0$, whereas our finer redshift bins show that the trend is less constant. We note that \citetalias{Pandya_2024} break the oblate-prolate degeneracy most effectively for low-mass galaxies at $z > 1$, where they infer $\mu_T \sim 1$ (favouring prolate geometries); as our sample extends only to $z < 1.2$, we do not reach the epoch in which this high-triaxiality, prolate-dominated population dominates. 

In the intermediate- and high-mass bins, our results show a dip in $\mu_T$ accompanied by a rise in $\sigma_T$ around $0.4 < z < 0.6$, with the high-mass bin retaining small $\mu_T$ for $z < 0.4$. Both parameters remain well constrained here and convergence is good throughout (Table \ref{tab:bin_conv_diag}), so this does not indicate a sampling artefact. In these bins $\mu_T$ is pinned against the $T = 0$ boundary and $\sigma_T$ remains large and tightly determined, while in the strongly oblate bins at lower redshift $\mu_T$ settles just above zero with a correspondingly small $\sigma_T$. The interpretation of this behaviour, and its bearing on the reliability of the inferred triaxialities in these bins, is discussed in Sect.\,\ref{ssc:Discussion-Limitations}.

\subsection{Evolution of 3D size \label{ssc:Results-Size_Evolution}}
\begin{figure*}[!t]
\sidecaption
\includegraphics[width=12cm]{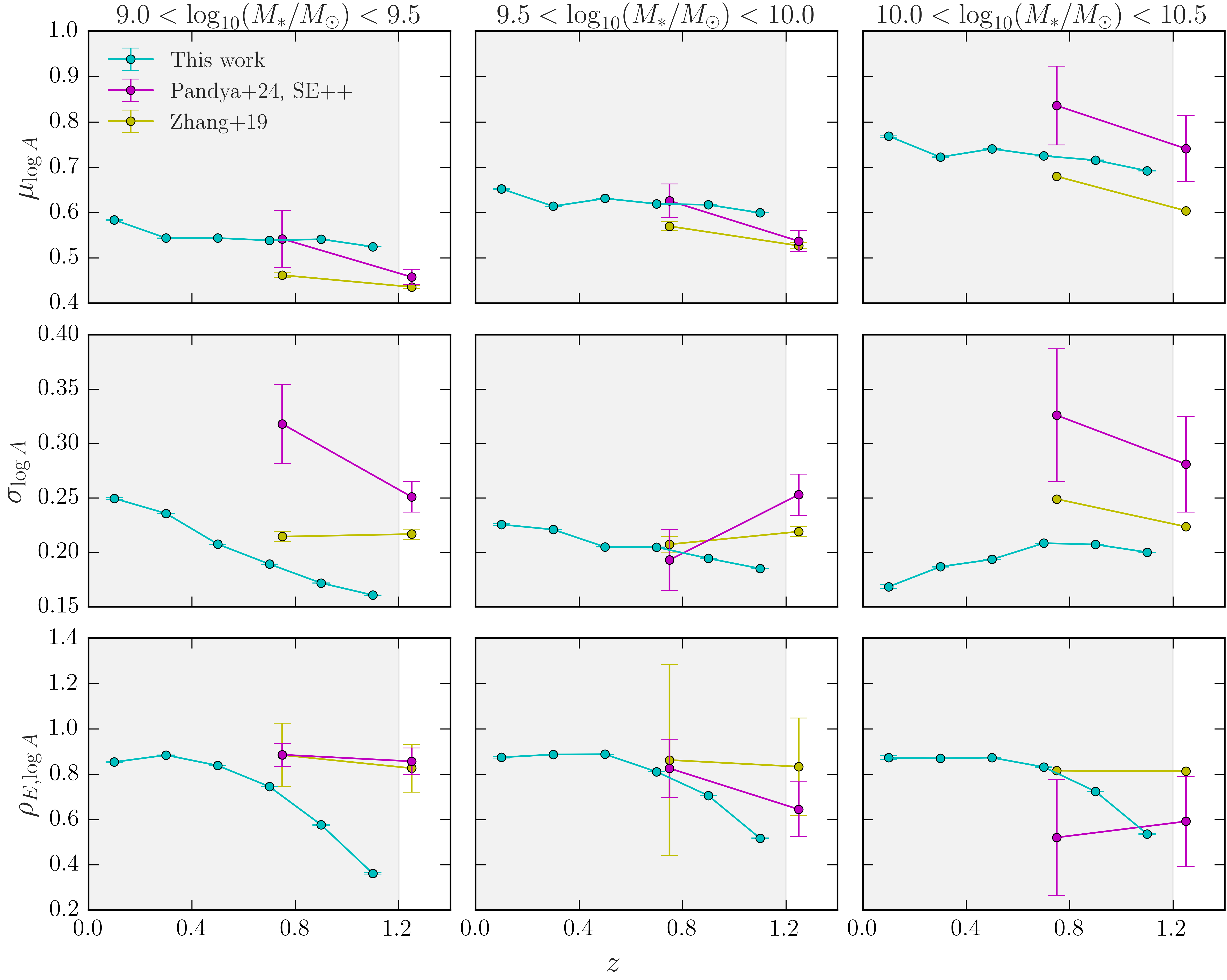}
\caption{Same as Fig.\,\ref{fig:3DE_evolution} but for the mean 3D size ($\mu_{\logten(A)}$; top row), size dispersion ($\sigma_{\logten(A)}$; middle row), and ellipticity--size correlation ($\rho_{E,\logten(A)}$; bottom row).}
\label{fig:3DlogArho_evolution}
\end{figure*}

The top and middle rows of Fig.\,\ref{fig:3DlogArho_evolution} show that the mean 3D size $\mu_{\logten(A)}$ follows the expected mass and redshift scaling, as in \citetalias{Zhang_2019} and \citetalias{Pandya_2024}: $\mu_{\logten(A)}$ increases moderately towards the present day, with our additional $0.0 < z<0.2$ and $0.2<z<0.4$ bins extending this trend to lower redshifts. Our values broadly agree with \citetalias{Pandya_2024} and \citetalias{Zhang_2019} in the low- and intermediate-mass bins; at higher masses, \citetalias{Pandya_2024} have values slightly higher than ours, while \citetalias{Zhang_2019} have values slightly lower. Our size dispersion $\sigma_{\logten(A)}$ values grow slightly towards the present day as mass decreases. In contrast, \citetalias{Zhang_2019} find little variation with redshift or mass, and \citetalias{Pandya_2024} report typically higher values that increase over time for low- and high-mass bins, but decrease at intermediate masses. As before, our uncertainties are markedly smaller: those for $\mu_{\logten(A)}$ are lower than \citetalias{Pandya_2024} by a factor of approximately 40--162 and \citetalias{Zhang_2019} by 5--29, and those for $\sigma_{\logten(A)}$ are lower by 34--172 and 4--56, respectively.

Our $\mu_{\logten(A)}$ values also follow the mild size evolution measured for late-type galaxies by \citet{VanDerWel_2014b}, who find $R_{\rm e} \propto (1+z)^{-0.75}$ at fixed stellar mass, and our $\sigma_{\logten(A)}$ values are of the same order as the intrinsic scatter of $\approx\,0.16$--$0.19\,{\rm dex}$ they measure in $\logten(R_{\rm e}/\si{\kpc})$ at fixed mass, which is nearly independent of both mass and redshift (noting that their scatter refers to projected sizes at fixed mass rather than intrinsic 3D sizes within a mass bin).

The bottom row of Fig.\,\ref{fig:3DlogArho_evolution} shows the evolution of $\rho_{E, \logten(A)}$, the correlation between 3D ellipticity and size. We find with small uncertainties that $\rho_{E, \logten(A)}$ increases towards the present day in all mass bins, reaching present-day values of $0.854\pm0.002$--$0.875\pm0.003$. In  \citetalias{Zhang_2019}, the correlation is roughly constant for all mass-redshift bins with values similar to our present-day ones (though with substantial uncertainties, particularly at intermediate masses), while \citetalias{Pandya_2024} find that this correlation becomes slightly stronger over time as mass decreases (with larger uncertainties at higher masses). This positive correlation indicates that larger galaxies tend to have higher intrinsic ellipticities (more flattened), consistent with galaxies growing in size and settling into progressively more disc-like configurations over time.

\subsection{Evolution of 3D axis ratios \label{ssc:Results-Pred_3D_Axis_Ratios}}
\begin{figure*}[htp!]
\centering
\includegraphics[angle=0,width=0.7\hsize]{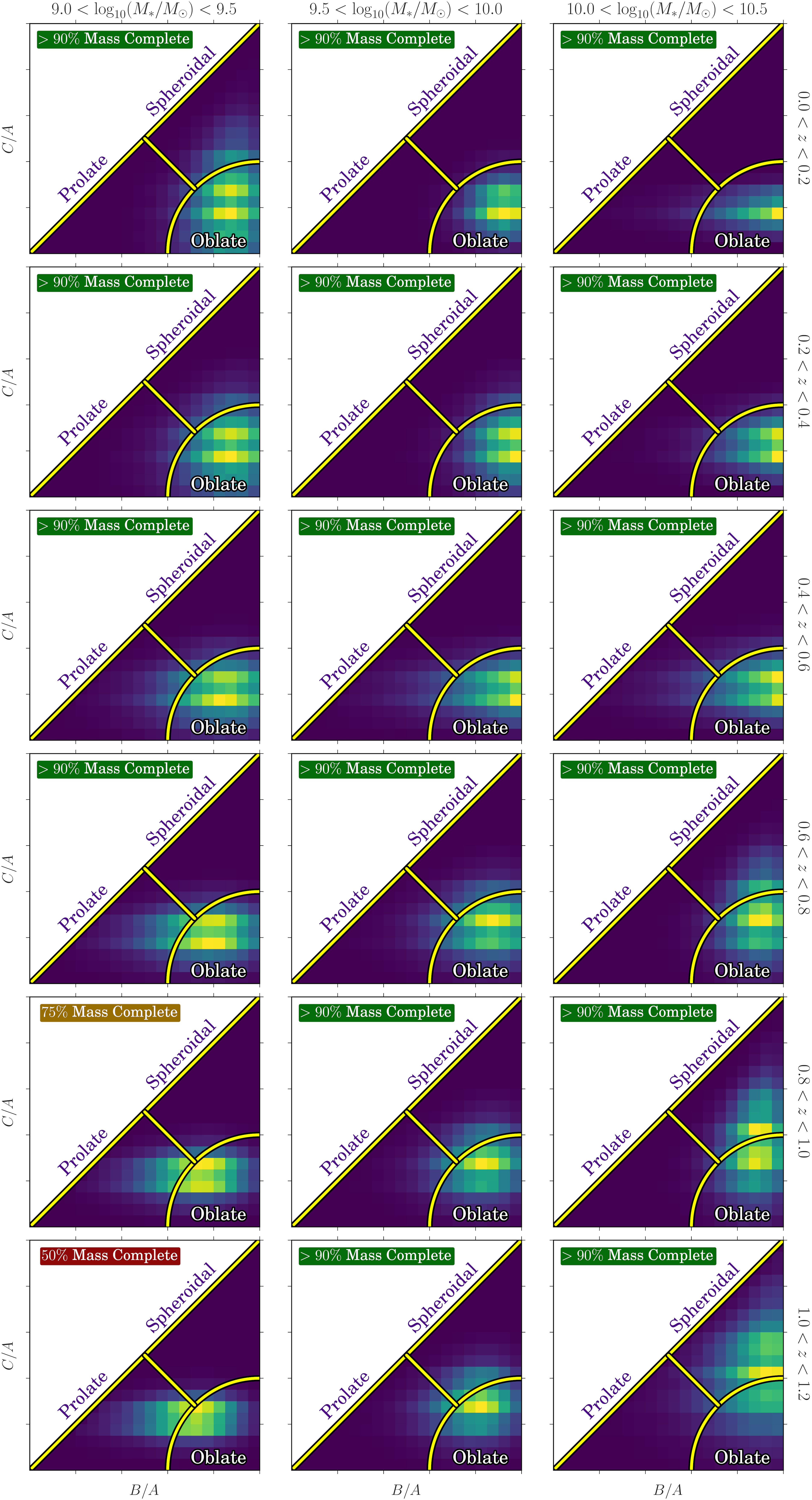}
\caption{Predicted $C/A$ versus $B/A$ axis-ratio distributions averaged from posterior draws for each mass-redshift bin. Intrinsic shape boundaries (in yellow) follow figure 1 of \citetalias{VanDerWel_2014a} and figure 2 of \citetalias{Zhang_2019}. Mass completeness is labelled for each bin (top left; Appendix \ref{app:Completeness}).}
\label{fig:avg_3D_axis_ratios_grid}
\end{figure*}

From the seven inferred population-level shape parameters, we drew random samples from the trivariate Gaussian for each mass–redshift bin and place them in 3D $C/A$--$B/A$ axis-ratio space. This is shown in Fig.\,\ref{fig:avg_3D_axis_ratios_grid}, where each distribution is constructed from the average of 500 random posterior draws, and the prolate, oblate, and spheroidal regions are delineated by the boundaries (in yellow) following figure 1 of \citetalias{Zhang_2019} and figure 2 of \citetalias{VanDerWel_2014a}. We then computed the fraction of samples that fall into each of the shape regions. Given that a different choice in boundaries would ultimately yield different fractions, these should therefore be read as a convenient summary of the inferred distributions rather than as directly constrained observables. The evolution described in the previous sections is also apparent here: all three mass bins converge towards the oblate region by the present day, but from distinct high-redshift starting points: near the prolate-oblate intersection at low mass, the spheroid-oblate boundary at intermediate mass, and a broad spread across spheroidal and oblate geometries at high mass.

\subsection{Evolution of 3D shape fractions \label{ssc:Results-Pred_3D_Fractions}}
\begin{figure*}[ht!]
\centering
\includegraphics[angle=0,width=0.72\hsize]{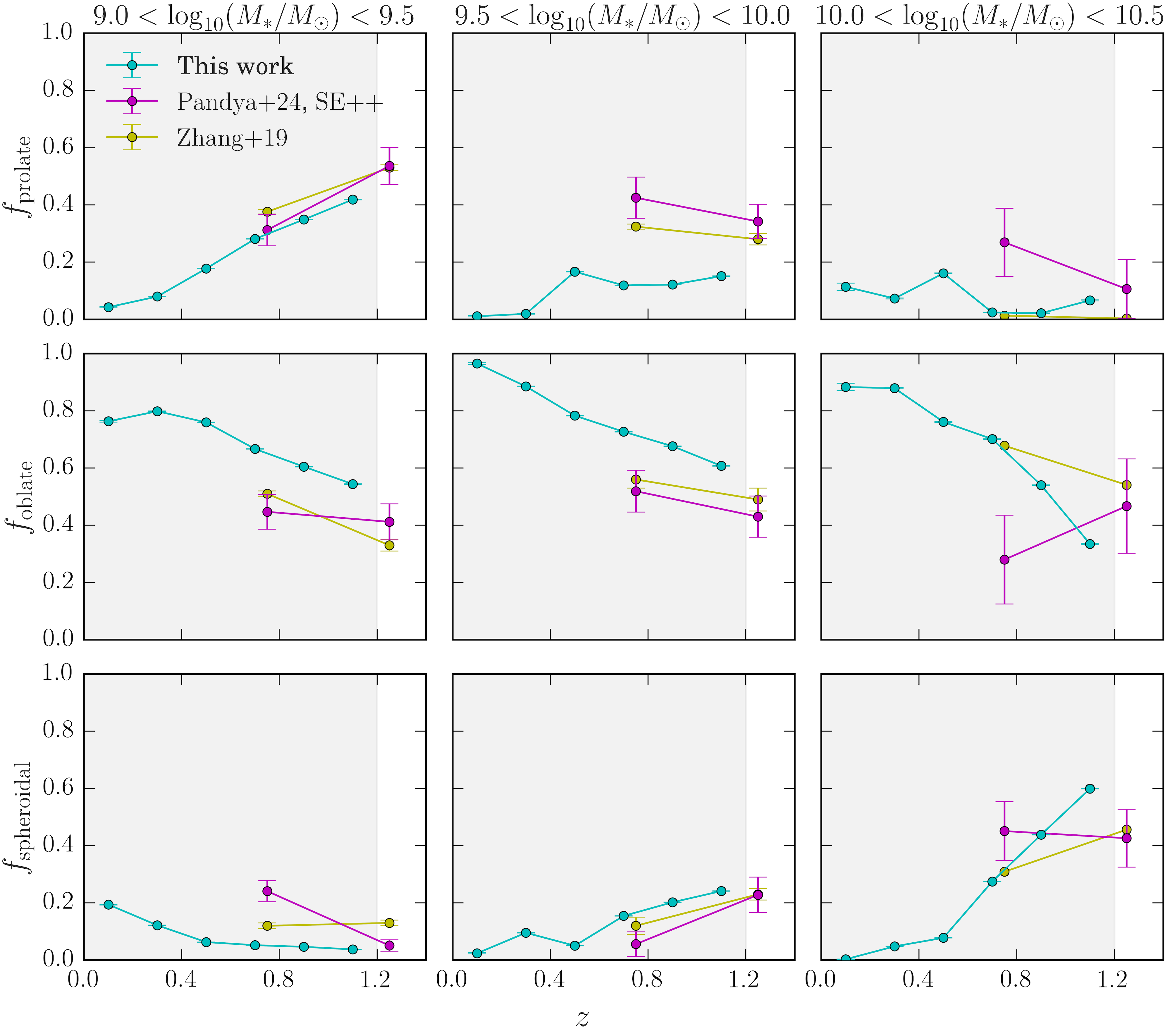}
\caption{Similar to Figs.\,\ref{fig:3DE_evolution}, \ref{fig:3DT_evolution}, and \ref{fig:3DlogArho_evolution} but for shape-class fractions derived from the inferred distributions, as a function of redshift and stellar mass.}
\label{fig:3Dfrac_evolution_random_draws}
\end{figure*}

\begin{figure*}[ht!]
\centering
\includegraphics[angle=0,width=0.91\hsize]{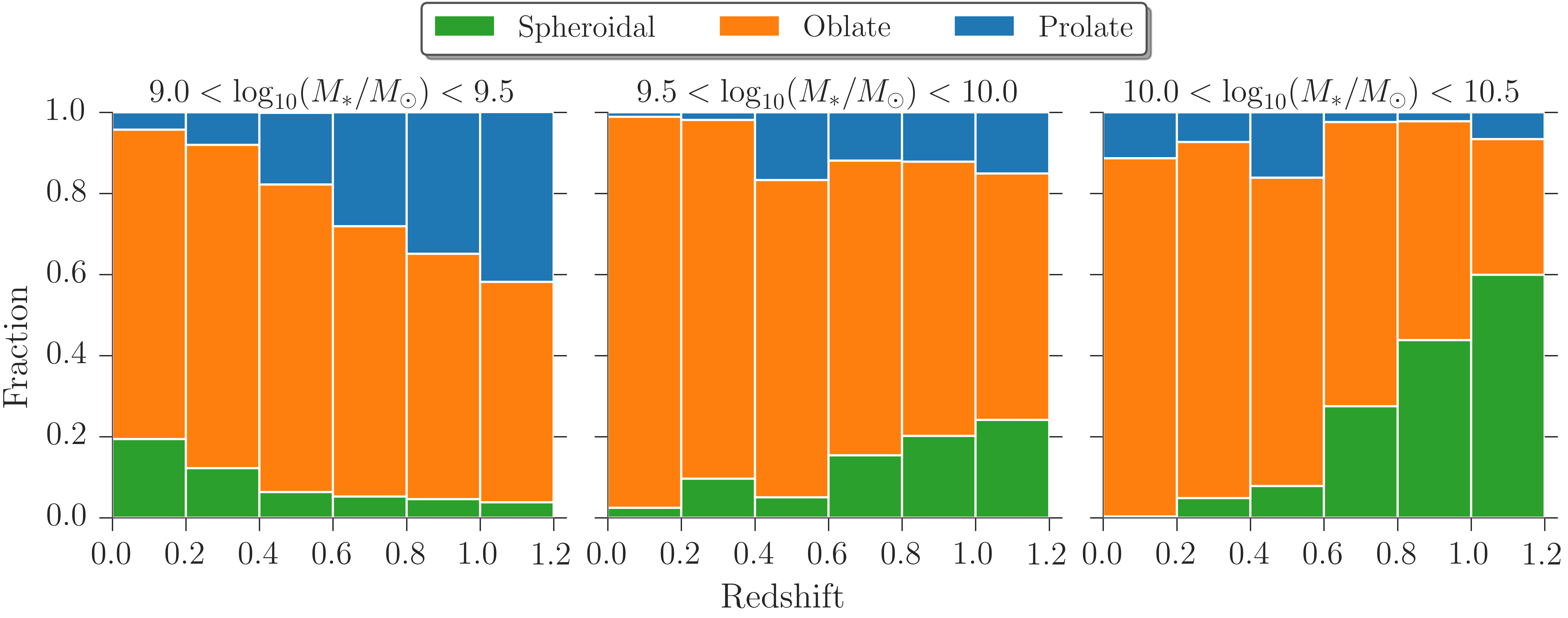}
\caption{Similar to Fig.\,\ref{fig:3Dfrac_evolution_random_draws} but with shape-class fractions computed from full posterior means rather than random draws.}
\label{fig:avg_3Dfrac_bar_chart}
\end{figure*}

The evolution of the inferred population-level 3D shape fractions for each mass–redshift bin, from random posterior draws and compared to \citetalias{Pandya_2024}, is shown in Fig.\,\ref{fig:3Dfrac_evolution_random_draws}. In addition, analogous to figure 4 of \citetalias{VanDerWel_2014a}, figure 13 of \citetalias{Zhang_2019}, and figure 16 of \citetalias{Pandya_2024}, our Fig.\,\ref{fig:avg_3Dfrac_bar_chart} shows stacked histograms of the inferred 3D shape fractions using the total posterior means rather than random draws.

The low-mass prolate fraction shows the best agreement with \citetalias{Zhang_2019} and \citetalias{Pandya_2024}, decreasing steadily towards the present day. Our additional $0.0 < z<0.2$ and $0.2<z<0.4$ bins show that this decline continues, approaching a near-negligible level by $z\sim0$. At intermediate masses, we find lower prolate fractions than in either study, and at high masses our prolate fraction is lower than \citetalias{Pandya_2024} but higher than the near-zero values of \citetalias{Zhang_2019}.

The oblate fractions show the inverse trend, steadily increasing towards the present day in all mass bins. Our values exceed those of \citetalias{Pandya_2024} across all mass bins and those of \citetalias{Zhang_2019} at low and intermediate masses. The results of \citetalias{Zhang_2019} and \citetalias{Pandya_2024} agree well with each other at low and intermediate masses, but both fall below our values. At $z\sim0.75$, our oblate fraction and that of \citetalias{Zhang_2019} are both around 70\%. The exception to the rising trend is the low-mass bin, where the oblate fraction dips slightly from $0.798\pm0.001$ at $0.2<z<0.4$ to $0.763\pm0.003$ at $0.0 < z<0.2$. In the intermediate-mass bin at $z\sim0$, nearly all galaxies are classified as oblate.

In the low-mass bin, our spheroid fractions rise towards the present day, from a near-negligible level to $0.194\pm0.002$ at $0.0<z<0.2$. Compared to \citetalias{Zhang_2019} and \citetalias{Pandya_2024}, our spheroid fractions are broadly consistent in intermediate masses and slightly lower in low masses; in high masses, they are consistent at $0.8<z<1.0$ but slightly higher than both studies at $1.0<z<1.2$. Notably, all three studies find elevated spheroid fractions at high masses for a nominally star-forming sample, despite the expectation that genuine star-forming spheroids should be rare (Sect.\,\ref{ssc:Discussion-Camparison_Previous_Studies}). We caution that these fractions are sensitive to the breadth of the inferred distribution rather than being directly constrained; because the classes are assigned a posteriori by sampling the trivariate Gaussian, a broad inferred distribution (large $\sigma_E$ or $\sigma_T$) scatters more samples into the spheroidal region of the $C/A$--$B/A$ plane, which can inflate the spheroid fraction even where the mean shape is disc-like. This is the same broadening responsible for the boundary behaviour of $\sigma_T$ discussed in Sect.\,\ref{ssc:Results-Triaxiality_Evolution}, and is most pronounced in the high-mass bins where the inferred dispersions are largest.

\section{Discussion \label{sc:Discussion}}
In this section, we further compare our findings with previous studies (Sect.\,\ref{ssc:Discussion-Camparison_Previous_Studies}), discuss the astrophysical implications (Sect.\,\ref{ssc:Discussion-Astrophysical_Implications}), acknowledge key limitations (Sect.\,\ref{ssc:Discussion-Limitations}), and outline recommendations for future work (Sect.\,\ref{ssc:Discussion-Future_Directions}).

\subsection{Comparison with previous studies \label{ssc:Discussion-Camparison_Previous_Studies}}
Having briefly compared our posterior-draw results with those from \citetalias{Pandya_2024} and \citetalias{Zhang_2019} in Sect.\,\ref{sc:Results}, we now focus on comparative results between this work, \citetalias{VanDerWel_2014a}, \citetalias{Zhang_2019}, and \citetalias{Pandya_2024} in more depth. Given that our redshift bins do not align with those used in these studies, the following comparisons are made at the trend level rather than bin by bin. 

For the low-mass bin, our \Euclid-based findings show a strongly oblate-dominated population at all redshifts, with the oblate fraction steadily increasing towards the present day, while the prolate fraction decreases. This matches the qualitative behaviour in \citetalias{VanDerWel_2014a}, \citetalias{Zhang_2019}, and \citetalias{Pandya_2024}. However, our oblate fraction is visibly higher, and our prolate fraction lower, than \citetalias{Zhang_2019} and \citetalias{Pandya_2024} in the overlapping redshift range; the oblate fraction in \citetalias{VanDerWel_2014a}, which extends to $0<z<0.5$, is greater than in the other studies, with near-zero prolate and spheroid fractions. As noted in Sect.\,\ref{ssc:Results-Triaxiality_Evolution}, \citetalias{Pandya_2024} break the oblate-prolate degeneracy most effectively for low-mass galaxies at $z > 1$, where the prolate-dominated population yields $\mu_T \sim 1$. Since our sample extends only to $z < 1.2$, our overlapping comparison samples the later epoch in which their inferred prolate fraction has already declined and discs have begun to dominate; our lower prolate fractions and triaxialities at low mass are therefore consistent with their evolutionary trend. In this sense, our results support the low-redshift, disc-settled end of the same prolate-to-oblate transition that \citetalias{Pandya_2024} trace towards its high-redshift, prolate-dominated origin. The spheroid fractions are very small, rising modestly towards the present day. In overlapping redshift bins, these fractions are fairly consistent with the small values seen in \citetalias{Pandya_2024} and \citetalias{Zhang_2019}.

For intermediate-mass galaxies, our population is again overwhelmingly oblate -- broadly consistent with the oblate dominance in the other studies -- increasing to nearly unity by the present day, with a similar ratio to \citetalias{Zhang_2019} and \citetalias{Pandya_2024} as in the low-mass bin. Prolate fractions remain lower than \citetalias{Zhang_2019} and \citetalias{Pandya_2024}, and are more comparable to the values of \citetalias{VanDerWel_2014a}. Our spheroid fractions are again low, decreasing towards the present day, similarly to \citetalias{VanDerWel_2014a}, \citetalias{Pandya_2024}, and \citetalias{Zhang_2019}. The high-mass bin also shows oblate dominance over time, and is the only bin in our grid where spheroids are the dominant class at higher redshifts. This substantial spheroid population is found in all three studies at higher masses. Similar to \citetalias{VanDerWel_2014a}, our $0.0 < z<0.2$ and $0.2<z<0.4$ bins reveal that the spheroid fraction continues to decline towards the present day, beyond where the bins in \citetalias{Zhang_2019} and \citetalias{Pandya_2024} end.

At the low-redshift end, our oblate fractions can also be compared with constraints on intrinsically flat galaxies in the local Universe. From the projected axis ratios of galaxies at $z<0.1$ in the Galaxy and Mass Assembly \citep[GAMA;][]{GAMA}, Dark Energy Spectroscopic Instrument \citep[DESI;][]{DESI_DR1}, and Arecibo Legacy Fast ALFA \citep[ALFALFA;][]{ALFALFA} surveys, \citet{Benavides} infer that approximately $70$--$90\%$ of galaxies with $\Mstellar=10^9$--$10^{10.5}\Msun$ are flatter than an intrinsic $c/a=0.3$. Our oblate fractions at $0.0 < z<0.2$ (0.76--0.97 across our three mass bins; Table\,\ref{tab:3D_shape_results}) are broadly consistent with these estimates, although the comparison is approximate: our oblate class is defined by a region of the $C/A$--$B/A$ plane rather than by a single $C/A$ threshold, our sample is restricted to star-forming galaxies, and their fractions are lower limits.

Overall, the most notable agreement is the prolate-to-oblate transition at low masses seen consistently in all studies. Our much larger sample suggests that this trend may be a genuine feature of star-forming galaxy evolution rather than an artefact of small-sample statistics. The increase in spheroid fractions with mass is also qualitatively consistent across all three studies. The offsets in central values for the seven shape parameters between this work and \citetalias{VanDerWel_2014a}, \citetalias{Zhang_2019}, and \citetalias{Pandya_2024} probably stem from a combination of differences in instrument, wavelength coverage, and selection function. The most important precision difference seems to be our substantially larger sample, which yields far tighter constraints on the inferred shape distributions. Our own experience illustrates this directly: applying the same method to a smaller sample of $929\,344$ star-forming galaxies from Euclid Quick Data Release 1 \citep[Q1;][]{Q1cite, Q1-TP001} produced posteriors that struggled to converge in many bins -- mainly for triaxiality -- whereas the DR1 Wide sample used here yields well-converged inferences across all bins. In a future paper, we plan to incorporate \Euclid near-infrared (NIR) imaging \citep[\YE, \JE, \HE;][]{EuclidSkyNISP, DR1-TP003} into this analysis, both to mitigate the surface-brightness and wavelength-dependent biases inherent to VIS-only shape measurements, and to test whether the agreement among studies persists towards higher redshifts, where rest-frame optical morphology shifts into the NIR.

\subsection{Astrophysical implications\label{ssc:Discussion-Astrophysical_Implications}}
The observed prolate-to-oblate evolutionary transition in low-mass galaxies is consistent with the theoretical framework of \citet{Pandya_2019} and with predictions from zoom-in cosmological simulations \citep{Ceverino_2015, Tomassetti_2016}. In this picture, high-redshift prolate dwarfs form within elongated dark matter haloes embedded in cosmic filaments, undergoing repeated mergers along the filament direction that maintain their elongated geometries and produce intrinsic alignments with large-scale structure. \citet{Ceverino_2015} and \citet{Tomassetti_2016} further predict that high-redshift galaxies undergo `compaction events', in which rapid gas inflows shift galaxy centres from dark-matter–dominated to baryon-dominated configurations capable of supporting stable discs \citep{Lapiner_2023}. \citet{Dekel_2020} show that gas discs are expected to persist only in haloes above a threshold of \SI{2e11}{\solarmass} (corresponding to a stellar mass of approximately $10^{9}\Msun$, with little redshift dependence), below which repeated mergers and dark-matter dominance disrupt early discs before they can stabilise. This threshold sits at the low-mass end of our sample and is consistent with the prolate phase persisting longest in our lowest-mass bins, where disc disruption is expected to dominate. As these systems evolve, gas accretion and compaction enable disc formation, driving the transition towards oblate-dominated populations. Our substantial disc fractions across all masses and redshifts, which increase towards the present day, indicate that disc structures begin to establish themselves relatively early, even as the transition to oblate dominance continues to low redshift. However, distinguishing genuine oblate discs from other morphologies remains a challenge. Recent work by \citet{Pandya_2025} found preliminary evidence for non-random orientations in JWST-CEERS data via weak lensing, supporting the filament-alignment scenario, while \citet{Hasan_Zhang_Pandya_2026} confirm, using TNG50 simulations, that prolate galaxies preferentially inhabit low-density, recently collapsing filaments with strong alignments, potentially detectable with the \textit{Roman} Space Telescope or \Euclid.

As in both \citetalias{Zhang_2019} and \citetalias{Pandya_2024}, our nominally prolate bins frequently occupy the triaxial rather than the extreme prolate region of $C/A$--$B/A$ space (Fig.\,\ref{fig:avg_3D_axis_ratios_grid}), indicating more oval flattened discs rather than purely elongated ellipsoids. This is consistent with a picture in which high-$z$ star-forming galaxies undergo disc settling along two axes simultaneously -- from both thick-to-thin and oval-to-circular -- as gas dissipates and angular momentum builds over time \citep{Elmegreen_2005, Law_2012, Pandya_2024}. Our positive $\rho_{E,\logten(A)}$ values link the size growth of star-forming galaxies at fixed stellar mass \citep{VanDerWel_2014b} directly to increasing intrinsic flattening, consistent with discs settling as they grow. As noted in Sect.\,\ref{ssc:Discussion-Camparison_Previous_Studies}, our predicted spheroid fractions and their trends appear to be broadly consistent with the other previous inference studies. Although HST-based observations suggest that star-forming spheroids should be rare \citep{Brennan_2015}, moderate spheroid fractions are found at higher masses at higher redshifts. As speculated in section 6.2 of \citetalias{Pandya_2024}, if these fractions are real, they may reflect the same compaction processes invoked above, but channelled towards spheroid formation rather than disc settling: gas-rich mergers, disc instabilities, or cold gas inflows drive a rapid increase in central mass surface density prior to quenching \citep{Dekel_Burkert_2014, Zolotov_2015, Tacchella_2016, Lapiner_2023}.

We also compared to predictions from the TNG50 hydrodynamical simulations at $z=0.5$ and $z=2$ \citep[figure 8 of][]{Pillepich_2019}. In agreement with our results, TNG50 predicts a higher prolate fraction at $z=2$ than at $z=0.5$ across all masses; however, it also predicts a correspondingly larger high-redshift spheroid fraction at all masses, whereas we recover this trend only for $\Mstarwun>9.5$. At $z=0.5$, \citet{Pillepich_2019} predict essentially no prolate galaxies, yet we find prolate fractions of approximately 16--18\% in every mass bin. This discrepancy echoes earlier findings from \citetalias{VanDerWel_2014a}, \citetalias{Zhang_2019}, and \citetalias{Pandya_2024}, and probably reflects limitations in simulation physics, observational systematics in how shapes are measured and classified, and genuine differences between simulated and real galaxy populations. This conclusion is also supported by \citet{Benavides}, who find that current cosmological simulations, including TNG50, produce far fewer intrinsically thin dwarf galaxies than observed.

\subsection{Limitations of the study \label{ssc:Discussion-Limitations}}
The larger statistical sample of DR1 enables tighter constraints on 3D shape distributions than previous inference studies. However, several methodological choices and observational limitations warrant careful consideration.

Our inference uses global single-Sérsic axis ratios and sizes, which describe the flux-weighted projected shape of the whole light distribution rather than that of individual galaxy components (e.g. bulges and discs, bars, spiral arms). For bulge+disc systems, the global axis ratio closely follows that of the disc for face-on galaxies and for a bulge-to-total light ratio ${\rm B/T}\lesssim0.1$, differing significantly only for edge-on discs with substantial bulges \citep{Quilley25}. In those cases, the bulge raises $b/a$ without reducing the projected size, which can shift the inference towards thicker oblate or spheroidal geometries rather than prolate ones, since prolate populations are identified mainly through round projections at small sizes. Our sample is restricted to star-forming galaxies, for which B/T is typically low \citep{Dimauro_2022, Shuntov_2026, Yang_2026}; the median Sérsic index in our lowest-mass bins is $n\approx1.1$. Sérsic axis ratios and sizes are typically recovered without significant bias down to a signal-to-noise ratio (S/N) of approximately 20, independent of depth and resolution \citep[e.g.][]{Haussler_2007, Guo_2009, vdW_2012, Bruce_2012, Sazonova_2026}, and our comparisons with HST and isophotal-based measurements (Appendix \ref{app:sersic_param_comparison}) show that the axis ratios remain reliable to $\IE=24.5$. Nevertheless, galaxies with several components may be poorly approximated by a single 3D ellipsoid, and non-parametric measures or multi-component fits could provide complementary constraints.

Single-source catalogues may also conflate multiple objects into one. Close pairs or small groups (whether physically associated or merely aligned in projection) along with interacting or merging systems, can be fit as a single galaxy, producing anomalous projected axis ratios and sizes that do not correspond to any single intrinsic ellipsoid. Because both the physical merger rate and the line-of-sight projection probability increase with redshift, the prevalence of such blends is expected to vary across our bins, potentially introducing a redshift-dependent systematic. This may already affect our highest-redshift bins, and becomes considerably more significant as we extend the analysis to higher redshifts in future work, where blending and merger confusion are increasingly common. Deep-learning morphology classifiers such as \texttt{Zoobot} \citep{Zoobot, Q1-SP047} could help identify mergers, interacting systems, and blended sources before inference.

As noted in Sect.\,\ref{ssc:MER_selection}, our choice to retain only galaxies larger than the PSF FWHM ensures that resolved sources dominate the analysis. However, it may systematically exclude rounder, more compact, high-surface-brightness objects, shifting the inferred sizes upwards and biasing the shape distributions -- particularly at high redshift and low masses, where such objects are most common.

Our reliance on the single \IE broadband filter also means that we probe different rest-frame wavelengths as a function of redshift, which is in large part why we restricted the present analysis to $0<z<1.2$. Over this range, \IE still samples at or near the optical rest-frame, where structural measurements best trace the underlying stellar-mass distribution. At higher redshifts, the band shifts into the rest-frame ultraviolet (UV), which increasingly traces clumpy, unobscured star formation rather than the bulk stellar light. Dust attenuation compounds this effect: because attenuation is strongest at short wavelengths and our star-forming sample is comparatively dust-rich, an inclination-dependent reddening can alter a galaxy's apparent ellipticity and size. An edge-on disc, for instance, suffers far more attenuation along the line of sight than the same disc seen face-on. As our rest-frame coverage shifts bluewards with increasing redshift, this attenuation grows, so dust may bias not only the measured shapes but also their apparent evolution with redshift \citep{Padilla_Strauss_2008, Zhang_2019, Zhang_2023}. In \citetalias{Pandya_2024}, this is mitigated by selecting JWST Near Infrared Camera (NIRCam) filters that maintain consistent rest-frame coverage across redshift. Incorporating NIR imaging in future work would maintain rest-frame optical coverage to higher redshift, ultimately enabling a robust extension of this analysis beyond $z\sim1$. This would also allow us to independently test the breaking of the oblate-prolate degeneracy that occurs at $z>1$ in \citetalias{Pandya_2024}.

The most fundamental assumption underlying our approach is that, within each bin of stellar mass and redshift, the galaxy population is well described by a single parent 3D shape distribution, of which the observed projected axis ratios and sizes are realisations (Sect.\,\ref{sc:Methodology}). The inference constrains the seven parameters of this distribution, but cannot uniquely recover the 3D shape of any individual galaxy, since a given projected $b/a$ and $\logten(a)$ is consistent with a range of intrinsic geometries viewed at different angles. The class fractions we report are likewise not inferred quantities, but computed by partitioning samples of the inferred distribution in the $C/A$--$B/A$ plane, and depend on the placement of class boundaries. They should therefore be read as summaries of the underlying shape parameters, not as independent constraints.

A related caution applies to the shape parameters themselves. Triaxiality is the most sensitive to our modelling assumptions, since it is recovered indirectly from the projected $b/a$--$\logten(a)$ distribution rather than constrained by any direct observable, and tight statistical constraints do not guarantee freedom from systematic bias by themselves. The single trivariate Gaussian enforces a unimodal triaxiality distribution within each bin; where the true population is instead a mixture of distinct geometric families -- for example a near-oblate ($T \approx 0$) component alongside a more triaxial one -- the inferred $\mu_T$ and $\sigma_T$ blend the two rather than resolving them. This is evident in intermediate-mass and high-mass bins, where $\mu_T$ is pinned at the $T=0$ boundary, while $\sigma_T$ is simultaneously large and tightly determined (Sect.\,\ref{ssc:Results-Triaxiality_Evolution}). The model can reproduce a broad, boundary hugging distribution only by placing its mean on the boundary and widening its dispersion. In these regimes, a precisely determined $\mu_T$ should not be read as an accurate one. We stress that this concerns the detailed triaxiality within these specific bins, not the robust oblate dominance itself, which is driven by the ellipticity and axis-ratio structure rather than by $\mu_T$. Sampling directly in $C/A$--$B/A$ space may avoid this, since the oblate region does not lie against a parameter boundary there; alternatively, a mixture model separating oblate and triaxial subpopulations would recover both components rather than averaging over them. Either change would require new model libraries and more complex models, for which methods more scalable than HMC may be necessary.

A key practical limitation is computational cost. Although HMC is far more efficient than traditional MCMC, inference can still take days per DR1 bin on a computing cluster. Scaling to larger datasets or performing extensive tests to address the aforementioned limitations would be computationally prohibitive, motivating the need for faster alternative methods.

\subsection{Future directions \label{ssc:Discussion-Future_Directions}}
This work recovers established evolutionary trends through improved statistics while highlighting several avenues for further progress, both scientific and methodological. Combining VIS and NIR imaging helps mitigate the biases inherent to single-band VIS shape measurements and extend the analysis to higher redshifts, yielding better-constrained shapes across the full sample. Future \Euclid releases reaching fainter magnitudes should enable finer redshift and mass binning, allowing evolutionary transitions to be traced more precisely. Connecting galaxy 3D shapes to the cosmic web would further reveal how these distributions vary with environment and large-scale structure. However, realising this potential across future \Euclid releases and other large surveys (e.g. LSST, \textit{Roman}) will require overcoming the computational cost per-bin that currently limits this approach. Simulation-based inference (SBI) offers a promising path: neural density estimators trained on simulated observations make posterior estimation nearly instantaneous, converting a recurring bottleneck into a one-off training cost. We plan to explore these in a forthcoming paper.

\section{Conclusions \label{sc:Conclusions}}
We applied the Bayesian inference framework of \citetalias{Pandya_2024} to \num{40864304} DR1 Wide star-forming galaxies with $\IE\leq24.5$ across bins of stellar mass $\Mstarwun=9.0$--$10.5$ at photometric redshifts $0<z<1.2$. Using Sérsic axis ratios and semi-major axes from VIS imaging, we constrained the population-level mean 3D ellipticity, triaxiality, and size, together with their covariance, and derived the fractions of galaxies classified as prolate, oblate, or spheroidal in each mass-redshift bin. We demonstrated that this methodology scales reliably to the tens of millions of galaxies in DR1 while recovering the evolutionary trends of previous studies. Our main conclusions are as follows.
\begin{itemize}
    \item Low-mass star-forming galaxies appear to evolve from predominantly prolate to oblate geometries towards the present day. This trend is consistent with comparable analyses from \citetalias{VanDerWel_2014a}, \citetalias{Zhang_2019}, and \citetalias{Pandya_2024}, despite differences in instrument, wavelength, and selection; this suggests that it is not an artefact of small-sample statistics, but potentially a genuine feature of star-forming galaxy evolution.
    
    \item Across all stellar masses, the oblate fraction increases towards $z=0$, eventually becoming the dominant class. For higher masses, spheroids constitute a substantial fraction at higher redshifts before declining towards the present day.
    
    \item Our DR1 sample significantly reduces the uncertainties in the inferred 3D shape parameters compared to \citetalias{Zhang_2019} and \citetalias{Pandya_2024}, and our additional bins at $z<0.5$ extend the observed trends seen in these studies to the present day.
    
    \item Even where galaxies are nominally classified as prolate, they frequently occupy the triaxial region of $C/A$--$B/A$ space, indicating more oval, flattened discs rather than purely elongated ellipsoids. This is consistent with a two-axis disc-settling picture for high-$z$ star-forming galaxies.
\end{itemize}
Future \Euclid releases reaching fainter magnitudes should improve stellar-mass completeness, enable finer mass-redshift binning, and facilitate detailed studies of morphology versus environment. Scaling to larger samples with more efficient inference methods, together with addressing the principal limitations identified here, is deferred to a forthcoming paper.

\begin{acknowledgements}
Funded by the European Union (MSCA Doctoral Network EDUCADO, GA 101119830 and Widening Participation, ExGal-Twin, GA 101158446). Views and opinions expressed are however those of the author(s) only and do not necessarily reflect those of the European Union. Neither the European Union nor the granting authority can be held responsible for them. JHK acknowledges grant PID2022-136505NB-I00 funded by MCIN/AEI/10.13039/501100011033 and EU, ERDF. This work is part of grant CEX2025-001609-S, awarded to the Instituto de Astrofísica de Canarias under the Severo Ochoa Centre of Excellence program and funded by MICIU/AEI/10.13039/501100011033.
\AckDRone 
\AckEC
\AckDatalabs
This work was made possible by utilising the CANDIDE cluster at the Institut d'Astrophysique de Paris. The cluster was funded through grants from the PNCG, CNES, DIM-ACAV, the Euclid Consortium, and the Danish National Research Foundation Cosmic Dawn Center (DNRF140). It is maintained by Stephane Rouberol. The author(s) also wish to acknowledge the contribution of Teide High-Performance Computing facilities to the results of this research. TeideHPC facilities are provided by the Instituto Tecnológico y de Energías Renovables (ITER, SA). URL: \url{https://teidehpc.iter.es}.
\end{acknowledgements}

\bibliography{refs}

@article{Pandya_2024, 
    title={Galaxies Going Bananas: Inferring the 3D Geometry of High-redshift Galaxies with JWST-CEERS},
    volume={963}, 
    ISSN={0004-637X, 1538-4357}, 
    url={https://iopscience.iop.org/article/10.3847/1538-4357/ad1a13}, 
    DOI={10.3847/1538-4357/ad1a13}, 
    number={1}, 
    journal={\apj}, 
    author={Pandya, Viraj and Zhang, Haowen and Huertas-Company, Marc and Iyer, Kartheik G. and McGrath, Elizabeth and Barro, Guillermo and Finkelstein, Steven L. and Kümmel, Martin and Hartley, William G. and Ferguson, Henry C. and Kartaltepe, Jeyhan S. and Primack, Joel and Dekel, Avishai and Faber, Sandra M. and Koo, David C. and Bryan, Greg L. and Somerville, Rachel S. and Amorín, Ricardo O. and Arrabal Haro, Pablo and Bagley, Micaela B. and Bell, Eric F. and Bertin, Emmanuel and Costantin, Luca and Davé, Romeel and Dickinson, Mark and Feldmann, Robert and Fontana, Adriano and Gavazzi, Raphael and Giavalisco, Mauro and Grazian, Andrea and Grogin, Norman A. and Guo, Yuchen and Hahn, ChangHoon and Holwerda, Benne W. and Kewley, Lisa J. and Kirkpatrick, Allison and Kocevski, Dale D. and Koekemoer, Anton M. and Lotz, Jennifer M. and Lucas, Ray A. and Papovich, Casey and Pentericci, Laura and Pérez-González, Pablo G. and Pirzkal, Nor and Ravindranath, Swara and Rose, Caitlin and Schefer, Marc and Simons, Raymond C. and Straughn, Amber N. and Tacchella, Sandro and Trump, Jonathan R. and De La Vega, Alexander and Wilkins, Stephen M. and Wuyts, Stijn and Yang, Guang and Yung, L. Y. Aaron},
    year={2024},
    month=mar, 
    pages={54}
}

@article{Pandya_2025, 
    title={Preliminary Evidence for Lensing-induced Alignments of High-redshift Galaxies in JWST-CEERS}, 
    volume={986}, 
    ISSN={0004-637X, 1538-4357}, 
    url={https://iopscience.iop.org/article/10.3847/1538-4357/adcd78}, 
    DOI={10.3847/1538-4357/adcd78}, 
    number={1}, 
    journal={\apj},
    author={Pandya, Viraj and Loeb, Abraham and McGrath, Elizabeth J. and Barro, Guillermo and Finkelstein, Steven L. and Ferguson, Henry C. and Grogin, Norman A. and Kartaltepe, Jeyhan S. and Koekemoer, Anton M. and Papovich, Casey and Pirzkal, Nor and Yung, L. Y. Aaron}, 
    year={2025},
    month={june},
    pages={72} 
}

@article{Hasan_Zhang_Pandya_2026, 
    author = {{Hasan}, Farhanul and {Zhang}, Haowen and {Pandya}, Viraj and {Rafelski}, Marc and {Burchett}, Joseph N. and {Hellinger}, Douglas and {Nedkova}, Kalina V. and {Goovaerts}, Ilias and {Mandelker}, Nir and {Nagai}, Daisuke and {Oyarz{\'u}n}, Grecco A. and {Primack}, Joel R. and {Woo}, Joanna},
        title = "{Impact of Cosmic Filaments on Galaxy Morphological Evolution and Predictions of Early Cosmic Web Structure for Roman}",
      journal = {\apj},
         year = 2026,
        month = jun,
       publisher = {The American Astronomical Society},
        volume = {1004},
        number = {1},
        pages = {74},
          doi = {10.3847/1538-4357/ae60f9},
          url = {https://doi.org/10.3847/1538-4357/ae60f9}
}

@article{VanDerWel_2014a, 
    title={GEOMETRY OF STAR-FORMING GALAXIES FROM SDSS, 3D-HST, AND CANDELS}, 
    volume={792}, 
    rights={http://iopscience.iop.org/info/page/text-and-data-mining}, 
    ISSN={2041-8213}, 
    url={https://iopscience.iop.org/article/10.1088/2041-8205/792/1/L6}, 
    DOI={10.1088/2041-8205/792/1/L6}, 
    number={1}, 
    journal={\apj},
    author={van der Wel, A. and Chang, Yu-Yen and Bell, E. F. and Holden, B. P. and Ferguson, H. C. and Giavalisco, M. and Rix, H.-W. and Skelton, R. and Whitaker, K. and Momcheva, I. and Brammer, G. and Kassin, S. A. and Martig, M. and Dekel, A. and Ceverino, D. and Koo, D. C. and Mozena, M. and Van Dokkum, P. G. and Franx, M. and Faber, S. M. and Primack, J.}, 
    year={2014}, 
    month=aug, 
    pages={L6} 
}

@article{VanDerWel_2014b, 
    title={3D-HST+CANDELS: THE EVOLUTION OF THE GALAXY SIZE-MASS DISTRIBUTION SINCE z = 3}, 
    volume={788}, 
    rights={http://iopscience.iop.org/info/page/text-and-data-mining}, 
    ISSN={0004-637X, 1538-4357},
    url={https://iopscience.iop.org/article/10.1088/0004-637X/788/1/28}, 
    DOI={10.1088/0004-637X/788/1/28},
    number={1}, 
    journal={\apj}, 
    author={van der Wel, A. and Franx, M. and Van Dokkum, P. G. and Skelton, R. E. and Momcheva, I. G. and Whitaker, K. E. and Brammer, G. B. and Bell, E. F. and Rix, H.-W. and Wuyts, S. and Ferguson, H. C. and Holden, B. P. and Barro, G. and Koekemoer, A. M. and Chang, Yu-Yen and McGrath, E. J. and Häussler, B. and Dekel, A. and Behroozi, P. and Fumagalli, M. and Leja, J. and Lundgren, B. F. and Maseda, M. V. and Nelson, E. J. and Wake, D. A. and Patel, S. G. and Labbé, I. and Faber, S. M. and Grogin, N. A. and Kocevski, D. D.},
    year={2014}, 
    month=may, 
    pages={28} 
}

@article{Zhang_2019, 
    title={The evolution of galaxy shapes in CANDELS: from prolate to discy}, 
    volume={484}, 
    rights={https://academic.oup.com/journals/pages/open_access/funder_policies/chorus/standard_publication_model}, 
    ISSN={0035-8711, 1365-2966}, 
    url={https://academic.oup.com/mnras/article/484/4/5170/5307087}, 
    DOI={10.1093/mnras/stz339}, 
    number={4}, 
    journal={\mnras}, 
    author={Zhang, Haowen and Primack, Joel R and Faber, S M and Koo, David C and Dekel, Avishai and Chen, Zhu and Ceverino, Daniel and Chang, Yu-Yen and Fang, Jerome J and Guo, Yicheng and Lin, Lin and van der Wel, Arjen}, 
    year={2019},
    month=apr, 
    pages={5170}
}

@article{Chang_2013, 
    title={STRUCTURAL EVOLUTION OF EARLY-TYPE GALAXIES TO z = 2.5 IN CANDELS}, 
    volume={773}, 
    rights={http://iopscience.iop.org/info/page/text-and-data-mining}, 
    ISSN={0004-637X, 1538-4357}, 
    url={https://iopscience.iop.org/article/10.1088/0004-637X/773/2/149}, 
    DOI={10.1088/0004-637X/773/2/149}, 
    number={2}, 
    journal={\apj}, 
    author={Chang, Yu-Yen and van der Wel, Arjen and Rix, Hans-Walter and Holden, Bradford and Bell, Eric F. and McGrath, Elizabeth J. and Wuyts, Stijn and Häussler, Boris and Barden, Marco and Faber, S. M. and Mozena, Mark and Ferguson, Henry C. and Guo, Yicheng and Galametz, Audrey and Grogin, Norman A. and Kocevski, Dale D. and Koekemoer, Anton M. and Dekel, Avishai and Huang, Kuang-Han and Hathi, Nimish P. and Donley, Jennifer}, 
    year={2013}, 
    month=aug, 
    pages={149} 
}

@article{Zhang_2023, 
 title={Dust attenuation, dust content, and geometry of star-forming galaxies}, 
 volume={524}, 
 rights={https://creativecommons.org/licenses/by/4.0/}, 
 ISSN={0035-8711, 1365-2966}, 
 url={https://academic.oup.com/mnras/article/524/3/4128/7226713}, 
 DOI={10.1093/mnras/stad2066}, 
 number={3}, 
 journal={\mnras}, 
 author={Zhang, Junkai and Wuyts, Stijn and Cutler, Sam E and Mowla, Lamiya A and Brammer, Gabriel B and Momcheva, Ivelina G and Whitaker, Katherine E and van•Dokkum, Pieter and Förster•Schreiber, Natascha M and Nelson, Erica J and Schady, Patricia and Villforth, Carolin and Wake, David and van der Wel, Arjen}, 
 year={2023}, 
 month={July},
 pages={4128}, 
 language={en} 
}

@article{Pandya_2017, 
    title={The nature of massive transition galaxies in CANDELS, GAMA and cosmological simulations}, 
    volume={472}, 
    ISSN={0035-8711, 1365-2966}, 
    url={http://academic.oup.com/mnras/article/472/2/2054/4082086/The-nature-of-massive-transition-galaxies-in},
    DOI={10.1093/mnras/stx2027}, 
    number={2}, 
    journal={\mnras}, 
    author={Pandya, Viraj and Brennan, Ryan and Somerville, Rachel S. and Choi, Ena and Barro, Guillermo and Wuyts, Stijn and Taylor, Edward N. and Behroozi, Peter and Kirkpatrick, Allison and Faber, Sandra M. and Primack, Joel and Koo, David C. and McIntosh, Daniel H. and Kocevski, Dale and Bell, Eric F. and Dekel, Avishai and Fang, Jerome J. and Ferguson, Henry C. and Grogin, Norman and Koekemoer, Anton M. and Lu, Yu and Mantha, Kameswara and Mobasher, Bahram and Newman, Jeffrey and Pacifici, Camilla and Papovich, Casey and van der Wel, Arjen and Yesuf, Hassen M.}, 
    year={2017}, 
    month=dec, 
    pages={2054}
}

@article{Pandya_2019, 
    title={Can intrinsic alignments of elongated low-mass galaxies be used to map the cosmic web at high redshift?}, 
    volume={488},
    rights={https://academic.oup.com/journals/pages/open_access/funder_policies/chorus/standard_publication_model}, 
    ISSN={0035-8711, 1365-2966}, 
    url={https://academic.oup.com/mnras/article/488/4/5580/5543231}, 
    DOI={10.1093/mnras/stz2129}, 
    number={4},
    journal={\mnras}, 
    author={Pandya, Viraj and Primack, Joel and Behroozi, Peter and Dekel, Avishai and Zhang, Haowen and Eckholm, Elliot and Faber, Sandra M and Ferguson, Henry C and Giavalisco, Mauro and Guo, Yicheng and Hathi, Nimish and Kodra, Dritan and Koekemoer, Anton M and Koo, David C and Newman, Jeffrey and van der Wel, Arjen},
    year={2019}, 
    month=oct, 
    pages={5580}, 
    language={en}
}

@article{Dekel_2020, 
    title={A mass threshold for galactic gas discs by spin flips}, 
    volume={493},
    rights={https://academic.oup.com/journals/pages/open_access/funder_policies/chorus/standard_publication_model}, 
    ISSN={0035-8711, 1365-2966}, 
    url={https://academic.oup.com/mnras/article/493/3/4126/5752134}, 
    DOI={10.1093/mnras/staa470}, 
    number={3}, 
    journal={\mnras},
    author={Dekel, Avishai and Ginzburg, Omri and Jiang, Fangzhou and Freundlich, Jonathan and Lapiner, Sharon and Ceverino, Daniel and Primack, Joel}, 
    year={2020}, 
    month=apr, 
    pages={4126},
    language={en} 
}

@article{Hoffmann_2022, 
    title={Redshift and stellar mass dependence of intrinsic shapes of disc-dominated galaxies from COSMOS observations below z = 1.0}, 
    volume={515}, 
    rights={https://creativecommons.org/licenses/by/4.0/}, 
    ISSN={0035-8711, 1365-2966}, 
    url={https://academic.oup.com/mnras/article/515/3/3603/6648832}, 
    DOI={10.1093/mnras/stac1988}, 
    number={3},
    journal={\mnras}, 
    author={Hoffmann, K and Laigle, C and Chisari, N E and Tallada-Crespí, P and Teyssier, R and Dubois, Y and Devriendt, J}, 
    year={2022},
    month=aug, 
    pages={3603}, 
    language={en} 
}

@article{Padilla_Strauss_2008, 
    title={The shapes of galaxies in the Sloan Digital Sky Survey}, 
    volume={388},
    ISSN={00358711, 13652966}, 
    url={https://academic.oup.com/mnras/article-lookup/doi/10.1111/j.1365-2966.2008.13480.x}, 
    DOI={10.1111/j.1365-2966.2008.13480.x}, 
    number={3},
    journal={\mnras},
    author={Padilla, Nelson D. and Strauss, Michael A.}, 
    year={2008}, 
    month=aug, 
    pages={1321}, 
    language={en} 
}

@article{VDV_2021, 
    title={Deprojecting Sérsic Profiles for Arbitrary Triaxial Shapes: Robust Measures of Intrinsic and Projected Galaxy Sizes},
    volume={914},
    publisher = {The American Astronomical Society},
    ISSN={0004-637X, 1538-4357},
    url={https://iopscience.iop.org/article/10.3847/1538-4357/abf047}, 
    DOI={10.3847/1538-4357/abf047},
     number={1}, 
    journal={\apj}, 
    author={van de Ven, Glenn and van der Wel, Arjen},
    year={2021}, 
    month = {jun},
    pages={45} 
}

@article{Hubble_1926, 
    title={Extragalactic nebulae.},
    volume={64}, 
    ISSN={0004-637X, 1538-4357}, 
    url={http://adsabs.harvard.edu/doi/10.1086/143018}, 
    DOI={10.1086/143018}, 
    journal={\apj}, 
    author={Hubble, E. P.}, 
    year={1926}, 
    month=dec,
    pages={321}, 
    language={en}
}

@article{De_Vaucouleurs_1959, 
     author = {{de Vaucouleurs}, Gerard},
        title = "{General Physical Properties of External Galaxies.}",
      journal = {HDP},
         year = 1959,
        month = jan,
       volume = {53},
        pages = {311},
          doi = {10.1007/978-3-642-45932-0_8},
       adsurl = {https://ui.adsabs.harvard.edu/abs/1959HDP....53..311D}
}

@book{Sandage_1961, 
    author = {{Sandage}, Allan},
        title = "{The Hubble Atlas of Galaxies}",
         year = 1961,
       adsurl = {https://ui.adsabs.harvard.edu/abs/1961hag..book.....S},
      publisher = {Carnegie Institution of Washington},
}

@article{Sandage_1970, 
    title={The Intrinsic Flattening of e, so, and Spiral Galaxies as Related to Galaxy Formation and Evolution}, 
    volume={160},
    ISSN={0004-637X, 1538-4357}, 
    url={http://adsabs.harvard.edu/doi/10.1086/150475},
    DOI={10.1086/150475}, 
    journal={\apj}, 
    author={Sandage, Allan and Freeman, Kenneth C. and Stokes, N. R.},
    year={1970}, 
    month={June}, 
    pages={831}, 
    language={en} 
}

@article{Binney_1978, 
    title={On the rotation of elliptical galaxies}, 
    volume={183}, 
    ISSN={0035-8711, 1365-2966}, 
    url={https://academic.oup.com/mnras/article-lookup/doi/10.1093/mnras/183.3.501}, 
    DOI={10.1093/mnras/183.3.501},
    number={3}, 
    journal={\mnras},
    author={Binney, J.}, 
    year={1978}, 
    month={July},
    pages={501}, 
    language={en} 
}

@article{Noerdlinger_1979, 
    title={The intrinsic flattening of galaxies.}, 
    volume={234}, 
    ISSN={0004-637X}, 
    url={https://ui.adsabs.harvard.edu/abs/1979ApJ...234..802N}, 
    DOI={10.1086/157559}, 
    journal={\apj}, 
    publisher={IOP}, 
    author={Noerdlinger, P. D.}, 
    year={1979}, 
    month=dec,
    pages={802}
}

@article{Binney_1985, 
    title={Testing for triaxiality with kinematic data.}, 
    volume={212}, 
    ISSN={0035-8711}, 
    url={https://ui.adsabs.harvard.edu/abs/1985MNRAS.212..767B},
    DOI={10.1093/mnras/212.4.767}, 
    journal={\mnras}, 
    publisher={OUP},
    author={Binney, J.}, 
    year={1985}, 
    month=feb,
    pages={767}
}

@article{Stark_1977, 
    title={Triaxial Models of the Bulge of M31}, 
    volume={213}, 
    ISSN={0004-637X}, 
    url={https://ui.adsabs.harvard.edu/abs/1977ApJ...213..368S}, 
    DOI={10.1086/155164}, 
    journal={\apj}, 
    publisher={IOP}, 
    author={Stark, Antony A.}, 
    year={1977},
    month=apr,
    pages={368}
}

@article{Contopoulos_1956, 
    author = {{Contopoulos}, George},
        title = "{On the Isophotes of Ellipsoidal Nebulae. With 1 Figure}",
      journal = {\zap},
         year = 1956,
        month = jan,
       volume = {39},
        pages = {126},
       adsurl = {https://ui.adsabs.harvard.edu/abs/1956ZA.....39..126C}
}

@article{RP_2013, 
    title={The intrinsic shape of galaxies in SDSS/Galaxy Zoo}, 
    volume={434},
    ISSN={0035-8711, 1365-2966}, 
    url={http://academic.oup.com/mnras/article/434/3/2153/1036270/The-intrinsic-shape-of-galaxies-in-SDSSGalaxy-Zoo},
    DOI={10.1093/mnras/stt1168}, 
    number={3}, 
    journal={\mnras},
    author={Rodríguez, Silvio and Padilla, Nelson D.}, 
    year={2013}, 
    month={September},
    pages={2153}, 
    language={en}
}

@article{Ryden_2004, 
    title={The Ellipticity of the Disks of Spiral Galaxies}, 
    volume={601},
    ISSN={0004-637X, 1538-4357}, 
    url={https://iopscience.iop.org/article/10.1086/380437},
    DOI={10.1086/380437},
    number={1}, 
    journal={\apj}, 
    author={Ryden, Barbara S.}, 
    year={2004},
    month={January}, 
    pages={214},
    language={en} 
}

@article{Vincent_Ryden_2005, 
    title={The Dependence of Galaxy Shape on Luminosity and Surface Brightness Profile}, 
    volume={623}, 
    ISSN={0004-637X, 1538-4357},
    url={https://iopscience.iop.org/article/10.1086/428765},
    DOI={10.1086/428765}, 
    number={1}, 
    journal={\apj}, 
    author={Vincent, R. Anthony and Ryden, Barbara S.}, 
    year={2005},
    month=apr, 
    pages={137}, 
    language={en} 
}

@article{Franx_1991, 
    title={The ordered nature of elliptical galaxies - Implications for their intrinsic angular momenta and shapes}, 
    volume={383}, 
    ISSN={0004-637X, 1538-4357}, 
    url={http://adsabs.harvard.edu/doi/10.1086/170769}, 
    DOI={10.1086/170769}, 
    journal={\apj},
    author={Franx, Marijn and Illingworth, Garth and De Zeeuw, Tim}, 
    year={1991}, 
    month=dec, 
    pages={112},
    language={en}
}

@article{Ryden_1992, 
    title={The Intrinsic Shapes of Elliptical Galaxies}, 
    volume={396},
    ISSN={0004-637X},
    url={https://ui.adsabs.harvard.edu/abs/1992ApJ...396..445R},
    DOI={10.1086/171731}, 
    journal={\apj},
    publisher={IOP}, 
    author={Ryden, Barbara}, 
    year={1992},
    month=sep, 
    pages={445}
}

@article{Tremblay_Merritt_1995, 
    title={The Frequency Function of Elliptical Galaxy Intrinsic Shapes}, 
    volume={110}, 
    ISSN={00046256},
    DOI={10.1086/117584}, 
    journal={\aj}, 
    author={Tremblay, Benoit and Merritt, David}, 
    year={1995}, 
    month=sep, 
    pages={1039} 
}

@article{Emsellem_2011,
    title={The ATLAS3D project - III. A census of the stellar angular momentum within the effective radius of early-type galaxies: unveiling the distribution of fast and slow rotators: Census of the stellar angular momentum in ETGs}, 
    volume={414}, 
    ISSN={00358711},
    url={https://academic.oup.com/mnras/article-lookup/doi/10.1111/j.1365-2966.2011.18496.x}, 
    DOI={10.1111/j.1365-2966.2011.18496.x}, 
    number={2}, 
    journal={\mnras}, 
    author={Emsellem, Eric and Cappellari, Michele and Krajnović, Davor and Alatalo, Katherine and Blitz, Leo and Bois, Maxime and Bournaud, Frédéric and Bureau, Martin and Davies, Roger L. and Davis, Timothy A. and De Zeeuw, P. T. and Khochfar, Sadegh and Kuntschner, Harald and Lablanche, Pierre-Yves and McDermid, Richard M. and Morganti, Raffaella and Naab, Thorsten and Oosterloo, Tom and Sarzi, Marc and Scott, Nicholas and Serra, Paolo and van de Ven, Glenn and Weijmans, Anne-Marie and Young, Lisa M.}, 
    year={2011}, 
    month={June}, 
    pages={888}, 
    language={en} 
}

@article{Cappellari_2016, 
    author = {{Cappellari}, Michele},
        title = "{Structure and Kinematics of Early-Type Galaxies from Integral Field Spectroscopy}",
      journal = {\araa},
         year = 2016,
        month = sep,
       volume = {54},
        pages = {597},
          doi = {10.1146/annurev-astro-082214-122432},
           url = {https://www.annualreviews.org/content/journals/10.1146/annurev-astro-082214-122432},
   publisher = {Annual Reviews}
}

@article{Lambas_1992, 
    title={On the true shapes of galaxies}, 
    volume={258}, 
    ISSN={0035-8711, 1365-2966}, 
    url={https://academic.oup.com/mnras/article-lookup/doi/10.1093/mnras/258.2.404}, 
    DOI={10.1093/mnras/258.2.404}, 
    number={2},
    journal={\mnras}, 
    author={Lambas, D. G. and Maddox, S. J. and Loveday, J.}, 
    year={1992}, 
    month=sep, 
    pages={404}, 
    language={en} 
}

@article{Alam_Ryden_2002, 
    title={The Shapes of Galaxies in the Sloan Digital Sky Survey}, 
    volume={570}, 
    ISSN={0004-637X, 1538-4357}, 
    url={https://iopscience.iop.org/article/10.1086/339790}, 
    DOI={10.1086/339790}, 
    number={2}, 
    journal={\apj}, 
    author={Alam, S. M. Khairul and Ryden, Barbara S.}, 
    year={2002},
    month=may, 
    pages={610}, 
    language={en} 
}

@article{Mendez-Abreu_2010,
    title={Structural properties of disk galaxies: II. Intrinsic shape of bulges}, 
    volume={521}, 
    ISSN={0004-6361, 1432-0746}, 
    url={http://www.aanda.org/10.1051/0004-6361/201014130}, 
    DOI={10.1051/0004-6361/201014130}, 
    journal={\aap}, 
    author={Méndez-Abreu, J. and Simonneau, E. and Aguerri, J. A. L. and Corsini, E. M.}, 
    year={2010}, 
    month=oct, 
    pages={A71}
}

@article{Costantin_2018, 
    title={The intrinsic shape of bulges in the CALIFA survey}, 
    volume={609}, 
    rights={https://www.edpsciences.org/en/authors/copyright-and-licensing}, 
    ISSN={0004-6361, 1432-0746}, 
    url={https://www.aanda.org/10.1051/0004-6361/201731823}, 
    DOI={10.1051/0004-6361/201731823},
    journal={\aap}, 
    author={Costantin, L. and Méndez-Abreu, J. and Corsini, E. M. and Eliche-Moral, M. C. and Tapia, T. and Morelli, L. and Dalla Bontà, E. and Pizzella, A.}, 
    year={2018},
    month=jan,
    pages={A132} 
}

@article{Caldwell_1983, 
    title={Structure and stellar content of dwarf elliptical galaxies}, 
    volume={88}, 
    ISSN={00046256}, 
    url={http://adsabs.harvard.edu/cgi-bin/bib_query?1983AJ.....88..804C},
    DOI={10.1086/113367}, 
    journal={\aj}, 
    author={Caldwell, N.}, 
    year={1983}, 
    month={June}, 
    pages={804} 
}

@article{Ichikawa_1986, 
    title={Surface photometry of dwarf elliptical galaxies in the Virgo cluster}, 
    volume={60}, 
    ISSN={0067-0049, 1538-4365}, 
    url={http://adsabs.harvard.edu/doi/10.1086/191094},
    DOI={10.1086/191094}, 
    journal={\apjs}, 
    author={Ichikawa, S.-I. and Wakamatsu, K.-I. and Okamura, S.}, 
    year={1986}, 
    month=feb, 
    pages={475}, 
    language={en}
}

@article{Ferguson_Sandage_1989, 
    author = {{Ferguson}, Henry C. and {Sandage}, Allan},
        title = "{The Spatial Distributions and Intrinsic Shapes of Dwarf Elliptical Galaxies in the Virgo and Fornax Clusters}",
      journal = {\apjl},
         year = 1989,
        month = nov,
       volume = {346},
        pages = {L53},
          doi = {10.1086/185577},
       adsurl = {https://ui.adsabs.harvard.edu/abs/1989ApJ...346L..53F}
}

@article{Ferguson_Binggeli_1994,
    author = {{Ferguson}, Henry C. and {Binggeli}, Bruno},
        title = "{Dwarf elliptical galaxies}",
      journal = {\aapr},
         year = 1994,
        month = nov,
       volume={6},
        pages={67-122},
       number = {1-2},
          doi = {10.1007/BF01208252},
}

@article{Sanchez-Janssen_2010, 
    author = {{S{\'a}nchez-Janssen}, R. and {M{\'e}ndez-Abreu}, J. and {Aguerri}, J.~A.~L.},
        title = "{Thin discs, thick dwarfs and the effects of stellar feedback}",
      journal = {\mnras},
     volume = {406},
    number = {1},
    year = {2010},
    pages = {L65-L69},
    doi = {https://doi.org/10.1111/j.1745-3933.2010.00883.x},
    url = {https://onlinelibrary.wiley.com/doi/abs/10.1111/j.1745-3933.2010.00883.x},
}

@article{Kado-Fong_2021, 
    title={The Intrinsic Shapes of Low Surface Brightness Galaxies (LSBGs): A Discriminant of LSBG Galaxy Formation Mechanisms}, 
    volume={920}, 
    ISSN={0004-637X, 1538-4357}, 
    url={https://iopscience.iop.org/article/10.3847/1538-4357/ac15f0}, 
    DOI={10.3847/1538-4357/ac15f0}, 
    number={2}, 
    journal={\apj}, 
    author={Kado-Fong, Erin and Petrescu, Mihai and Mohammad, Majid and Greco, Johnny and Greene, Jenny E. and Adams, Elizabeth A. K. and Huang, Song and Leisman, Lukas and Munshi, Ferah and Tanoglidis, Dimitrios and Van Nest, Jordan}, 
    year={2021}, 
    month=oct, 
    pages={72} 
}

@article{Cowie_1995, 
    title={Faintest Galaxy Morphologies From HST WFPC2 Imaging of the Hawaii Survey Fields}, 
    volume={110},
    ISSN={00046256}, 
    url={http://adsabs.harvard.edu/cgi-bin/bib_query?1995AJ....110.1576C}, 
    DOI={10.1086/117631}, 
    journal={\aj}, 
    author={Cowie, Lennox L. and Hu, Esther M. and Songaila, Antoinette}, 
    year={1995}, 
    month=oct, 
    pages={1576} 
}

@article{Ravindranath_2006, 
    title={The Morphological Diversities among Star‐forming Galaxies at High Redshifts in the Great Observatories Origins Deep Survey}, 
    volume={652},
    ISSN={0004-637X, 1538-4357},
    url={https://iopscience.iop.org/article/10.1086/507016}, 
    DOI={10.1086/507016}, 
    number={2},
    journal={\apj}, 
    author={Ravindranath, Swara and Giavalisco, Mauro and Ferguson, Henry C. and Conselice, Christopher and Katz, Neal and Weinberg, Martin and Lotz, Jennifer and Dickinson, Mark and Fall, S. Michael and Mobasher, Bahram and Papovich, Casey}, 
    year={2006}, 
    month=dec, 
    pages={963},
    language={en}
}

@article{Law_2012, 
    title={AN HST /WFC3-IR MORPHOLOGICAL SURVEY OF GALAXIES AT z = 1.5-3.6. I. SURVEY DESCRIPTION AND MORPHOLOGICAL PROPERTIES OF STAR-FORMING GALAXIES}, 
    volume={745},
    ISSN={0004-637X, 1538-4357},
    url={https://iopscience.iop.org/article/10.1088/0004-637X/745/1/85}, 
    DOI={10.1088/0004-637X/745/1/85}, 
    number={1}, 
    journal={\apj},
    author={Law, David R. and Steidel, Charles C. and Shapley, Alice E. and Nagy, Sarah R. and Reddy, Naveen A. and Erb, Dawn K.}, 
    year={2012}, 
    month=jan, 
    pages={85}
}

@article{Yuma_2012, 
    title={INTRINSIC SHAPE OF STAR-FORMING BzK GALAXIES. II. REST-FRAME ULTRAVIOLET AND OPTICAL STRUCTURES IN GOODS-SOUTH AND SXDS}, 
    volume={761}, 
    ISSN={0004-637X, 1538-4357}, 
    url={https://iopscience.iop.org/article/10.1088/0004-637X/761/1/19}, 
    DOI={10.1088/0004-637X/761/1/19}, 
    number={1}, 
    journal={\apj},
    author={Yuma, Suraphong and Ohta, Kouji and Yabe, Kiyoto},
    year={2012},
    month=dec, 
    pages={19} 
}

@article{Zhang_2022, 
    title={3D intrinsic shapes of quiescent galaxies in observations and simulations}, 
    volume={513}, 
    rights={https://academic.oup.com/journals/pages/open_access/funder_policies/chorus/standard_publication_model}, 
    ISSN={0035-8711, 1365-2966}, 
    url={https://academic.oup.com/mnras/article/513/4/4814/6590677}, 
    DOI={10.1093/mnras/stac1083}, 
    number={4}, 
    journal={\mnras},
    author={Zhang, Junkai and Wuyts, Stijn and Witten, Callum and Avery, Charlotte R and Hao, Lei and Sharma, Raman and Shen, Juntai and Toshikawa, Jun and Villforth, Carolin}, 
    year={2022}, 
    month=may, 
    pages={4814}, 
    language={en} 
}

@article{Hamilton-Campos_2023, 
    title={The Physical Thickness of Stellar Disks to z ∼ 2}, 
    volume={956}, 
    ISSN={0004-637X, 1538-4357}, 
    url={https://iopscience.iop.org/article/10.3847/1538-4357/acf211}, 
    DOI={10.3847/1538-4357/acf211}, 
    number={2}, 
    journal={\apj},
    author={Hamilton-Campos, Kathleen A. and Simons, Raymond C. and Peeples, Molly S. and Snyder, Gregory F. and Heckman, Timothy M.}, 
    year={2023}, 
    month=oct, 
    pages={147} 
}

@article{Elmegreen_2004A, 
    title={Discovery of Face-on Counterparts of Chain Galaxies in the Tadpole Advanced Camera for Surveys Field}, 
    volume={604}, 
    ISSN={0004-637X, 1538-4357}, 
    url={https://iopscience.iop.org/article/10.1086/383312}, 
    DOI={10.1086/383312}, 
    number={1}, 
    journal={\apj}, 
    author={Elmegreen, Debra Meloy and Elmegreen, Bruce G. and Hirst, Amelia C.}, 
    year={2004}, 
    month=feb, 
    pages={L21}, 
    language={en} 
}

@article{Elmegreen_2005, 
    title={Galaxy Morphologies in the Hubble Ultra Deep Field: Dominance of Linear Structures at the Detection Limit}, 
    volume={631}, 
    ISSN={0004-637X, 1538-4357},
    url={https://iopscience.iop.org/article/10.1086/432502},
    DOI={10.1086/432502}, 
    number={1},
    journal={\apj}, 
    author={Elmegreen, Debra Meloy and Elmegreen, Bruce G. and Rubin, Douglas S. and Schaffer, Meredith A.}, 
    year={2005},
    month=sep, 
    pages={85},
    language={en} 
}

@article{Elmegreen_2004B, 
    title={Chain Galaxies in the Tadpole Advanced Camera for Surveys Field}, 
    volume={603}, 
    ISSN={0004-637X, 1538-4357}, 
    url={https://iopscience.iop.org/article/10.1086/381357}, 
    DOI={10.1086/381357}, 
    number={1},
    journal={\apj}, 
    author={Elmegreen, Debra Meloy and Elmegreen, Bruce G. and Sheets, Clara M.}, 
    year={2004}, 
    month=nov, 
    pages={74},
    language={en} 
}

@article{Bruce_2012, 
    title={The morphologies of massive galaxies at 1 < z < 3 in the CANDELS-UDS field: compact bulges, and the rise and fall of massive discs: Morphologies of massive galaxies at 1 < z < 3}, 
    volume={427}, 
    ISSN={00358711, 13652966}, 
    url={https://academic.oup.com/mnras/article-lookup/doi/10.1111/j.1365-2966.2012.22087.x}, 
    DOI={10.1111/j.1365-2966.2012.22087.x}, 
    number={2}, 
    journal={\mnras}, 
    author={Bruce, V. A. and Dunlop, J. S. and Cirasuolo, M. and McLure, R. J. and Targett, T. A. and Bell, E. F. and Croton, D. J. and Dekel, A. and Faber, S. M. and Ferguson, H. C. and Grogin, N. A. and Kocevski, D. D. and Koekemoer, A. M. and Koo, D. C. and Lai, K. and Lotz, J. M. and McGrath, E. J. and Newman, J. A. and Van Der Wel, A.}, 
    year={2012}, 
    month=dec, 
    pages={1666},
    language={en}
}

@article{Guo_2009, 
    title={Structural properties of central galaxies in groups and clusters}, 
    volume={398}, 
    ISSN={00358711, 13652966}, 
    url={https://academic.oup.com/mnras/article-lookup/doi/10.1111/j.1365-2966.2009.15223.x}, 
    DOI={10.1111/j.1365-2966.2009.15223.x}, 
    number={3}, 
    journal={\mnras}, 
    author={Guo, Yicheng and McIntosh, Daniel H. and Mo, H. J. and Katz, Neal and Van Den Bosch, Frank C. and Weinberg, Martin and Weinmann, Simone M. and Pasquali, Anna and Yang, Xiaohu}, 
    year={2009},
    month={September}, 
    pages={1129},
    language={en} 
}

@article{vdW_2012, 
    title={STRUCTURAL PARAMETERS OF GALAXIES IN CANDELS}, 
    volume={203}, 
    ISSN={0067-0049, 1538-4365}, 
    url={https://iopscience.iop.org/article/10.1088/0067-0049/203/2/24}, 
    DOI={10.1088/0067-0049/203/2/24}, 
    number={2}, 
    journal={\apjs}, 
    author={van Der Wel, A. and Bell, E. F. and Häussler, B. and McGrath, E. J. and Chang, Yu-Yen and Guo, Yicheng and McIntosh, D. H. and Rix, H.-W. and Barden, M. and Cheung, E. and Faber, S. M. and Ferguson, H. C. and Galametz, A. and Grogin, N. A. and Hartley, W. and Kartaltepe, J. S. and Kocevski, D. D. and Koekemoer, A. M. and Lotz, J. and Mozena, M. and Peth, M. A. and Peng, Chien Y.}, 
    year={2012}, 
    month=dec, 
    pages={24}
}

@article{Sazonova_2026, 
    title={statmorph-lsst: Quantifying and correcting morphological biases in galaxy surveys}, 
    volume={9}, 
    DOI={10.33232/001c.160562}, 
    journal={OJAp}, 
    author={Sazonova, Elizaveta and Morgan, Cameron R. and Balogh, Michael and Blaña, Matías and Bornancini, Carlos G. and Cotter, Aidan P. and Donevski, Darko and Graham, Alister W. and Toledo, Hector M. Hernandez and Holwerda, Benne W. and Kartaltepe, Jeyhan S. and Martin, Garreth and Pearson, William J. and Ragusa, Rossella and Rodriguez-Gomez, Vicente and Rutkowski, Michael J. and Vázquez-Mata, Jose Antonio and Windhorst, Rogier A. and Yuzovitskiy, Jacob}, 
    year={2026}, 
    month=apr, 
    language={en},
	publisher = {Maynooth Academic Publishing},
}

@article{Shuntov_2026, 
    title={The stellar mass function of quiescent and star-forming galaxies and its dependence on morphology in COSMOS-Web}, 
    volume={707}, 
    rights={https://creativecommons.org/licenses/by/4.0},
    ISSN={0004-6361, 1432-0746},
    url={https://www.aanda.org/10.1051/0004-6361/202558022}, 
    DOI={10.1051/0004-6361/202558022},
    journal={\aap}, 
    author={Shuntov, Marko and Ilbert, Olivier and Del P. Lagos, Claudia and Toft, Sune and Valentino, Francesco and Mercier, Wilfried and Akins, Hollis B. and Binh, Nguyen and Brinch, Malte and Casey, Caitlin M. and Franco, Maximilien and Gentile, Fabrizio and Gozaliasl, Ghassem and Haghjoo, Aryana and Harish, Santosh and Hirschmann, Michaela and Huertas-Company, Marc and Jin, Shuowen and Kartaltepe, Jeyhan S. and Koekemoer, Anton M. and Laigle, Clotilde and Lewis, Joseph S. W. and Magdis, Georgios E. and Joy McCracken, Henry and Mobasher, Bahram and Moutard, Thibaud and Oesch, Pascal A. and Paquereau, Louise and Renzini, Alvio and Rich, Michael R. and Sanders, David B. and Toni, Greta and Tresse, Laurence and Weibel, Andrea and Weaver, John R. and Yang, Lilan}, 
    year={2026},
    month=mar, 
    pages={A391} 
}

@article{Dimauro_2022, 
    title={Coincidence between morphology and star formation activity through cosmic time: the impact of the bulge growth}, 
    volume={513}, 
    rights={https://academic.oup.com/journals/pages/open_access/funder_policies/chorus/standard_publication_model}, 
    ISSN={0035-8711, 1365-2966}, 
    url={https://academic.oup.com/mnras/article/513/1/256/6563709}, 
    DOI={10.1093/mnras/stac884}, 
    number={1}, 
    journal={\mnras}, 
    author={Dimauro, Paola and Daddi, Emanuele and Shankar, Francesco and Cattaneo, Andrea and Huertas-Company, Marc and Bernardi, Mariangela and Caro, Fernando and Dupke, Renato and Häußler, Boris and Johnston, Evelyn and Cortesi, Arianna and Mei, Simona and Peletier, Reynier},
    year={2022},
    month=apr,
    pages={256}, 
    language={en} 
}

@article{Yang_2026,
       author = {{Yang}, Lilan and {Kartaltepe}, Jeyhan S. and {Drakos}, Nicole E. and {Faisst}, Andreas L. and {Flayhart}, Carter and {Franco}, Maximilien and {Koekemoer}, Anton M. and {Ilbert}, Olivier and {Akins}, Hollis B. and {Casey}, Caitlin M. and et al.},
        title = "{COSMOS-Web: A Multi-wavelength Morphological Catalog of \raisebox{-0.5ex}\textasciitilde780,000 Galaxies}",
         year = 2026,
        month = jun,
        journal = {\apjs},
          eid = {arXiv:2606.14869},
        pages = {arXiv:2606.14869},
          doi = {10.48550/arXiv.2606.14869},
archivePrefix = {arXiv},
       eprint = {2606.14869},
 primaryClass = {astro-ph.GA},
       adsurl = {https://ui.adsabs.harvard.edu/abs/2026arXiv260614869Y}
}

@article{Haussler_2007, 
    title={GEMS: Galaxy Fitting Catalogs and Testing Parametric Galaxy Fitting Codes: GALFIT and GIM2D},
    volume={172}, 
    ISSN={0067-0049, 1538-4365}, 
    url={https://iopscience.iop.org/article/10.1086/518836},
    DOI={10.1086/518836}, 
    number={2}, 
    journal={\apjs}, 
    author={Häussler, Boris and McIntosh, Daniel H. and Barden, Marco and Bell, Eric F. and Rix, Hans‐Walter and Borch, Andrea and Beckwith, Steven V. W. and Caldwell, John A. R. and Heymans, Catherine and Jahnke, Knud and Jogee, Shardha and Koposov, Sergey E. and Meisenheimer, Klaus and Sanchez, Sebastian F. and Somerville, Rachel S. and Wisotzki, Lutz and Wolf, Christian}, 
    year={2007}, 
    month=oct, 
    pages={615}, 
    language={en}
}

@ARTICLE{Benavides,
       author = {{Benavides}, Jose and {Sales}, Laura V. and {Navarro}, Julio F. and {White}, Simon D.~M. and {Frenk}, Carlos S. and {Oman}, Kyle A. and {Cole}, Shaun},
        title = "{The Abundance of Thin Dwarf Galaxies: a Challenge for Cosmological Simulations}",
      journal = {OJAp},
         year = 2026,
        month = may,
       volume = {9},
        pages = {62091},
          doi = {10.33232/001c.162091},
archivePrefix = {arXiv},
       eprint = {2512.11035},
 primaryClass = {astro-ph.GA},
       adsurl = {https://ui.adsabs.harvard.edu/abs/2026OJAp....962091B},
}

@article{PyMC, 
    title={Probabilistic programming in Python using PyMC3}, 
    volume={2}, 
    rights={http://creativecommons.org/licenses/by/4.0/}, 
    ISSN={2376-5992}, 
    url={https://peerj.com/articles/cs-55}, 
    DOI={10.7717/peerj-cs.55}, 
    journal={PeerJ~Comput.~Sci.}, 
    author={Salvatier, John and Wiecki, Thomas V. and Fonnesbeck, Christopher}, 
    year={2016}, 
    month=apr,
    pages={e55},
    language={en}
}

@inproceedings{SE++1, 
    author = {{Bertin}, E. and {Schefer}, M. and {Apostolakos}, N. and {{\'A}lvarez-Ayll{\'o}n}, A. and {Dubath}, P. and {K{\"u}mmel}, M.},
        title = "{The SourceXtractor++ Software}",
    booktitle = {Astronomical Data Analysis Software and Systems XXIX},
         year = 2020,
       editor = {{Pizzo}, R. and {Deul}, E.~R. and {Mol}, J.~D. and {de Plaa}, J. and {Verkouter}, H.},
       series = {ASP Conf. Ser.},
       volume = {527},
        month = jan,
        pages = {461},
       adsurl = {https://ui.adsabs.harvard.edu/abs/2020ASPC..527..461B}
}

@article{SE++2,
      author = {{K{\"u}mmel}, M. and {{\'A}lvarez-Ayll{\'o}n}, A. and {Bertin}, E. and {Dubath}, P. and {Gavazzi}, R. and {Hartley}, W. and {Schefer}, M.},
        title = "{Using the SourceXtractor++ package for data reduction}",
         year = 2022,
        month = dec,
          eid = {arXiv:2212.02428},
          pages = {arXiv:2212.02428},
archivePrefix = {arXiv},
       eprint = {2212.02428},
 primaryClass = {astro-ph.IM},
       adsurl = {https://ui.adsabs.harvard.edu/abs/2022arXiv221202428K}
}

@article{GALAPAGOS, 
    title={galapagos: from pixels to parameters}, 
    volume={422}, 
    ISSN={0035-8711}, 
    url={https://doi.org/10.1111/j.1365-2966.2012.20619.x}, 
    DOI={10.1111/j.1365-2966.2012.20619.x}, 
    number={1}, 
    journal={\mnras}, 
    author={Barden, Marco and Häußler, Boris and Peng, Chien Y. and McIntosh, Daniel H. and Guo, Yicheng}, 
    year={2012}, 
    month=may,
    pages={449} 
}

@article{Cutana,
    author = {{G{\'o}mez}, Pablo and {Ruhberg}, Laslo Erik and {Anett Remmelgas}, Kristin and {O'Ryan}, David},
        title = "{Cutana: A High-Performance Tool for Astronomical Image Cutout Generation at Petabyte Scale}",
         year = 2025,
        month = nov,
        eid = {arXiv:2511.04429},
        pages = {arXiv:2511.04429},
        doi = {10.48550/arXiv.2511.04429},
archivePrefix = {arXiv},
       eprint = {2511.04429},
 primaryClass = {astro-ph.IM},
       adsurl = {https://ui.adsabs.harvard.edu/abs/2025arXiv251104429G}
}

@article{corner,
      doi = {10.21105/joss.00024},
      url = {https://doi.org/10.21105/joss.00024},
      year  = {2016},
      month = {jun},
      publisher = {The Open Journal},
      volume = {1},
      number = {2},
      pages = {24},
      author = {Daniel Foreman-Mackey},
      title = {corner.py: Scatterplot matrices in Python},
      journal = {JOSS}
    }

@article{Sersic_1963, 
    author = {{S{\'e}rsic}, J.~L.},
        title = "{Influence of the atmospheric and instrumental dispersion on the brightness distribution in a galaxy}",
      journal = {BAAA},
         year = 1963,
        month = feb,
       volume = {6},
        pages = {41},
       adsurl = {https://ui.adsabs.harvard.edu/abs/1963BAAA....6...41S}
}

@article{Pozzetti_2010, 
    title={zCOSMOS – 10k-bright spectroscopic sample: The bimodality in the galaxy stellar mass function: exploring its evolution with redshift⋆}, 
    volume={523}, 
    ISSN={0004-6361, 1432-0746}, 
    url={http://www.aanda.org/10.1051/0004-6361/200913020}, 
    DOI={10.1051/0004-6361/200913020}, 
    journal={\aap}, 
    author={Pozzetti, L. and Bolzonella, M. and Zucca, E. and Zamorani, G. and Lilly, S. and Renzini, A. and Moresco, M. and Mignoli, M. and Cassata, P. and Tasca, L. and Lamareille, F. and Maier, C. and Meneux, B. and Halliday, C. and Oesch, P. and Vergani, D. and Caputi, K. and Kovač, K. and Cimatti, A. and Cucciati, O. and Iovino, A. and Peng, Y. and Carollo, M. and Contini, T. and Kneib, J.-P. and Le Févre, O. and Mainieri, V. and Scodeggio, M. and Bardelli, S. and Bongiorno, A. and Coppa, G. and De La Torre, S. and De Ravel, L. and Franzetti, P. and Garilli, B. and Kampczyk, P. and Knobel, C. and Le Borgne, J.-F. and Le Brun, V. and Pellò, R. and Perez Montero, E. and Ricciardelli, E. and Silverman, J. D. and Tanaka, M. and Tresse, L. and Abbas, U. and Bottini, D. and Cappi, A. and Guzzo, L. and Koekemoer, A. M. and Leauthaud, A. and Maccagni, D. and Marinoni, C. and McCracken, H. J. and Memeo, P. and Porciani, C. and Scaramella, R. and Scarlata, C. and Scoville, N.}, 
    year={2010}, 
    month=nov, 
    pages={A13}
}

@article{Oke_1974, 
    title={Absolute Spectral Energy Distributions for White Dwarfs}, 
    volume={27}, 
    ISSN={0067-0049}, 
    url={https://ui.adsabs.harvard.edu/abs/1974ApJS...27...21O},
    DOI={10.1086/190287},  
    journal={\apjs}, 
    publisher={IOP}, 
    author={Oke, J. B.}, 
    year={1974}, 
    month=feb, 
    pages={21} 
}

@ARTICLE{2016A&A...594A..13P,
       author = {{Planck Collaboration} and {Ade}, P.~A.~R. and {Aghanim}, N. and {Arnaud}, M. and {Ashdown}, M. and {Aumont}, J. and {Baccigalupi}, C. and {Banday}, A.~J. and {Barreiro}, R.~B. and {Bartlett}, J.~G. and {Bartolo}, N. and {Battaner}, E. and {Battye}, R. and {Benabed}, K. and {Beno{\^\i}t}, A. and {Benoit-L{\'e}vy}, A. and {Bernard}, J.-P. and {Bersanelli}, M. and {Bielewicz}, P. and {Bock}, J.~J. and {Bonaldi}, A. and {Bonavera}, L. and {Bond}, J.~R. and {Borrill}, J. and {Bouchet}, F.~R. and {Boulanger}, F. and {Bucher}, M. and {Burigana}, C. and {Butler}, R.~C. and {Calabrese}, E. and {Cardoso}, J.-F. and {Catalano}, A. and {Challinor}, A. and {Chamballu}, A. and {Chary}, R.-R. and {Chiang}, H.~C. and {Chluba}, J. and {Christensen}, P.~R. and {Church}, S. and {Clements}, D.~L. and {Colombi}, S. and {Colombo}, L.~P.~L. and {Combet}, C. and {Coulais}, A. and {Crill}, B.~P. and {Curto}, A. and {Cuttaia}, F. and {Danese}, L. and {Davies}, R.~D. and {Davis}, R.~J. and {de Bernardis}, P. and {de Rosa}, A. and {de Zotti}, G. and {Delabrouille}, J. and {D{\'e}sert}, F.-X. and {Di Valentino}, E. and {Dickinson}, C. and {Diego}, J.~M. and {Dolag}, K. and {Dole}, H. and {Donzelli}, S. and {Dor{\'e}}, O. and {Douspis}, M. and {Ducout}, A. and {Dunkley}, J. and {Dupac}, X. and {Efstathiou}, G. and {Elsner}, F. and {En{\ss}lin}, T.~A. and {Eriksen}, H.~K. and {Farhang}, M. and {Fergusson}, J. and {Finelli}, F. and {Forni}, O. and {Frailis}, M. and {Fraisse}, A.~A. and {Franceschi}, E. and {Frejsel}, A. and {Galeotta}, S. and {Galli}, S. and {Ganga}, K. and {Gauthier}, C. and {Gerbino}, M. and {Ghosh}, T. and {Giard}, M. and {Giraud-H{\'e}raud}, Y. and {Giusarma}, E. and {Gjerl{\o}w}, E. and {Gonz{\'a}lez-Nuevo}, J. and {G{\'o}rski}, K.~M. and {Gratton}, S. and {Gregorio}, A. and {Gruppuso}, A. and {Gudmundsson}, J.~E. and {Hamann}, J. and {Hansen}, F.~K. and {Hanson}, D. and {Harrison}, D.~L. and {Helou}, G. and {Henrot-Versill{\'e}}, S. and {Hern{\'a}ndez-Monteagudo}, C. and {Herranz}, D. and {Hildebrandt}, S.~R. and {Hivon}, E. and {Hobson}, M. and {Holmes}, W.~A. and {Hornstrup}, A. and {Hovest}, W. and {Huang}, Z. and {Huffenberger}, K.~M. and {Hurier}, G. and {Jaffe}, A.~H. and {Jaffe}, T.~R. and {Jones}, W.~C. and {Juvela}, M. and {Keih{\"a}nen}, E. and {Keskitalo}, R. and {Kisner}, T.~S. and {Kneissl}, R. and {Knoche}, J. and {Knox}, L. and {Kunz}, M. and {Kurki-Suonio}, H. and {Lagache}, G. and {L{\"a}hteenm{\"a}ki}, A. and {Lamarre}, J.-M. and {Lasenby}, A. and {Lattanzi}, M. and {Lawrence}, C.~R. and {Leahy}, J.~P. and {Leonardi}, R. and {Lesgourgues}, J. and {Levrier}, F. and {Lewis}, A. and {Liguori}, M. and {Lilje}, P.~B. and {Linden-V{\o}rnle}, M. and {L{\'o}pez-Caniego}, M. and {Lubin}, P.~M. and {Mac{\'\i}as-P{\'e}rez}, J.~F. and {Maggio}, G. and {Maino}, D. and {Mandolesi}, N. and {Mangilli}, A. and {Marchini}, A. and {Maris}, M. and {Martin}, P.~G. and {Martinelli}, M. and {Mart{\'\i}nez-Gonz{\'a}lez}, E. and {Masi}, S. and {Matarrese}, S. and {McGehee}, P. and {Meinhold}, P.~R. and {Melchiorri}, A. and {Melin}, J.-B. and {Mendes}, L. and {Mennella}, A. and {Migliaccio}, M. and {Millea}, M. and {Mitra}, S. and {Miville-Desch{\^e}nes}, M.-A. and {Moneti}, A. and {Montier}, L. and {Morgante}, G. and {Mortlock}, D. and {Moss}, A. and {Munshi}, D. and {Murphy}, J.~A. and {Naselsky}, P. and {Nati}, F. and {Natoli}, P. and {Netterfield}, C.~B. and {N{\o}rgaard-Nielsen}, H.~U. and {Noviello}, F. and {Novikov}, D. and {Novikov}, I. and {Oxborrow}, C.~A. and {Paci}, F. and {Pagano}, L. and {Pajot}, F. and {Paladini}, R. and {Paoletti}, D. and {Partridge}, B. and {Pasian}, F. and {Patanchon}, G. and {Pearson}, T.~J. and {Perdereau}, O. and {Perotto}, L. and {Perrotta}, F. and {Pettorino}, V. and {Piacentini}, F. and {Piat}, M. and {Pierpaoli}, E. and {Pietrobon}, D. and {Plaszczynski}, S. and {Pointecouteau}, E. and {Polenta}, G. and {Popa}, L. and {Pratt}, G.~W. and {Pr{\'e}zeau}, G.},
        title = "{Planck 2015 results. XIII. Cosmological parameters}",
      journal = {\aap},
         year = 2016,
        month = sep,
       volume = {594},
        pages = {63},
          doi = {10.1051/0004-6361/201525830},
       adsurl = {https://ui.adsabs.harvard.edu/abs/2016A&A...594A..13P}
}

@article{Van_Dokkum_2015, 
    title={FORMING COMPACT MASSIVE GALAXIES},
    volume={813},
    rights={http://iopscience.iop.org/info/page/text-and-data-mining}, 
    ISSN={1538-4357}, 
    url={https://iopscience.iop.org/article/10.1088/0004-637X/813/1/23}, 
    DOI={10.1088/0004-637X/813/1/23}, 
    number={1}, 
    journal={\apj}, 
    author={Van Dokkum, Pieter G. and Nelson, Erica June and Franx, Marijn and Oesch, Pascal and Momcheva, Ivelina and Brammer, Gabriel and Schreiber, Natascha M. Förster and Skelton, Rosalind E. and Whitaker, Katherine E. and van der Wel, Arjen and Bezanson, Rachel and Fumagalli, Mattia and Illingworth, Garth D. and Kriek, Mariska and Leja, Joel and Wuyts, Stijn}, 
    year={2015}, 
    month=oct, 
    pages={23} 
}

@article{Rix_Zaritsky_1995, 
    title={Nonaxisymmetric Structures in the Stellar Disks of Galaxies}, 
    volume={447}, 
    ISSN={0004-637X, 1538-4357}, 
    url={http://adsabs.harvard.edu/doi/10.1086/175858},
    DOI={10.1086/175858}, 
    journal={\apj},
    publisher={IOP},
    author={Rix, Hans-Walter and Zaritsky, Dennis}, 
    year={1995}, 
    month={July},
    pages={82}, 
    language={en}
}

@article{Brennan_2015, 
    title={Quenching and morphological transformation in semi-analytic models and CANDELS}, 
    volume={451}, 
    ISSN={1365-2966, 0035-8711}, 
    url={http://academic.oup.com/mnras/article/451/3/2933/1178833/Quenching-and-morphological-transformation-in}, 
    DOI={10.1093/mnras/stv1007}, 
    number={3}, 
    journal={\mnras}, 
    author={Brennan, Ryan and Pandya, Viraj and Somerville, Rachel S. and Barro, Guillermo and Taylor, Edward N. and Wuyts, Stijn and Bell, Eric F. and Dekel, Avishai and Ferguson, Henry C. and McIntosh, Daniel H. and Papovich, Casey and Primack, Joel},
    year={2015}, 
    month=aug, 
    pages={2933}, 
    language={en}
}

@article{Dekel_Burkert_2014, 
    title={Wet disc contraction to galactic blue nuggets and quenching to red nuggets}, 
    volume={438}, 
    ISSN={0035-8711, 1365-2966},
    url={http://academic.oup.com/mnras/article/438/2/1870/1017947/Wet-disc-contraction-to-galactic-blue-nuggets-and}, 
    DOI={10.1093/mnras/stt2331}, 
    number={2}, 
    journal={\mnras}, 
    author={Dekel, A. and Burkert, A.},
    year={2014}, 
    month=feb, 
    pages={1870}, 
    language={en} 
}

@article{Zolotov_2015, 
    title={Compaction and quenching of high-z galaxies in cosmological simulations: blue and red nuggets}, 
    volume={450}, 
    ISSN={0035-8711, 1365-2966},
    url={http://academic.oup.com/mnras/article/450/3/2327/1057925/Compaction-and-quenching-of-highz-galaxies-in}, 
    DOI={10.1093/mnras/stv740}, 
    number={3}, 
    journal={\mnras}, 
    author={Zolotov, Adi and Dekel, Avishai and Mandelker, Nir and Tweed, Dylan and Inoue, Shigeki and DeGraf, Colin and Ceverino, Daniel and Primack, Joel R. and Barro, Guillermo and Faber, Sandra M.}, 
    year={2015}, 
    month={july},
    pages={2327},
    language={en} 
}

@article{Tacchella_2016, 
    title={Evolution of density profiles in high- z galaxies: compaction and quenching inside-out}, 
    volume={458}, 
    ISSN={0035-8711, 1365-2966},
    url={https://academic.oup.com/mnras/article-lookup/doi/10.1093/mnras/stw303}, 
    DOI={10.1093/mnras/stw303}, 
    number={1}, 
    journal={\mnras},
    author={Tacchella, Sandro and Dekel, Avishai and Carollo, C. Marcella and Ceverino, Daniel and DeGraf, Colin and Lapiner, Sharon and Mandelker, Nir and Primack, Joel R.}, 
    year={2016},
    month=may, 
    pages={242}, 
    language={en} 
}

@article{Lapiner_2023, 
    author = {{Lapiner}, Sharon and {Dekel}, Avishai and {Freundlich}, Jonathan and {Ginzburg}, Omri and {Jiang}, Fangzhou and {Kretschmer}, Michael and {Tacchella}, Sandro and {Ceverino}, Daniel and {Primack}, Joel},
        title = "{Wet compaction to a blue nugget: a critical phase in galaxy evolution}",
      journal = {\mnras},
         year = 2023,
        month = jul,
       volume = {522},
       number = {3},
        pages={4515},
          ISSN={0035-8711}, 
          url={https://doi.org/10.1093/mnras/stad1263}, 
          DOI={10.1093/mnras/stad1263},
       adsurl = {https://ui.adsabs.harvard.edu/abs/2023MNRAS.522.4515L}
}

@article{NUTS, 
    author  = {Matthew D. Hoffman and Andrew Gelman},
    title   = {The No-U-Turn Sampler: Adaptively Setting Path Lengths in Hamiltonian Monte Carlo},
    journal = {JMLR},
    year    = {2014},
    volume  = {15},
    number  = {47},
    pages   = {1593},
    url     = {http://jmlr.org/papers/v15/hoffman14a.html}
}

@article{Zoobot, 
    title={Zoobot: Adaptable Deep Learning Models for Galaxy Morphology}, 
    volume={8}, 
    rights={http://creativecommons.org/licenses/by/4.0/}, 
    ISSN={2475-9066}, 
    url={https://joss.theoj.org/papers/10.21105/joss.05312},
    DOI={10.21105/joss.05312}, 
    number={85}, 
    journal={JOSS}, 
    author={Walmsley, Mike and Allen, Campbell and Aussel, Ben and Bowles, Micah and Gregorowicz, Kasia and Slijepcevic, Inigo Val and Lintott, Chris J. and Scaife, Anna M. M. and Jabłońska, Maja and Karchev, Kosio and Lanzieri, Denise and Mohan, Devina and O’Ryan, David and Saiguhan, Bharath and Suárez, Crisel and Guerra-Varas, Nicolás and Velu, Renuka}, 
    year={2023}, 
    month=may, 
    pages={5312}
}

@article{Rhat, 
    author = {{Vehtari}, Aki and {Gelman}, Andrew and {Simpson}, Daniel and {Carpenter}, Bob and {B{\"u}rkner}, Paul-Christian},
    title = {{Rank-Normalization, Folding, and Localization: An Improved $\widehat{R}$ for Assessing Convergence of MCMC (with Discussion)}},
    volume = {16},
    journal = {Bayesian~Anal.},
    number = {2},
    publisher = {International Society for Bayesian Analysis},
    pages = {667-718},
    year = {2021},
    doi = {10.1214/20-BA1221},
    URL = {https://doi.org/10.1214/20-BA1221}
}

@book{Gelman_2013, 
    title={Bayesian Data Analysis},
    author={Gelman, Andrew and Carlin, John B and Stern, Hal S and Dunson, David B and Vehtari, Aki and Rubin, Donald B},
    edition={3rd},
    year={2013},
    publisher={CRC Press},
    address={Boca Raton, FL},
    isbn={9781439840955}
}

@article{Vats_2019,
    author = {{Vats}, Dootika and {Flegal}, James M. and {Jones}, Galin L.},
        title = "{Multivariate output analysis for Markov chain Monte Carlo}",
      journal = {Biometrika},
         year = 2019,
        month = jun,
       volume = {106},
       number = {2},
        pages = {321-337},
          doi = {10.1093/biomet/asz002},
       ISSN={0006-3444}, 
       url={https://doi.org/10.1093/biomet/asz002},
}

@article{Gong_Flegal_2016, 
    title={A Practical Sequential Stopping Rule for High-Dimensional Markov Chain Monte Carlo}, 
    volume={25}, 
    ISSN={1061-8600, 1537-2715}, 
    url={https://www.tandfonline.com/doi/full/10.1080/10618600.2015.1044092}, DOI={10.1080/10618600.2015.1044092}, 
    number={3}, 
    journal={JCJS},
    author={Gong, Lei and Flegal, James M.},
    year={2016},
    month={July},
    pages={684–700}, 
    language={en} 
}

@article{Flegal_Jones_2010, 
    title={Batch means and spectral variance estimators in Markov chain Monte Carlo}, 
    volume={38}, 
    ISSN={0090-5364}, 
    url={https://projecteuclid.org/journals/annals-of-statistics/volume-38/issue-2/Batch-means-and-spectral-variance-estimators-in-Markov-chain-Monte/10.1214/09-AOS735.full}, 
    DOI={10.1214/09-AOS735}, 
    number={2}, 
    journal={Ann.~Stat.}, 
    author={Flegal, James M. and Jones, Galin L.}, 
    year={2010}, 
    month=apr
}

@inbook{Flegal_Jones_2011, 
    author = {{Flegal}, J.~M. and {Jones}, G.~L.},
    title = "{Implementing MCMC: Estimating with confidence}",
booktitle = {Handbook of Markov Chain Monte Carlo},
     year = 2011,
   editor = {{Brooks}, S. and {Gelman}, A. and {Jones}, G. and {Meng}, X.-L.},
publisher = {CRC Press},
  address = {Boca Raton},
    pages = {175}
}

@article{BFMI1, 
       author = {{Betancourt}, Michael},
        title = "{Diagnosing Suboptimal Cotangent Disintegrations in Hamiltonian Monte Carlo}",
         year = 2016,
        month = apr,
          eid = {arXiv:1604.00695},
          pages = {arXiv:1604.00695},
archivePrefix = {arXiv},
       eprint = {1604.00695},
 primaryClass = {stat.ME},
       adsurl = {https://ui.adsabs.harvard.edu/abs/2016arXiv160400695B}
}

@article{Ceverino_2015, 
    title={Formation of elongated galaxies with low masses at high redshift}, 
    volume={453}, 
    ISSN={0035-8711, 1365-2966}, 
    url={https://academic.oup.com/mnras/article-lookup/doi/10.1093/mnras/stv1603}, 
    DOI={10.1093/mnras/stv1603}, 
    number={1}, 
    journal={\mnras}, 
    author={Ceverino, Daniel and Primack, Joel and Dekel, Avishai}, 
    year={2015},
    month=oct, 
    pages={408}, 
    language={en} 
}

@article{Tomassetti_2016, 
    title={Evolution of galaxy shapes from prolate to oblate through compaction events}, 
    volume={458}, 
    ISSN={0035-8711, 1365-2966}, 
    url={https://academic.oup.com/mnras/article-lookup/doi/10.1093/mnras/stw606}, 
    DOI={10.1093/mnras/stw606}, 
    number={4}, 
    journal={\mnras}, 
    author={Tomassetti, Matteo and Dekel, Avishai and Mandelker, Nir and Ceverino, Daniel and Lapiner, Sharon and Faber, Sandra and Kneller, Omer and Primack, Joel and Sai, Tanmayi}, 
    year={2016}, 
    month={June}, 
    pages={4477}, 
    language={en} 
}

@article{Pillepich_2019, 
    title={First results from the TNG50 simulation: the evolution of stellar and gaseous discs across cosmic time}, 
    volume={490}, 
    rights={https://academic.oup.com/journals/pages/open_access/funder_policies/chorus/standard_publication_model}, 
    ISSN={0035-8711, 1365-2966}, 
    url={https://academic.oup.com/mnras/article/490/3/3196/5566345}, 
    DOI={10.1093/mnras/stz2338}, 
    number={3}, 
    journal={\mnras}, 
    author={Pillepich, Annalisa and Nelson, Dylan and Springel, Volker and Pakmor, Rüdiger and Torrey, Paul and Weinberger, Rainer and Vogelsberger, Mark and Marinacci, Federico and Genel, Shy and van der Wel, Arjen and Hernquist, Lars}, 
    year={2019}, 
    month=dec, 
    pages={3196}, 
    language={en}
}

@article{COSMOS, 
    title={The Cosmic Evolution Survey (COSMOS): Overview}, 
    volume={172}, 
    ISSN={0067-0049, 1538-4365}, 
    url={https://iopscience.iop.org/article/10.1086/516585}, 
    DOI={10.1086/516585}, 
    number={1}, 
    journal={\apjs}, 
    author={Scoville, N. and Aussel, H. and Brusa, M. and Capak, P. and Carollo, C. M. and Elvis, M. and Giavalisco, M. and Guzzo, L. and Hasinger, G. and Impey, C. and Kneib, J.‐P. and LeFevre, O. and Lilly, S. J. and Mobasher, B. and Renzini, A. and Rich, R. M. and Sanders, D. B. and Schinnerer, E. and Schminovich, D. and Shopbell, P. and Taniguchi, Y. and Tyson, N. D.}, 
    year={2007}, 
    month={September}, 
    pages={1}, 
    language={en} 
}

@article{CANDELS1, 
    title={CANDELS: THE COSMIC ASSEMBLY NEAR-INFRARED DEEP EXTRAGALACTIC LEGACY SURVEY}, 
    volume={197}, 
    ISSN={0067-0049, 1538-4365}, 
    url={https://iopscience.iop.org/article/10.1088/0067-0049/197/2/35}, 
    DOI={10.1088/0067-0049/197/2/35}, 
    number={2}, 
    journal={\apjs}, 
    author={Grogin, Norman A. and Kocevski, Dale D. and Faber, S. M. and Ferguson, Henry C. and Koekemoer, Anton M. and Riess, Adam G. and Acquaviva, Viviana and Alexander, David M. and Almaini, Omar and Ashby, Matthew L. N. and Barden, Marco and Bell, Eric F. and Bournaud, Frédéric and Brown, Thomas M. and Caputi, Karina I. and Casertano, Stefano and Cassata, Paolo and Castellano, Marco and Challis, Peter and Chary, Ranga-Ram and Cheung, Edmond and Cirasuolo, Michele and Conselice, Christopher J. and Cooray, Asantha Roshan and Croton, Darren J. and Daddi, Emanuele and Dahlen, Tomas and Davé, Romeel and De Mello, Duília F. and Dekel, Avishai and Dickinson, Mark and Dolch, Timothy and Donley, Jennifer L. and Dunlop, James S. and Dutton, Aaron A. and Elbaz, David and Fazio, Giovanni G. and Filippenko, Alexei V. and Finkelstein, Steven L. and Fontana, Adriano and Gardner, Jonathan P. and Garnavich, Peter M. and Gawiser, Eric and Giavalisco, Mauro and Grazian, Andrea and Guo, Yicheng and Hathi, Nimish P. and Häussler, Boris and Hopkins, Philip F. and Huang, Jia-Sheng and Huang, Kuang-Han and Jha, Saurabh W. and Kartaltepe, Jeyhan S. and Kirshner, Robert P. and Koo, David C. and Lai, Kamson and Lee, Kyoung-Soo and Li, Weidong and Lotz, Jennifer M. and Lucas, Ray A. and Madau, Piero and McCarthy, Patrick J. and McGrath, Elizabeth J. and McIntosh, Daniel H. and McLure, Ross J. and Mobasher, Bahram and Moustakas, Leonidas A. and Mozena, Mark and Nandra, Kirpal and Newman, Jeffrey A. and Niemi, Sami-Matias and Noeske, Kai G. and Papovich, Casey J. and Pentericci, Laura and Pope, Alexandra and Primack, Joel R. and Rajan, Abhijith and Ravindranath, Swara and Reddy, Naveen A. and Renzini, Alvio and Rix, Hans-Walter and Robaina, Aday R. and Rodney, Steven A. and Rosario, David J. and Rosati, Piero and Salimbeni, Sara and Scarlata, Claudia and Siana, Brian and Simard, Luc and Smidt, Joseph and Somerville, Rachel S. and Spinrad, Hyron and Straughn, Amber N. and Strolger, Louis-Gregory and Telford, Olivia and Teplitz, Harry I. and Trump, Jonathan R. and van der Wel, Arjen and Villforth, Carolin and Wechsler, Risa H. and Weiner, Benjamin J. and Wiklind, Tommy and Wild, Vivienne and Wilson, Grant and Wuyts, Stijn and Yan, Hao-Jing and Yun, Min S.}, 
    year={2011}, 
    month=dec,
    pages={35}
}

@article{CANDELS2, 
    title={CANDELS: THE COSMIC ASSEMBLY NEAR-INFRARED DEEP EXTRAGALACTIC LEGACY SURVEY—THE HUBBLE SPACE TELESCOPE OBSERVATIONS, IMAGING DATA PRODUCTS, AND MOSAICS}, 
    volume={197}, 
    ISSN={0067-0049, 1538-4365}, 
    url={https://iopscience.iop.org/article/10.1088/0067-0049/197/2/36}, 
    DOI={10.1088/0067-0049/197/2/36}, 
    number={2}, 
    journal={\apjs}, 
    author={Koekemoer, Anton M. and Faber, S. M. and Ferguson, Henry C. and Grogin, Norman A. and Kocevski, Dale D. and Koo, David C. and Lai, Kamson and Lotz, Jennifer M. and Lucas, Ray A. and McGrath, Elizabeth J. and Ogaz, Sara and Rajan, Abhijith and Riess, Adam G. and Rodney, Steve A. and Strolger, Louis and Casertano, Stefano and Castellano, Marco and Dahlen, Tomas and Dickinson, Mark and Dolch, Timothy and Fontana, Adriano and Giavalisco, Mauro and Grazian, Andrea and Guo, Yicheng and Hathi, Nimish P. and Huang, Kuang-Han and van der Wel, Arjen and Yan, Hao-Jing and Acquaviva, Viviana and Alexander, David M. and Almaini, Omar and Ashby, Matthew L. N. and Barden, Marco and Bell, Eric F. and Bournaud, Frédéric and Brown, Thomas M. and Caputi, Karina I. and Cassata, Paolo and Challis, Peter J. and Chary, Ranga-Ram and Cheung, Edmond and Cirasuolo, Michele and Conselice, Christopher J. and Cooray, Asantha Roshan and Croton, Darren J. and Daddi, Emanuele and Davé, Romeel and De Mello, Duilia F. and De Ravel, Loic and Dekel, Avishai and Donley, Jennifer L. and Dunlop, James S. and Dutton, Aaron A. and Elbaz, David and Fazio, Giovanni G. and Filippenko, Alexei V. and Finkelstein, Steven L. and Frazer, Chris and Gardner, Jonathan P. and Garnavich, Peter M. and Gawiser, Eric and Gruetzbauch, Ruth and Hartley, Will G. and Häussler, Boris and Herrington, Jessica and Hopkins, Philip F. and Huang, Jia-Sheng and Jha, Saurabh W. and Johnson, Andrew and Kartaltepe, Jeyhan S. and Khostovan, Ali A. and Kirshner, Robert P. and Lani, Caterina and Lee, Kyoung-Soo and Li, Weidong and Madau, Piero and McCarthy, Patrick J. and McIntosh, Daniel H. and McLure, Ross J. and McPartland, Conor and Mobasher, Bahram and Moreira, Heidi and Mortlock, Alice and Moustakas, Leonidas A. and Mozena, Mark and Nandra, Kirpal and Newman, Jeffrey A. and Nielsen, Jennifer L. and Niemi, Sami and Noeske, Kai G. and Papovich, Casey J. and Pentericci, Laura and Pope, Alexandra and Primack, Joel R. and Ravindranath, Swara and Reddy, Naveen A. and Renzini, Alvio and Rix, Hans-Walter and Robaina, Aday R. and Rosario, David J. and Rosati, Piero and Salimbeni, Sara and Scarlata, Claudia and Siana, Brian and Simard, Luc and Smidt, Joseph and Snyder, Diana and Somerville, Rachel S. and Spinrad, Hyron and Straughn, Amber N. and Telford, Olivia and Teplitz, Harry I. and Trump, Jonathan R. and Vargas, Carlos and Villforth, Carolin and Wagner, Cory R. and Wandro, Pat and Wechsler, Risa H. and Weiner, Benjamin J. and Wiklind, Tommy and Wild, Vivienne and Wilson, Grant and Wuyts, Stijn and Yun, Min S.}, 
    year={2011}, 
    month=dec, 
    pages={36} 
}

@article{Griffith_2012,
    doi = {10.1088/0067-0049/200/1/9},
    url = {https://doi.org/10.1088/0067-0049/200/1/9},
    year = {2012},
    month = {may},
    publisher = {The American Astronomical Society},
    volume = {200},
    number = {1},
    pages = {9},
    author = {Griffith, Roger L. and Cooper, Michael C. and Newman, Jeffrey A. and Moustakas, Leonidas A. and Stern, Daniel and Comerford, Julia M. and Davis, Marc and Lotz, Jennifer M. and Barden, Marco and Conselice, Christopher J. and Capak, Peter L. and Faber, S. M. and Kirkpatrick, J. Davy and Koekemoer, Anton M. and Koo, David C. and Noeske, Kai G. and Scoville, Nick and Sheth, Kartik and Shopbell, Patrick and Willmer, Christopher N. A. and Weiner, Benjamin},
    title = {THE ADVANCED CAMERA FOR SURVEYS GENERAL CATALOG: STRUCTURAL PARAMETERS FOR APPROXIMATELY HALF A MILLION GALAXIES},
    journal = {\apjs},
}

@article{GEMS, 
    title={GEMS: Galaxy Evolution from Morphologies and SEDs}, 
    volume={152}, 
    ISSN={0067-0049}, 
    url={https://ui.adsabs.harvard.edu/abs/2004ApJS..152..163R}, 
    DOI={10.1086/420885}, 
    journal={\apjs},
    publisher={IOP},
    author={Rix, Hans-Walter and Barden, Marco and Beckwith, Steven V. W. and Bell, Eric F. and Borch, Andrea and Caldwell, John A. R. and Häussler, Boris and Jahnke, Knud and Jogee, Shardha and McIntosh, Daniel H. and Meisenheimer, Klaus and Peng, Chien Y. and Sanchez, Sebastian F. and Somerville, Rachel S. and Wisotzki, Lutz and Wolf, Christian}, 
    year={2004},
    month={June}, 
    pages={163}
}

@article{CEERS, 
    title={CEERS Key Paper. I. An Early Look into the First 500 Myr of Galaxy Formation with JWST}, 
    volume={946}, 
    ISSN={2041-8205, 2041-8213}, 
    url={https://iopscience.iop.org/article/10.3847/2041-8213/acade4}, 
    DOI={10.3847/2041-8213/acade4},
    number={1}, 
    journal={\apjl}, 
    author={Finkelstein, Steven L. and Bagley, Micaela B. and Ferguson, Henry C. and Wilkins, Stephen M. and Kartaltepe, Jeyhan S. and Papovich, Casey and Yung, L. Y. Aaron and Arrabal Haro, Pablo and Behroozi, Peter and Dickinson, Mark and Kocevski, Dale D. and Koekemoer, Anton M. and Larson, Rebecca L. and Le Bail, Aurélien and Morales, Alexa M. and Pérez-González, Pablo G. and Burgarella, Denis and Davé, Romeel and Hirschmann, Michaela and Somerville, Rachel S. and Wuyts, Stijn and Bromm, Volker and Casey, Caitlin M. and Fontana, Adriano and Fujimoto, Seiji and Gardner, Jonathan P. and Giavalisco, Mauro and Grazian, Andrea and Grogin, Norman A. and Hathi, Nimish P. and Hutchison, Taylor A. and Jha, Saurabh W. and Jogee, Shardha and Kewley, Lisa J. and Kirkpatrick, Allison and Long, Arianna S. and Lotz, Jennifer M. and Pentericci, Laura and Pierel, Justin D. R. and Pirzkal, Nor and Ravindranath, Swara and Ryan, Russell E. and Trump, Jonathan R. and Yang, Guang and Bhatawdekar, Rachana and Bisigello, Laura and Buat, Véronique and Calabrò, Antonello and Castellano, Marco and Cleri, Nikko J. and Cooper, M. C. and Croton, Darren and Daddi, Emanuele and Dekel, Avishai and Elbaz, David and Franco, Maximilien and Gawiser, Eric and Holwerda, Benne W. and Huertas-Company, Marc and Jaskot, Anne E. and Leung, Gene C. K. and Lucas, Ray A. and Mobasher, Bahram and Pandya, Viraj and Tacchella, Sandro and Weiner, Benjamin J. and Zavala, Jorge A.}, 
    year={2023}, 
    month=mar, 
    pages={L13}
}

@article{GAMA, 
    title={GAMA: towards a physical understanding of galaxy formation}, 
    volume={50}, 
    ISSN={1366-8781}, 
    url={https://doi.org/10.1111/j.1468-4004.2009.50512.x}, 
    DOI={10.1111/j.1468-4004.2009.50512.x}, 
    number={5}, 
    journal={A\&G}, 
    author={Driver, Simon P and Norberg, Peder and Baldry, Ivan K and Bamford, Steven P and Hopkins, Andrew M and Liske, Jochen and Loveday, Jon and Peacock, John A and below), GAMA Team (listed}, 
    year={2009}, 
    month=oct, 
    pages={5.12} 
}

@ARTICLE{ALFALFA,
       author = {{Giovanelli}, Riccardo and {Haynes}, Martha P. and {Kent}, Brian R. and {Perillat}, Philip and {Saintonge}, Amelie and {Brosch}, Noah and {Catinella}, Barbara and {Hoffman}, G. Lyle and {Stierwalt}, Sabrina and {Spekkens}, Kristine and et al.},
        title = "{The Arecibo Legacy Fast ALFA Survey. I. Science Goals, Survey Design, and Strategy}",
      journal = {\aj},
         year = 2005,
        month = dec,
       volume = {130},
       number = {6},
        pages = {2598},
          doi = {10.1086/497431},
archivePrefix = {arXiv},
       eprint = {astro-ph/0508301},
 primaryClass = {astro-ph},
       adsurl = {https://ui.adsabs.harvard.edu/abs/2005AJ....130.2598G}
}

@ARTICLE{DESI_DR1,
       author = {{DESI Collaboration} and {Abdul Karim}, M. and {Adame}, A.~G. and {Aguado}, D. and {Aguilar}, J. and {Ahlen}, S. and {Alam}, S. and {Aldering}, G. and {Alexander}, D.~M. and {Alfarsy}, R. and et al.},
        title = "{Data Release 1 of the Dark Energy Spectroscopic Instrument}",
      journal = {\aj},
         year = 2026,
        month = may,
       volume = {171},
       number = {5},
          eid = {285},
        pages = {285},
          doi = {10.3847/1538-3881/ae4c43},
archivePrefix = {arXiv},
       eprint = {2503.14745},
 primaryClass = {astro-ph.CO},
       adsurl = {https://ui.adsabs.harvard.edu/abs/2026AJ....171..285D}
}

@ARTICLE{DR1-TP001,
       author = {{Euclid Collaboration: Aussel}, H. and others},
        title = "{Euclid Data Release 1 (DR1). Data release overview}",
      journal = {arXiv e-prints},
         year = 2026,
        month = nov,
          eid = {arXiv:2611.16510},
        pages = {arXiv:2611.16510},
archivePrefix = {arXiv},
       eprint = {2611.16510},
 primaryClass = {astro-ph.CO},
       adsurl = {https://ui.adsabs.harvard.edu/abs/2026arXiv261116510E}
}

@ARTICLE{DR1-TP002,
       author = {{Euclid Collaboration: McCracken}, H.~J. and {Benson}, K. and {Cropper}, M. and others},
        title = "{Euclid Data Release 1 (DR1): VIS processing and data products}",
      journal = {A\&A, submitted},
         year = 2026,
        month = nov,
          eid = {arXiv:2611.15890},
        pages = {arXiv:2611.15890},
archivePrefix = {arXiv},
       eprint = {2611.15890},
 primaryClass = {astro-ph.CO},
       adsurl = {https://ui.adsabs.harvard.edu/abs/2026arXiv261115890E}
}

@ARTICLE{DR1-TP003,
       author = {{Euclid Collaboration: Polenta}, G. and {Frailis}, M. and {Alavi}, A. and others},
        title = "{Euclid Data Release 1 (DR1). Near-infrared processing and data products}",
      journal = {A\&A, submitted},
         year = 2026,
        month = nov,
          eid = {arXiv:2611.16119},
        pages = {arXiv:2611.16119},
archivePrefix = {arXiv},
       eprint = {2611.16119},
 primaryClass = {astro-ph.IM},
       adsurl = {https://ui.adsabs.harvard.edu/abs/2026arXiv261116119E}
}

@ARTICLE{DR1-TP009,
       author = {{Euclid Collaboration: K{\"u}mmel}, M. and {Romelli}, E. and {Gracia-Carpio}, J. and others},
       title = "{Euclid Data Release 1 (DR1): The MERge processing}",
      journal = {A\&A, submitted},
         year = 2026,
        month = nov,
          eid = {arXiv:2611.17832},
        pages = {arXiv:2611.17832},
archivePrefix = {arXiv},
       eprint = {2611.17832},
 primaryClass = {astro-ph.IM},
       adsurl = {https://ui.adsabs.harvard.edu/abs/2026arXiv261117832E}
}

@ARTICLE{DR1-TP011,
       author = {{Euclid Collaboration: Hartley}, W.~G. and {Paltani}, S. and {Dubath}, F. and others},
        title = "{Euclid Data Release 1 (DR1): Photometric-redshift pipeline overview and validation}",
      journal = {A\&A, submitted},
         year = 2026,
        month = nov,
          eid = {arXiv:2611.15987},
        pages = {arXiv:2611.15987},
archivePrefix = {arXiv},
       eprint = {2611.15987},
 primaryClass = {astro-ph.CO},
       adsurl = {https://ui.adsabs.harvard.edu/abs/2026arXiv261115987E}
}

@ARTICLE{DR1-TP012,
       author = {{Euclid Collaboration: Enia}, A. and {Hartley}, W.~G. and {Bolzonella}, M. and others},
        title = "{Euclid Data Release 1 (DR1). The PHZ processing function for the physical properties of galaxies}",
      journal = {A\&A, submitted},
         year = 2026,
        month = nov,
          eid = {arXiv:2611.16444},
        pages = {arXiv:2611.16444},
archivePrefix = {arXiv},
       eprint = {2611.16444},
 primaryClass = {astro-ph.GA},
       adsurl = {https://ui.adsabs.harvard.edu/abs/2026arXiv261116444E}
}

@ARTICLE{DR1-TP027,
       author = {{Euclid Collaboration: Tereno}, I. and {Dinis}, J. and {Dupac}, X. and others},
       title = "{Euclid Data Release 1 (DR1). The Euclid surveys at DR1}",
      journal = {arXiv preprint},
         year = 2026,
        month = nov,
          eid = {arXiv:2611.17040},
        pages = {arXiv:2611.17040},
archivePrefix = {arXiv},
       eprint = {2611.17040},
 primaryClass = {astro-ph.CO},
       adsurl = {https://ui.adsabs.harvard.edu/abs/2026arXiv261117040E}
}

@ARTICLE{Q1-TP001,
       author = {{Euclid Collaboration: Aussel}, H. and {Tereno}, I. and {Schirmer}, M. and others},
       title = "{Euclid Quick Data Release (Q1). I. Data release overview}",
      journal = {\aap},
         year = 2026,
        month = jun,
       volume = {711},
          eid = {A1},
        pages = {A1},
          doi = {10.1051/0004-6361/202554610},
       adsurl = {https://ui.adsabs.harvard.edu/abs/2026A&A...711A...1E}
}

@ARTICLE{Q1-SP040,
       author = {{Euclid Collaboration: Quilley}, L. and {Damjanov}, I. and {de Lapparent}, V. and others},
        title = "{Euclid Quick Data Release (Q1). VIII. Exploring galaxy morphology across cosmic time through S{\'e}rsic fits}",
      journal = {\aap},
         year = 2026,
        month = jun,
       volume = {711},
          eid = {A8},
        pages = {A8},
          doi = {10.1051/0004-6361/202554585},
       adsurl = {https://ui.adsabs.harvard.edu/abs/2026A&A...711A...8E}
}

@ARTICLE{Q1-SP047,
       author = {{Euclid Collaboration: Walmsley}, M. and {Huertas-Company}, M. and {Quilley}, L. and others},
        title = "{Euclid Quick Data Release (Q1): IX. First visual morphology catalogue}",
      journal = {\aap},
         year = 2026,
        month = jun,
       volume = {711},
          eid = {A9},
        pages = {A9},
          doi = {10.1051/0004-6361/202555140},
       adsurl = {https://ui.adsabs.harvard.edu/abs/2026A&A...711A...9E}
}

@ARTICLE{EuclidSkyOverview,
author = {{Euclid Collaboration: Mellier}, Y. and {Abdurro'uf} and {Acevedo~Barroso}, J.A. and others},
	title = {Euclid - I. Overview of the Euclid mission},
	DOI= "10.1051/0004-6361/202450810",
	url= "https://doi.org/10.1051/0004-6361/202450810",
	journal = {A\&A},
	year = 2025,
	volume = 697,
	pages = "A1",
}

@ARTICLE{EuclidSkyVIS,
author = {{Euclid Collaboration: Cropper}, M. and {Al-Bahlawan}, A. and {Amiaux}, J. and others},
	title = {Euclid - II. The VIS instrument},
	DOI= "10.1051/0004-6361/202450996",
	url= "https://doi.org/10.1051/0004-6361/202450996",
	journal = {A\&A},
	year = 2025,
	volume = 697,
	pages = "A2",
}

@ARTICLE{EuclidSkyNISP,
author = {{Euclid Collaboration: Jahnke}, K. and {Gillard}, W. and {Schirmer}, M. and others},
	title = {Euclid - III. The NISP Instrument},
	DOI= "10.1051/0004-6361/202450786",
	url= "https://doi.org/10.1051/0004-6361/202450786",
	journal = {A\&A},
	year = 2025,
	volume = 697,
	pages = "A3",
}

@misc{Q1cite,
author = "{Euclid Quick Release Q1}",
howpublished = "\url{https://doi.org/10.57780/esa-2853f3b}",
year = 2025
}

@misc{DR1cite,
author = "{Euclid Data Release 1}",
howpublished = "\url{https://doi.org/10.57780/esa-d079c51}",
year = 2026
}

@INCOLLECTION{Datalabscite,
       author = {{Navarro}, V. and {del Rio}, S. and {Angel Diego}, M. and {Lopez-Caniego}, M. and {Marinic}, F. and {Kruk}, S. and {Reerink}, J. and {Arviset}, C.},
    title = "{ESA Datalabs: Digital Innovation in Space Science}",
booktitle = {Space Data Management},
     year = 2024,
   series = {Studies in Big Data},
   volume = {141},
   editor = {{Cortesi}, A.},
publisher = {Springer},
  address = {Singapore},
    pages = {1}
}

@ARTICLE{Bretonniere-EP26,
       author = {{Euclid Collaboration: Bretonni{\`e}re}, H. and {Kuchner}, U. and {Huertas-Company}, M. and others},
        title = "{Euclid preparation. XXVI. The Euclid Morphology Challenge: Towards structural parameters for billions of galaxies}",
      journal = {\aap},
         year = 2023,
        month = mar,
       volume = {671},
          eid = {A102},
        pages = {A102},
          doi = {10.1051/0004-6361/202245042},
archivePrefix = {arXiv},
       eprint = {2209.12907},
 primaryClass = {astro-ph.GA},
       adsurl = {https://ui.adsabs.harvard.edu/abs/2023A&A...671A.102E}
}

@ARTICLE{Merlin-EP25,
       author = {{Euclid Collaboration: Merlin}, E. and {Castellano}, M. and {Bretonni{\`e}re}, H. and others},
        title = "{Euclid preparation. XXV. The Euclid Morphology Challenge: Towards model-fitting photometry for billions of galaxies}",
      journal = {\aap},
         year = 2023,
        month = mar,
       volume = {671},
          eid = {A101},
        pages = {A101},
          doi = {10.1051/0004-6361/202245041},
archivePrefix = {arXiv},
       eprint = {2209.12906},
 primaryClass = {astro-ph.GA},
       adsurl = {https://ui.adsabs.harvard.edu/abs/2023A&A...671A.101E}
}

@ARTICLE{Scaramella-EP1,
       author = {{Euclid Collaboration: Scaramella}, R. and {Amiaux}, J. and {Mellier}, Y. and others},
        title = "{Euclid preparation. I. The Euclid Wide Survey}",
      journal = {\aap},
         year = 2022,
        month = jun,
       volume = {662},
          eid = {A112},
        pages = {A112},
          doi = {10.1051/0004-6361/202141938},
archivePrefix = {arXiv},
       eprint = {2108.01201},
 primaryClass = {astro-ph.CO},
       adsurl = {https://ui.adsabs.harvard.edu/abs/2022A&A...662A.112E}
}

@ARTICLE{Quilley25,
       author = {{Quilley}, L. and {de Lapparent}, V. and {Bolzonella}, M. and others},
        title = "{Euclid: Galaxy morphology and photometry from bulge-disc decomposition of Early Release Observations}",
      journal = {\aap},
         year = 2026,
        month = jun,
       volume = {710},
          eid = {A277},
        pages = {A277},
          doi = {10.1051/0004-6361/202554219},
       adsurl = {https://ui.adsabs.harvard.edu/abs/2026A&A...710A.277Q}
}

\begin{appendix}

\FloatBarrier
\section{DR1 selection cuts \label{app:DR1_Selection_Cuts}}
We used the following selections to build our magnitude-limited sample from the DR1 catalogue:
\bi
    \item $\IE\leq24.5$, implemented by making a cut in the sum of the flux in the segmented pixels
    $\texttt{FLUX$\_$SEGMENTATION} \gtrsim 0.575$;
    \item $\texttt{VIS$\_$DET} = 1$ to ensure sources are detected in VIS mosaic tiles;
    \item $\texttt{SPURIOUS$\_$FLAGS} = 0$ to prevent spurious sources from contaminating the sample;
    \item $\texttt{POINT$\_$LIKE$\_$PROB} \leq 0.1$ to discard point-like sources;
    \item $\texttt{PHYS$\_$PARAM$\_$FLAGS} = 0$ indicating good physical parameter processing from the \texttt{PHZ} processing function;
    \item $\texttt{PHZ$\_$CLASSIFICATION} = 2$, indicating the source is accepted as a galaxy by the \texttt{PHZ} processing function;
    \item $\texttt{DET$\_$QUALITY$\_$FLAG} < 4$ to remove poor-quality detections;
    \item $\texttt{MUMAX$\_$MINUS$\_$MAG} > -2.6$ to remove extended sources.
\ei
We then applied quality cuts, including those recommended by \citet{Q1-SP040} and \citet[][their section 3]{DR1-TP012}, to our star-forming sample to ensure reliable measurements and physical parameters:
\bi
    \item $\lvert\texttt{PHZ$\_$PP$\_$MEDIAN$\_$Z}-\texttt{PHZ$\_$MEDIAN}\rvert < 0.2$;
    \item $\lvert\texttt{PHZ$\_$PP$\_$MEDIAN$\_$Z}-\texttt{PHZ$\_$PP$\_$MODE$\_$Z}\rvert < 0.2$;
    \item $0.01a < \texttt{SERSIC$\_$SERSIC$\_$VIS$\_$RADIUS} < 20a$, where $a$ in this case is the isophotal semi-major axis in units of VIS pixels;
    \item $0.302 < \texttt{SERSIC$\_$SERSIC$\_$VIS$\_$INDEX} < 5.45$ to limit the bounds of Sérsic indices;
    \item $0.05 < \texttt{SERSIC$\_$SERSIC$\_$VIS$\_$AXIS$\_$RATIO} < 1.0$ to limit the bounds of Sérsic axis ratios;
    \item $\texttt{SERSIC$\_$SERSIC$\_$VIS$\_$AXIS$\_$RATIO$\_$ERR} < 0.10$ to exclude sources with large errors in Sérsic axis ratio;
    \item $\texttt{SERSIC$\_$SERSIC$\_$VIS$\_$RADIUS$\_$ERR} < 0.10$ to exclude sources with large errors in angular Sérsic effective radius;
    \item $\texttt{SERSIC$\_$SERSIC$\_$VIS$\_$RADIUS} > \ang{;;0.16}$ to ensure angular Sérsic effective radii are larger than the VIS PSF FWHM;
    \item Finite values for \texttt{SERSIC\_SERSIC\_VIS\_RADIUS},
    \texttt{SERSIC\_SERSIC\_VIS\_AXIS\_RATIO},
    \texttt{PHZ\_PP\_MEDIAN\_STELLARMASS}, \texttt{PHZ\_PP\_MEDIAN\_SFR},
    and \texttt{PHZ\_PP\_MEDIAN\_REDSHIFT}.
\ei

\FloatBarrier
\section{Sérsic parameter comparison \label{app:sersic_param_comparison}}
\begin{figure*}[!t]
\centering
\includegraphics[angle=0,width=\hsize]{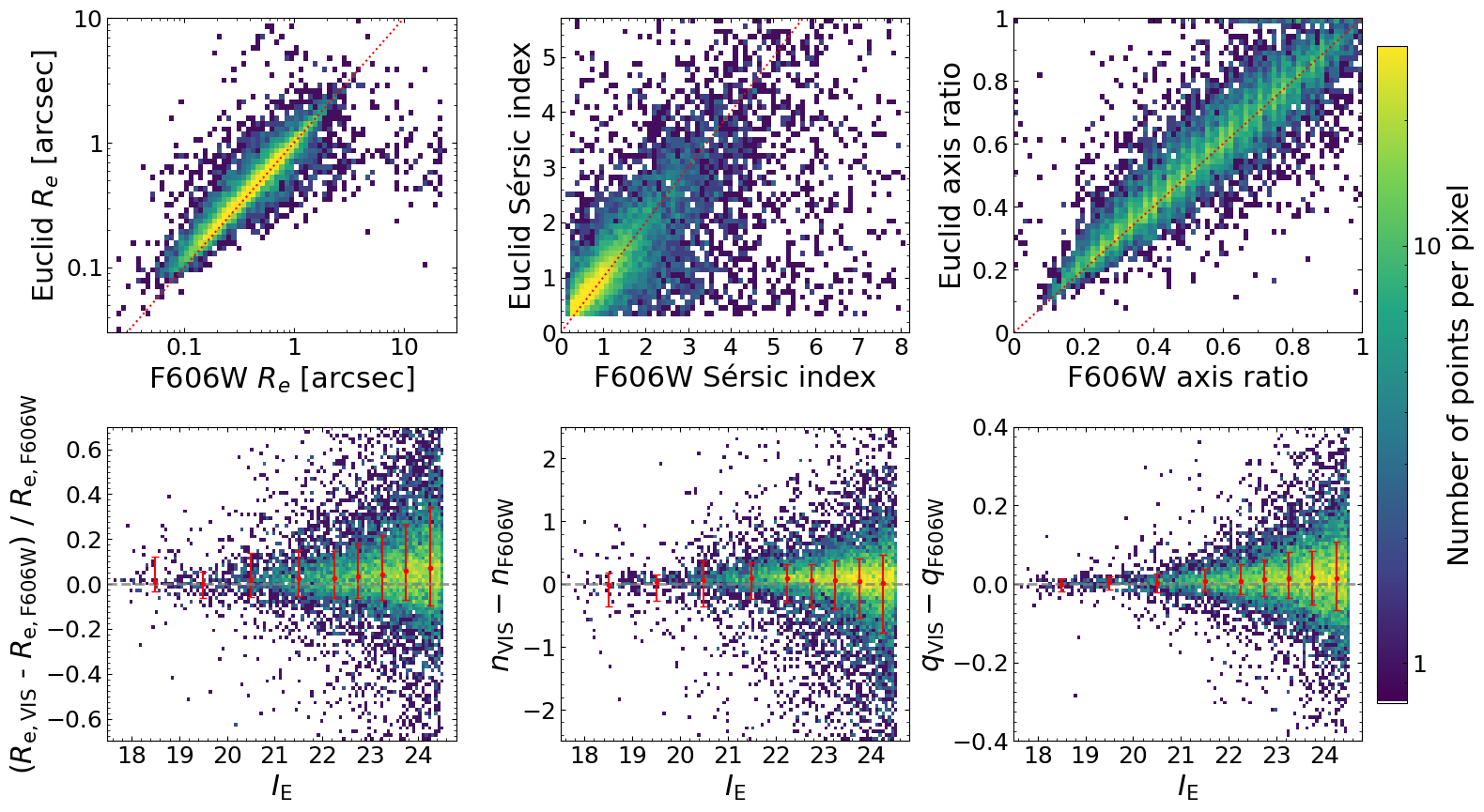} 
\caption{Comparison of Sérsic parameters for cross-matched galaxies with
$\IE\leq24.5$ between ACS-GC using HST F606W imaging, and single-visit DR1 EDF-F measurements at EWS depth. Top row: direct comparison of effective radius $R_{\rm e}$ (in arcseconds), Sérsic index $n$, and axis ratio $q$. Bottom row: differences (or relative differences) in each parameter as a function of \IE. Red points show the median per magnitude bin; error bars show the 16th and 84th percentiles of the distribution.}
\label{fig:sersic_params_comparison}
\end{figure*}

\begin{figure*}[!t]
\includegraphics[angle=0,width=\hsize]{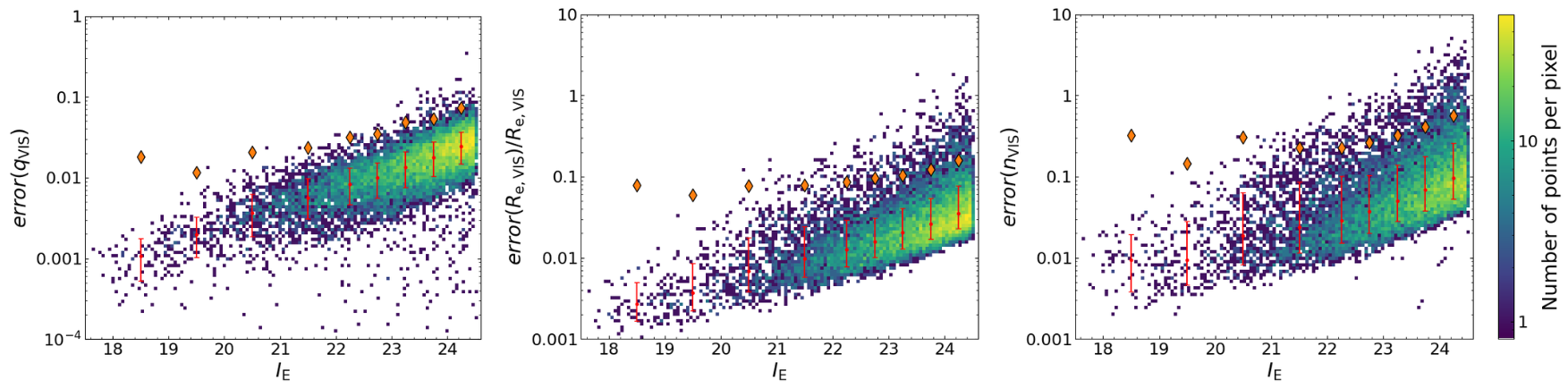} 
\caption{Uncertainties in $q$, $R_{\rm e}$, and $n$ as a function of \IE for the same cross-matched sample as in Fig.\,\ref{fig:sersic_params_comparison}. Red points and error bars show the median, and 16th/84th percentiles per magnitude bin. Orange diamonds show dispersion of differences between \Euclid and ACS-GC from Fig.\,\ref{fig:sersic_params_comparison}.}
\label{fig:sersic_error_comparison}
\end{figure*}

Following \citet[][section 4]{Q1-SP040}, we compared Sérsic parameters for cross-matched galaxies with $\IE\leq24.5$ in the GEMS field between the Advanced Camera for Surveys General Catalog \citep[ACS-GC;][]{Griffith_2012}, whose fits were performed with the \texttt{GALAPAGOS} code \citep{GALAPAGOS} on HST F606W images, and a single DR1 EDF-F visit at EWS depth (Fig.\,\ref{fig:sersic_params_comparison}). The top row compares Sérsic parameters (effective radius $R_{\rm e}$, in arcseconds; index $n$; and axis ratio $q$) between \Euclid and ACS-GC, while the bottom row shows the difference (or relative difference) for each parameter as a function of magnitude (based on MER \texttt{FLUX\_SEGMENTATION}), with red points indicating the median, and error bars indicating the 16th and 84th percentiles of the distribution, per magnitude bin.

Figure \ref{fig:sersic_error_comparison} shows how the measured errors for $q$, $R_{\rm e}$, and $n$ vary with magnitude, again with the per-bin median, 16th, and 84th percentiles in red; the orange diamonds indicate the dispersion around the median, defined as the 68th percentile of $|y - \mathrm{median}(y)|$. The different bandpasses in HST and \Euclid introduce systematic offsets between the two datasets, so ACS-GC measurements do not constitute a perfect `ground truth' for this comparison. This bandpass mismatch also contributes to the dispersion itself: galaxies with different colour gradients suffer different biases, adding scatter beyond the purely statistical measurement uncertainty. Quantitatively, the dispersion for $R_{\rm e}$ increases steadily towards fainter magnitudes, rising from $\lesssim\,10\%$ at $\IE<23$ to around $16\%$ in the bin $24.0<\IE<24.5$, while for $q$ it remains smaller throughout, increasing from $\lesssim\,3\%$ to around $7\%$. The corresponding systematic offsets are modest, with $R_{\rm e}$ biased high by approximately $7\%$ and $q$ by $\lesssim$\,2\% at the faint limit. Sérsic index $n$ is by far the least well constrained of the three parameters, with a dispersion several times larger than that of $R_{\rm e}$ or $q$ ($D\approx0.57$ by $\IE=24.5$). These trends closely follow those found for the comparison between Q1 and ACS-GC in \citet[][figure 6]{Q1-SP040}. We note that the DR1 sample Sérsic parameter error selection cut (Appendix \ref{app:DR1_Selection_Cuts}) only applies to Sérsic $q$ and $R_{\rm e}$, not Sérsic $n$.

We also compared the single-Sérsic structural parameters of our final sample with the isophotal-based measurements ($a_{\rm iso}=\texttt{SEMIMAJOR\_AXIS}$, $q_{\rm iso}=1-\texttt{ELLIPTICITY}$) in the MER catalogue. We converted $a_{\rm iso}$ from pixels to arcseconds. These measurements are not PSF-corrected, whereas the Sérsic parameters come from PSF-convolved model fits and are therefore corrected for the PSF. Medians and scatters are computed over all galaxies with both $a_{\rm iso}$ and $a_{\rm S\acute{e}rsic}$ larger than the VIS PSF FWHM, with the scatter defined as half the difference between the 16th and 84th percentiles of the distribution (equivalent to $1\sigma$ for a Gaussian distribution). The two size measures are tightly correlated (Fig.\,\ref{fig:sersic_iso_comparison}, left) with a median $\logten(a_{\rm S\acute{e}rsic}/a_{\rm iso})=0.081$ and a scatter of $0.074\,{\rm dex}$. Sérsic radii are slightly larger on average, as expected for two different definitions of size, and extend to smaller values because they are corrected for the PSF. The axis ratios (Fig.\,\ref{fig:sersic_iso_comparison}, right) show a median offset $q_{\rm S\acute{e}rsic}-q_{\rm iso}=-0.093$ with a scatter of 0.091, consistent with the Q1 result of \citet{Q1-SP040}, because the isophotal-based shapes are rounded by the PSF. Looking at the distributions of axis ratios as a function of Sérsic index, this offset is present at all Sérsic indices (Fig.\,\ref{fig:q_vs_n}). As in \citet[][figures 2 and 3]{Q1-SP040}, pure disc galaxies with $n\approx1$ span the full range of axis ratios, $0.1\lesssim q\leq1$, as expected for discs seen at all inclinations. The Sérsic axis ratios recover more of these discs at very low $q$, since PSF rounding prevents the isophotal-based measurements from reaching the axis ratios of thin, edge-on systems.

\FloatBarrier

\twocolumn[{
\capfig
\sidecaption
\begin{minipage}[b]{12.5cm}
\includegraphics[width=0.49\linewidth]{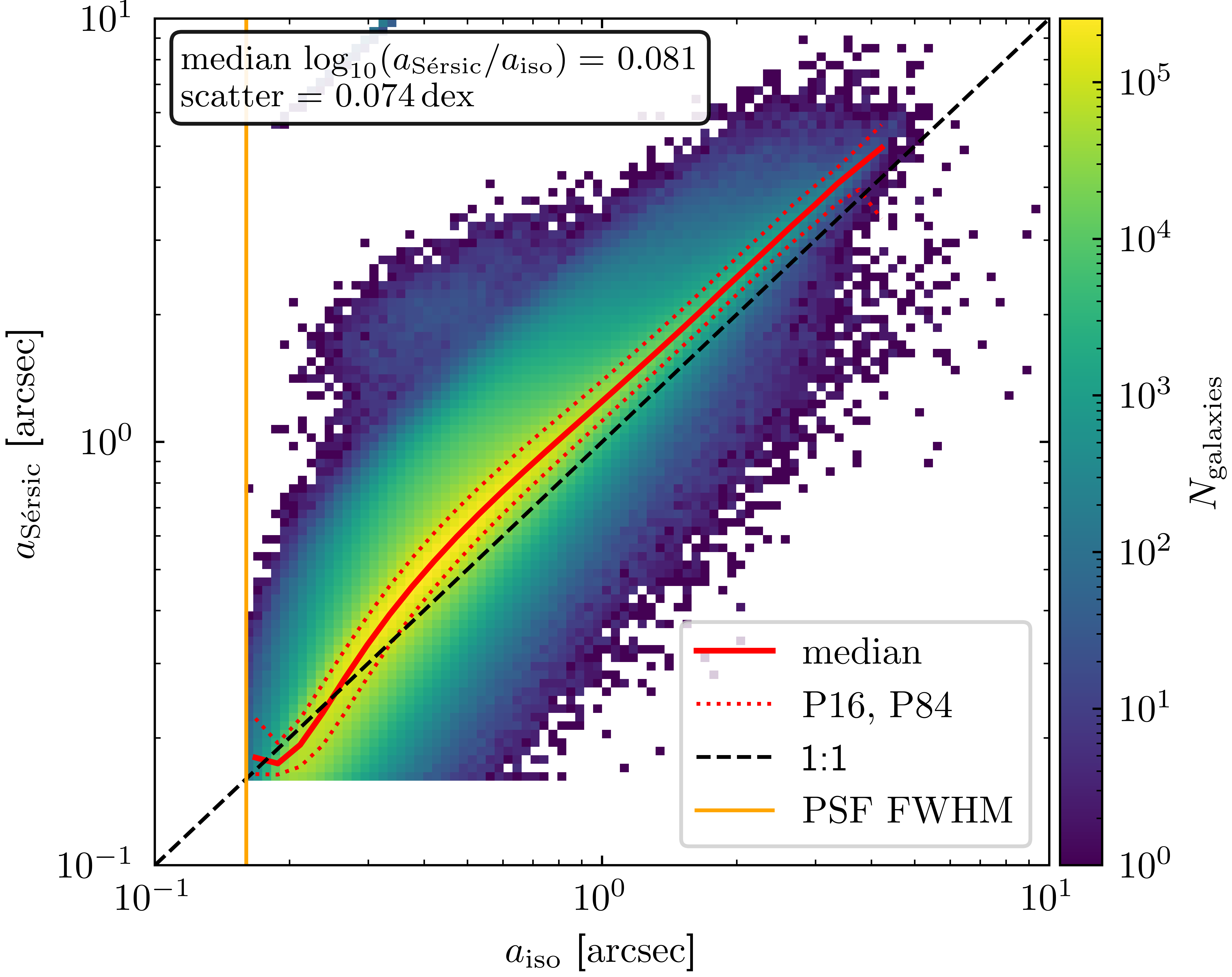}\hfill
\includegraphics[width=0.49\linewidth]{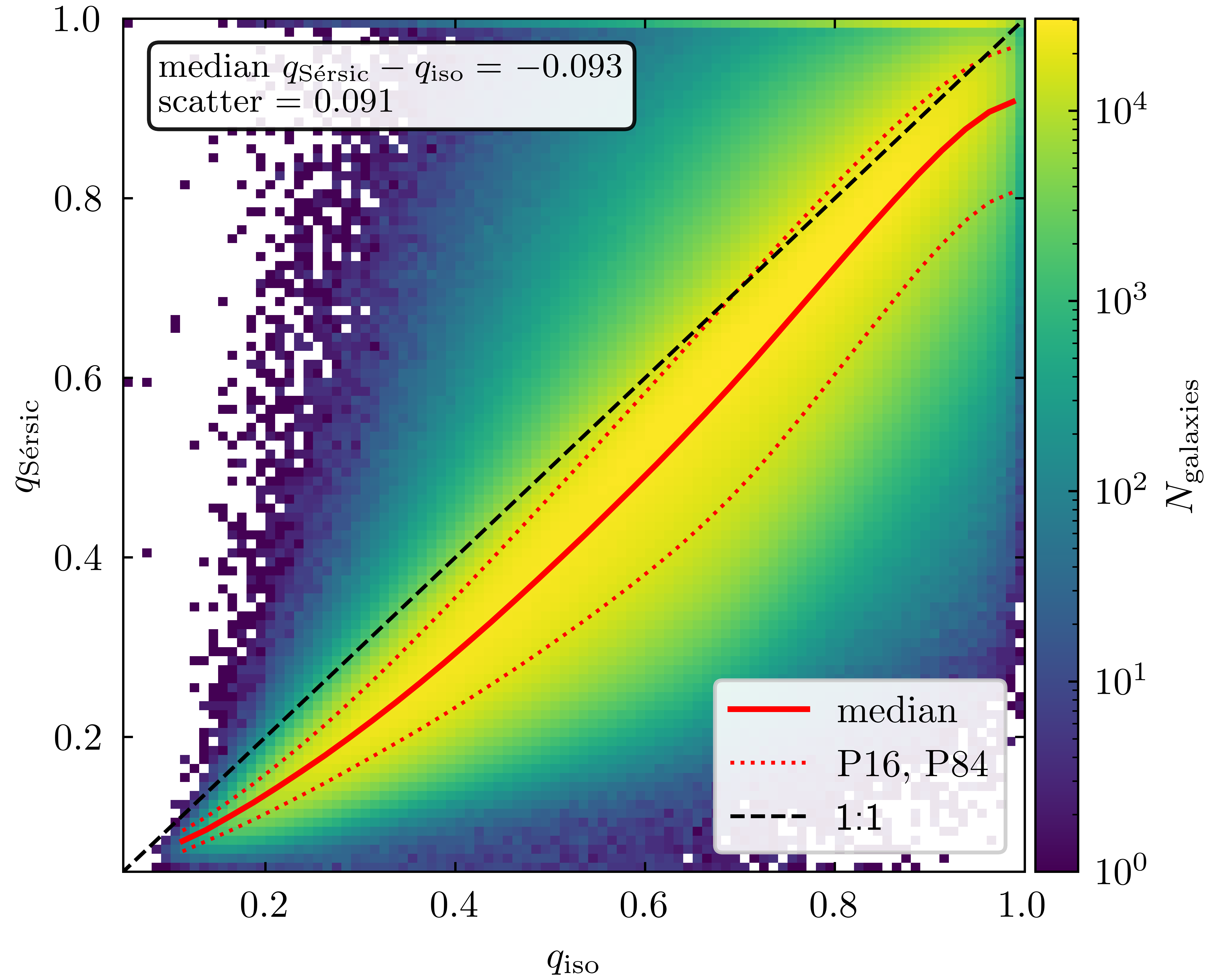}
\end{minipage}
\caption{Comparison of single-Sérsic and isophotal-based structural parameters for our final sample with both $a_{\rm iso}$ and $a_{\rm S\acute{e}rsic}$ larger than the VIS PSF FWHM (\ang{;;0.16}). Left: Sérsic effective radius $a_{\rm S\acute{e}rsic}$ versus isophotal semi-major axis $a_{\rm iso}$, in arcseconds; the orange line marks the PSF FWHM. Right: Sérsic axis ratio $q_{\rm S\acute{e}rsic}$ versus isophotal axis ratio $q_{\rm iso}$. Colours show the number of galaxies per pixel (logarithmic scale). Solid and dotted red lines show the running median, and the 16th and 84th percentiles in bins of the $x$-axis quantity. The dashed black line marks the 1:1 relation.}
\label{fig:sersic_iso_comparison}
\par\bigskip
\sidecaption
\begin{minipage}[b]{12.5cm}
\includegraphics[width=0.49\linewidth]{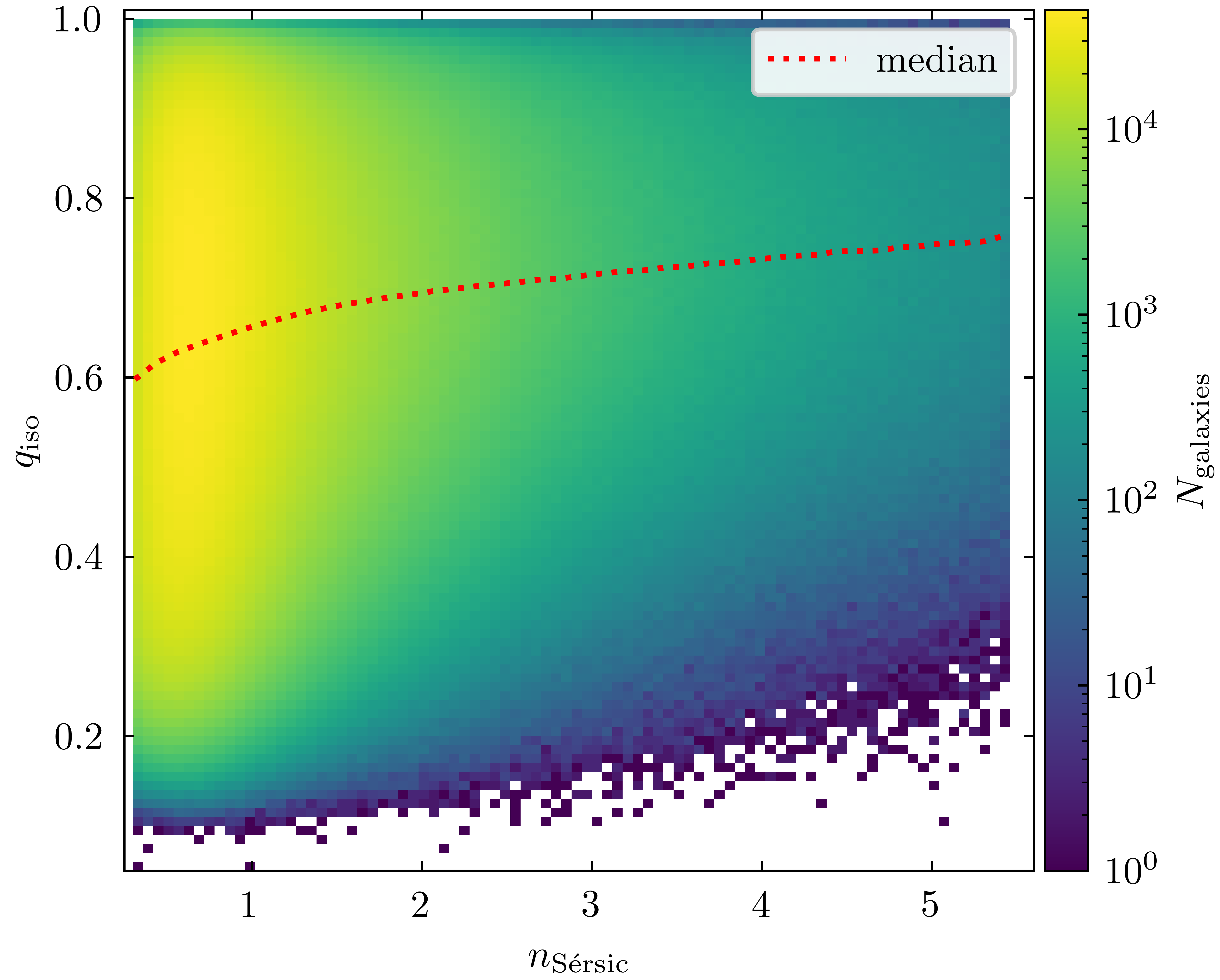}\hfill
\includegraphics[width=0.49\linewidth]{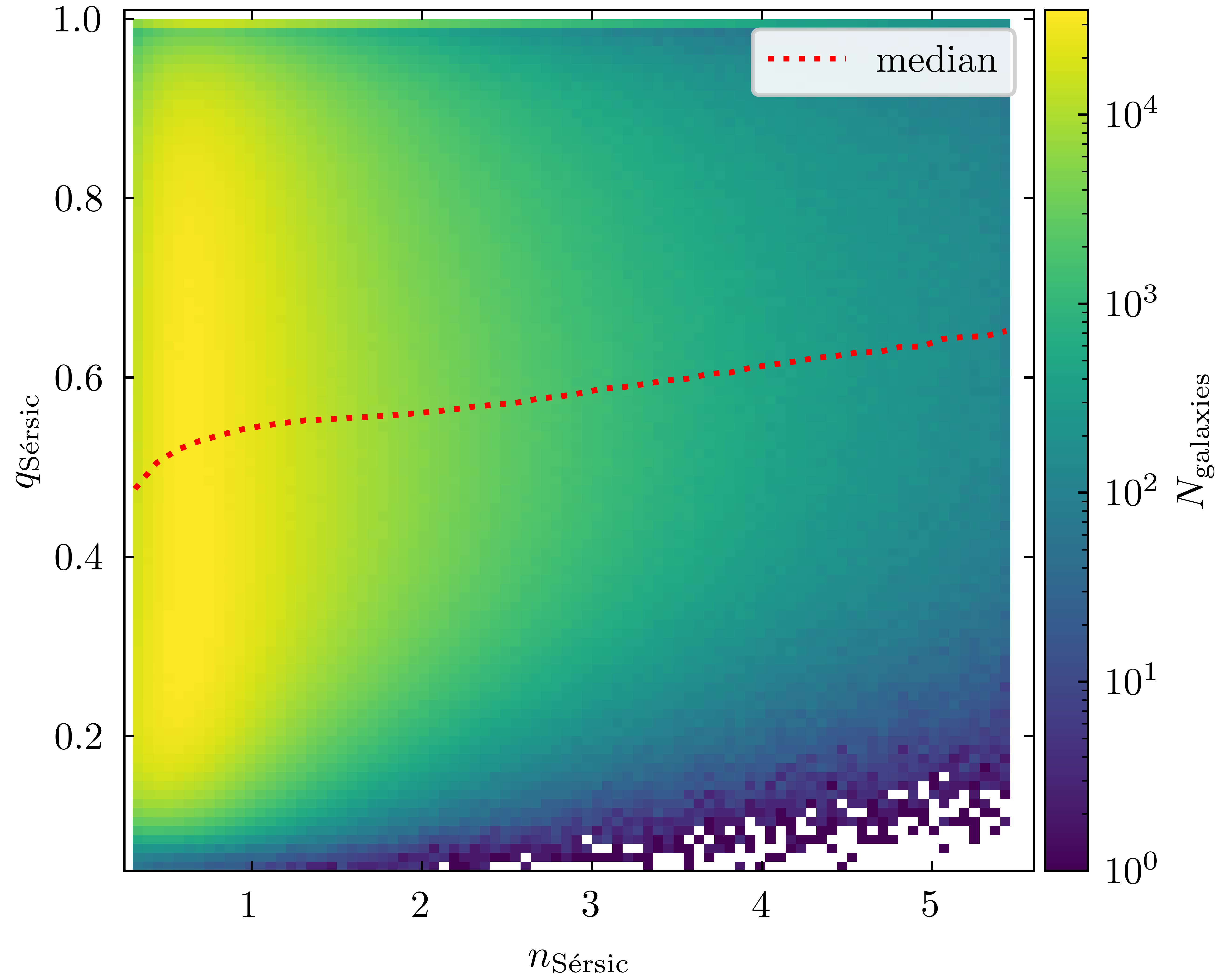}
\end{minipage}
\caption{Axis ratio as a function of Sérsic index $n$ for the same galaxies as Fig.\,\ref{fig:sersic_iso_comparison}, using the isophotal axis ratio $q_{\rm iso}$ (left) and the Sérsic axis ratio $q_{\rm S\acute{e}rsic}$ (right). Colours show the number of galaxies per pixel (logarithmic scale); dotted red lines show the running median $q$ in bins of $\Delta n = 0.125$.}
\label{fig:q_vs_n}
\par\bigskip
\section{Star-forming galaxy selection \label{app:SF_Selection}}
Figure \ref{fig:SFMS_diagram_app} shows the sSFR-based classification described in Sect.\,\ref{ssc:sSFR_selection}, and Fig.\,\ref{fig:UVJ_Av_diagram_app} the colour-based classification described in Sect.\,\ref{ssc:Colour_selection}.
\par\bigskip
\sidecaption
\includegraphics[width=12.5cm]{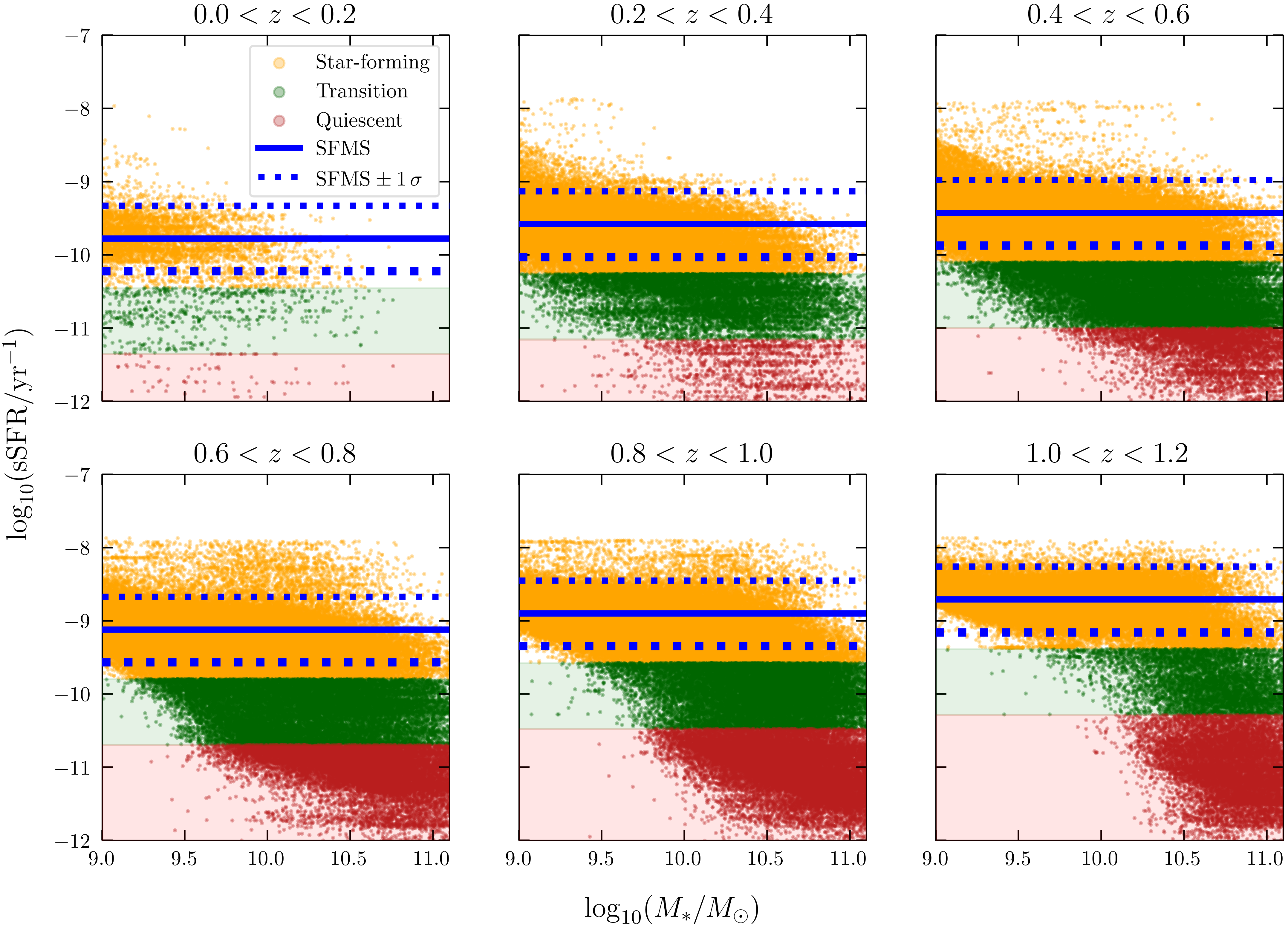}
\caption{sSFR--$\Mstellar$ diagrams per redshift slice. The solid blue line indicates the sSFR of the zero-slope median SFMS; dotted blue lines indicate the $0.45\,{\rm dex}$ ($\pm1\,\sigma$) width of the SFMS. Galaxies are classified as star-forming if they are located above the lower dotted blue line, transition if located in the green shaded region ($1.5$--$3.5\,\sigma$ below the SFMS), and quiescent if located in the red shaded region ($>\,3.5\,\sigma$ below the SFMS).}
\label{fig:SFMS_diagram_app}
}]

\twocolumn[{
\capfig
\centering
\includegraphics[angle=0,width=0.78\hsize]{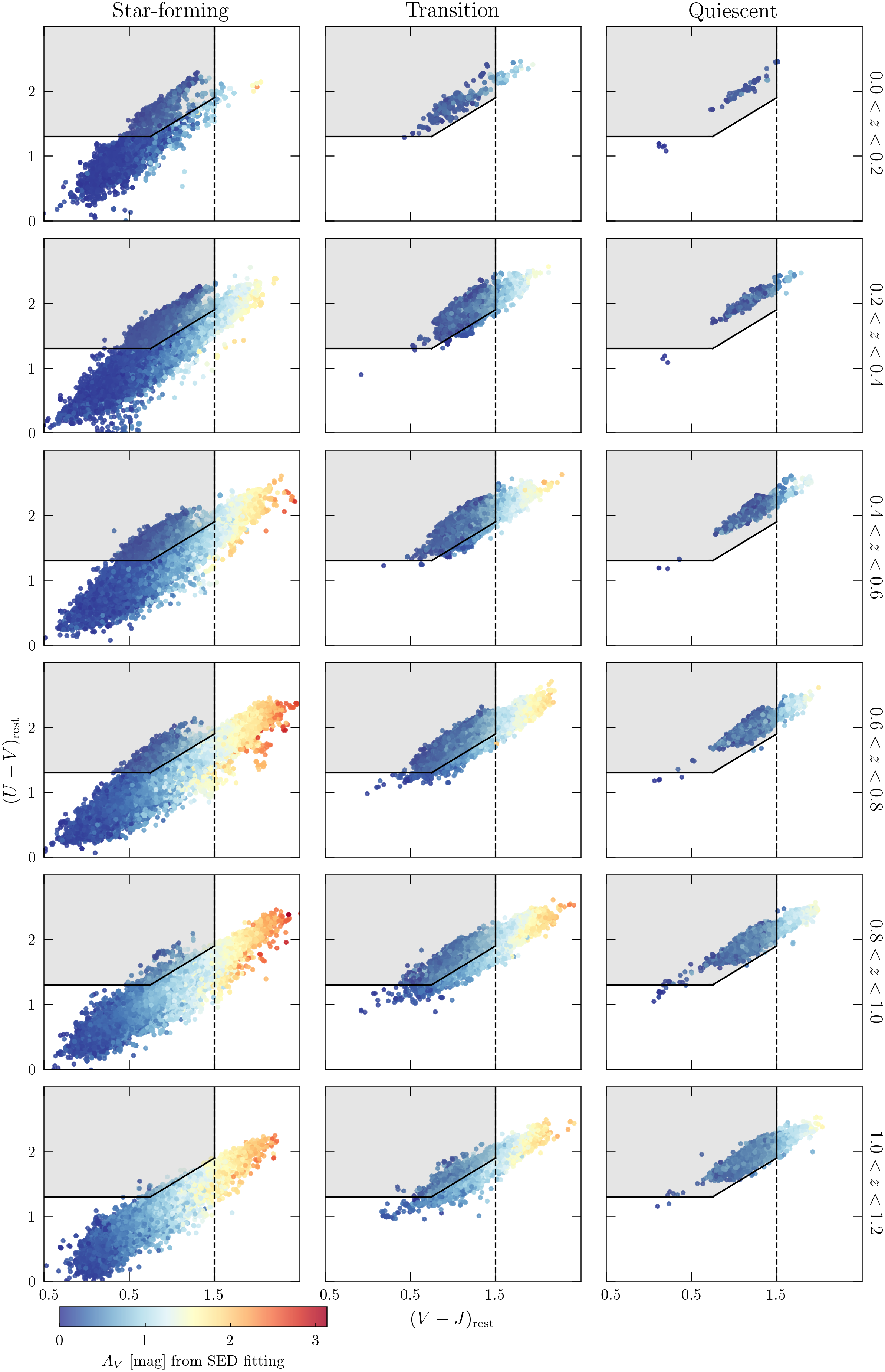} 
\caption{Rest-frame $(U-V)$ vs. $(V-J)$ diagrams for each redshift slice, divided into columns of star-forming, transition, and quiescent galaxy populations, and colour-coded by optical dust attenuation $A_V$ from SED fitting by \citet{DR1-TP012}. Quiescent wedges (shaded regions bounded by solid black lines) contain galaxies satisfying the relations in Eq.\,(\ref{eq:UVJ_quiescent}).}
\label{fig:UVJ_Av_diagram_app}
}]

\FloatBarrier
\twocolumn[{
\section{Mock HMC parameter recovery tests \label{app:Mock_Tests}}
\capfig
\centering
\includegraphics[angle=0,width=0.95\hsize]{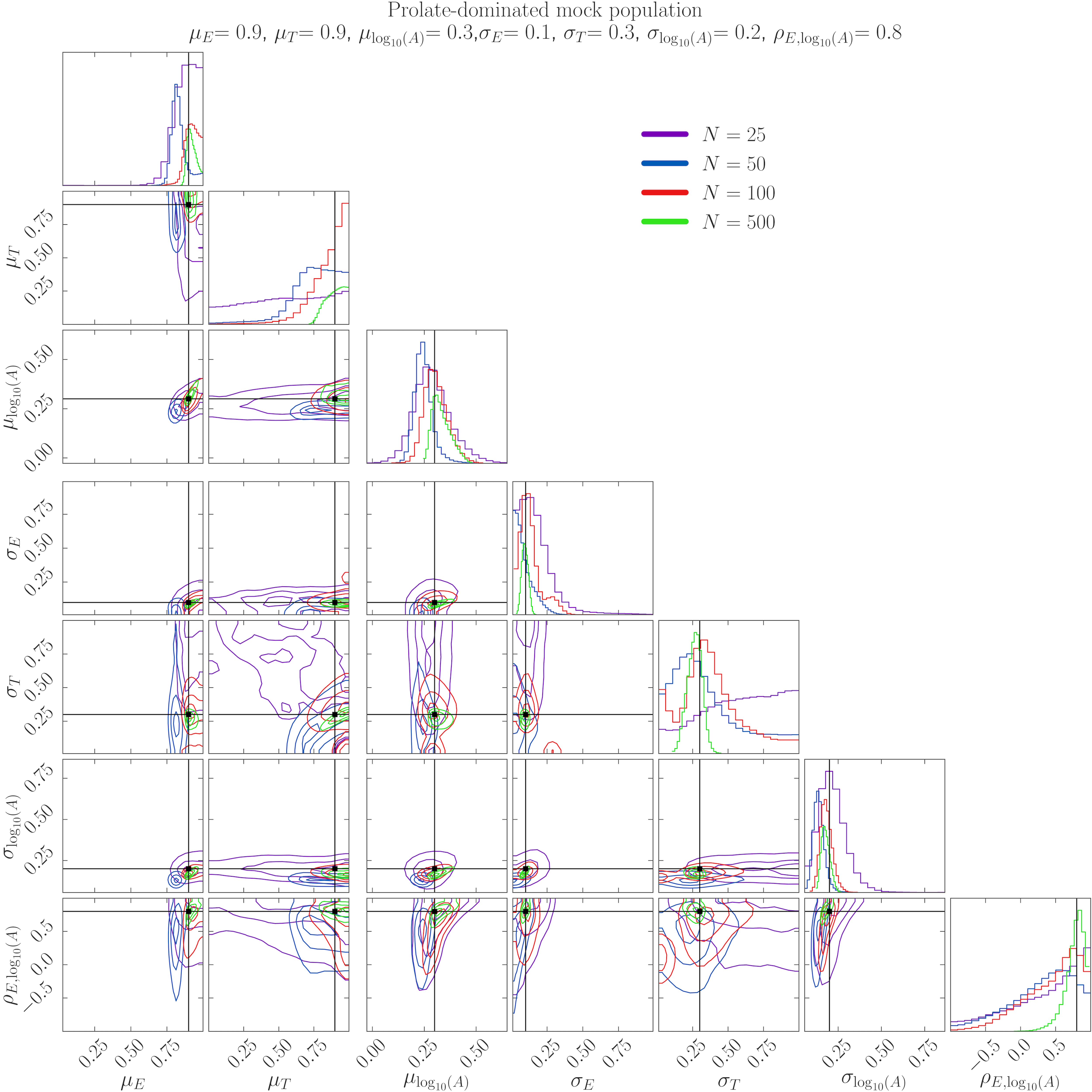}
\caption{HMC recovery of shape parameters for a prolate-dominated mock population with varying sample size: $N=500$ (green); 100 (red); 50 (blue); and 25 (purple). Contours show the 11.8\% ($0.5\,\sigma$), 39.3\% ($1\,\sigma$), and 67.5\% ($1.5\,\sigma$) credible levels; black lines mark ground-truth values (in title).}
\label{fig:mocks_prolate}
}]

We performed mock parameter-recovery tests following the procedure of \citetalias{Pandya_2024} to assess the robustness of our model and the HMC NUTS sampler. We generated synthetic galaxy populations corresponding to three intrinsic shape regimes by varying only the mean ellipticity and triaxiality: prolate-dominated ($\mu_E=0.9$, $\mu_T=0.9$), oblate-dominated ($\mu_E=0.9$, $\mu_T=0.1$), and spheroid-dominated ($\mu_E=0.1$, $\mu_T=0.1$). All other parameters are at fixed values; $\mu_{\logten(A)}=0.3$, $\sigma_E=0.1$, $\sigma_T=0.3$, $\sigma_{\logten(A)}=0.2$, and $\rho_{E,\logten(A)}=0.8$. For each shape class, we generated mock 2D $b/a$ versus $\logten(a)$ histograms with varying sample sizes ($N=25, 50, 100, 500$) to study how the fidelity of posterior recovery scales with the number of observed galaxies. We then performed inference with \texttt{PyMC} using the same HMC sampling setup, and compared the inferred posterior means to the ground-truth parameter values. Figure \ref{fig:mocks_prolate} shows a corner plot \citep{corner} for a parameter recovery test for an example prolate-dominated mock ellipsoid population for different sample sizes ($N=500, 100, 50,$ and $25$). As in \citetalias{Pandya_2024}, the HMC model can successfully recover the true parameters for samples as small as $N=50$, but struggles at lower sample sizes -- particularly for $\mu_T$ and $\sigma_T$ -- making it difficult to reliably distinguish prolate from oblate populations.

\FloatBarrier
\twocolumn[{
\section{Data sample completeness \label{app:Completeness}}
\capfig
\centering
\includegraphics[angle=0,width=\hsize]{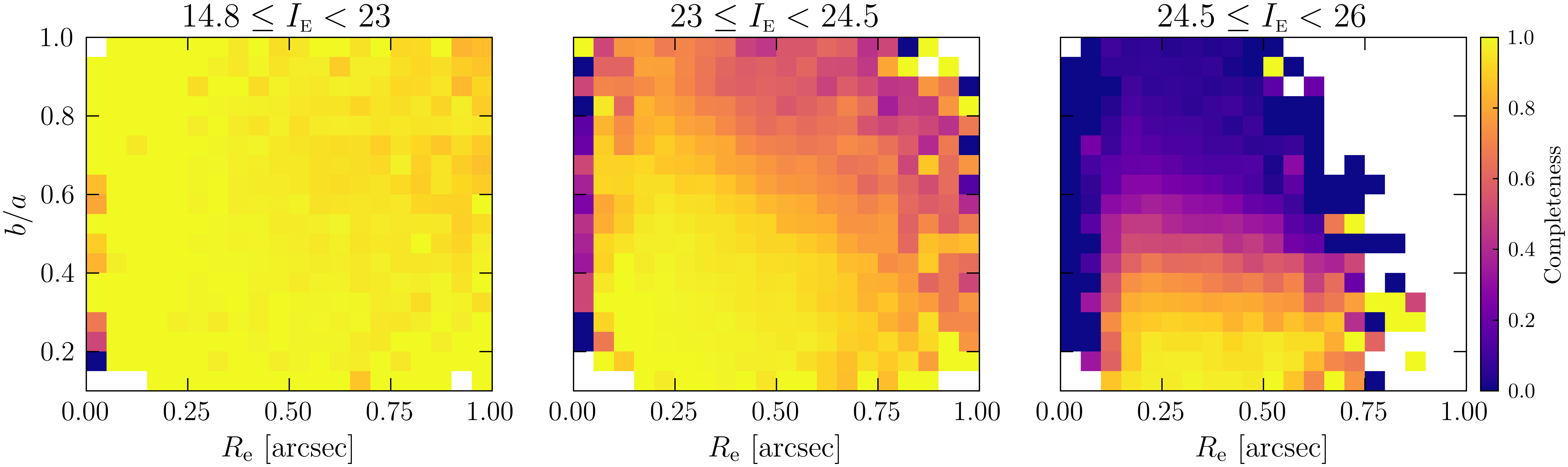}
\caption{Detection fraction across the $b/a$--$\mathrm{size}$ plane based on mock single-Sérsic fits to VIS images from the \Euclid Morphology Challenge. Left: mock galaxies with $14.8\leq\IE<23$; Middle: mock galaxies with $23\leq\IE<24.5$; Right: mock galaxies with $24.5\leq\IE<26$.}
\label{fig:completeness-EMC}
}]
\par\vspace*{1pt}

We first evaluated the morphological completeness across the $b/a$--$\mathrm{size}$ plane for our sample using mock galaxies from the \Euclid Morphology Challenge \citep{Merlin-EP25, Bretonniere-EP26}. For both the mock sources and our magnitude-limited sample, we used the axis ratios $b/a$ and effective (half-light) radii ($R_{\rm e}$, in arcseconds) from single-component Sérsic fits to VIS images with \texttt{SourceXtractor++}, and compared the two distributions to assess the detection fraction. Figure \ref{fig:completeness-EMC} shows the detection fraction for three different bins of brightness: (1) `bright' mock sources with $14.8\leq\IE<23$; (2) `moderately faint' mock sources with $23\leq\IE<24.5$; and (3) `very faint' mock sources with $24.5\leq\IE<26$. Crucially, across the magnitude range of our sample ($\IE\leq24.5$), completeness remains high in the upper right of the $b/a$--$\mathrm{size}$ plane, where large face-on discs (high $b/a$, large $R_{\rm e}$) reside. These are the objects most vulnerable to surface-brightness dimming and thus most likely to be lost from a magnitude-limited sample; their retention indicates that our inferred shape distributions are not strongly biased against intrinsically flat, extended systems. The most incomplete region is instead confined to faint, compact sources (low $b/a$, low $R_{\rm e}$) beyond our magnitude limit, which contribute least to the population-level shape constraints.

\begin{figure}[!ht]
\par\vspace*{12pt}
\resizebox{\hsize}{!}{\includegraphics{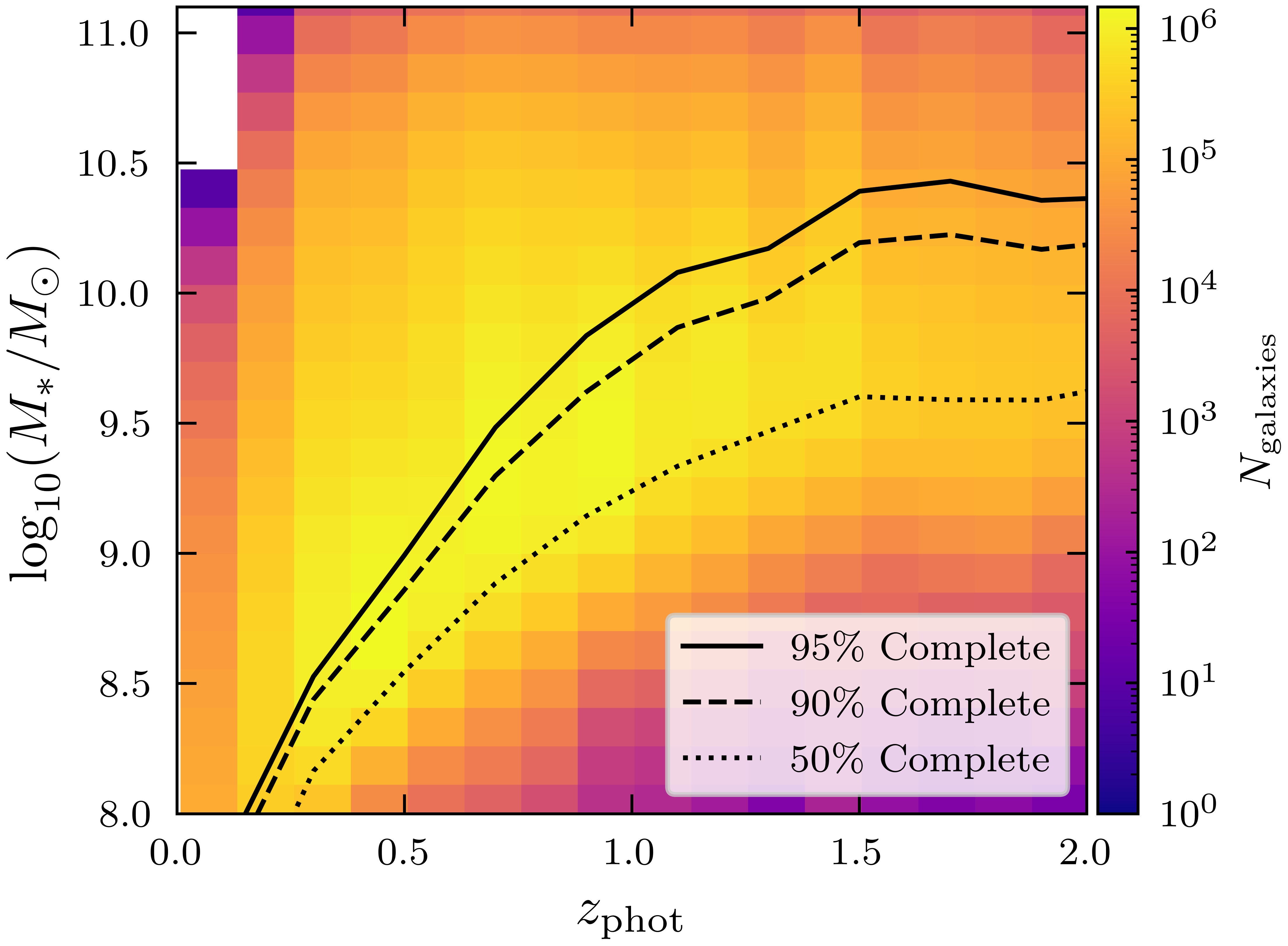}}
\caption{Mass completeness of the magnitude-limited sample; 95\%, 90\%, and 50\% limits (black lines) are derived following \citet{Pozzetti_2010}.}
\label{fig:mass_completeness}
\end{figure}

We also assessed stellar-mass completeness by applying the method of \citet{Pozzetti_2010} to our DR1 magnitude-limited ($\IE\leq24.5$) star-forming sample. We computed the stellar masses at which the sample reaches 95\%, 90\%, and 50\% completeness, indicated by the black lines in Fig.\,\ref{fig:mass_completeness}. We find that the sample is at least 90\% complete for all bins except the $\Mstarwun=9.0$--$9.5$, $0.8<z<1.0$ and $\Mstarwun=9.0$--$9.5$, $1.0<z<1.2$ bin, which reach up to approximately 75\% and 50\% completeness, respectively. Nevertheless, Figs.\,\ref{fig:3DE_evolution}--\ref{fig:3DlogArho_evolution} show that there is no significant increase in shape parameter uncertainties at higher redshifts. Given this, and the fact that our mock detection fraction remains nearly complete across the $b/a$--$\mathrm{size}$ plane, mass completeness should not strongly bias morphological classifications. We reiterate that future \Euclid releases should push completeness limits to lower masses.

\FloatBarrier
\twocolumn[{
\section{3D shape modelling results \label{app:Model-Results}}
Table \ref{tab:bin_conv_diag} provides basic convergence diagnostic statistics (described in Sect.\,\ref{sc:Methodology}) per bin from \texttt{PyMC} NUTS sampling across all chains. Table \ref{tab:3D_shape_results} tabulates the results of our inference model from random draws of the posterior for each mass-redshift bin.
\par\bigskip
\captab
\caption{Per-bin convergence diagnostics.}
\label{tab:bin_conv_diag}
\centering
\begin{tabular}{ccccccc}
\hline
\noalign{\vskip 2pt}
{$z$} & {$\Mstarwun$} & {$N_{\rm galaxies}$} & {$\hat{R}_{\rm max}$} & {ESS$_{\rm bulk,\,min}$} & {ESS$_{\rm tail,\,min}$} & {BFMI$_{\rm min}$} \\ [0.5ex]
\hline
\noalign{\vskip 2pt}
$0.0$--$0.2$ & $9.0$--$9.5$   & \num{306909}  & 1.003 & 1989.517 & 2581.169 & 1.017 \\ [0.5ex]
$0.0$--$0.2$ & $9.5$--$10.0$  & \num{109780}  & 1.001 & 4693.701 & 4701.812 & 0.961 \\ [0.5ex]
$0.0$--$0.2$ & $10.0$--$10.5$ & \num{21000}   & 1.001 & 4218.635 & 2996.148 & 0.940 \\ [0.5ex]
$0.2$--$0.4$ & $9.0$--$9.5$   & \num{2694631} & 1.002 & 2458.077 & 3020.175 & 0.978 \\ [0.5ex]
$0.2$--$0.4$ & $9.5$--$10.0$  & \num{1462850} & 1.002 & 3501.545 & 3608.021 & 0.975 \\ [0.5ex]
$0.2$--$0.4$ & $10.0$--$10.5$ & \num{694841}  & 1.001 & 5245.822 & 3371.312 & 0.964 \\ [0.5ex]
$0.4$--$0.6$ & $9.0$--$9.5$   & \num{3983560} & 1.001 & 5329.382 & 5143.545 & 1.012 \\ [0.5ex]
$0.4$--$0.6$ & $9.5$--$10.0$  & \num{2386934} & 1.001 & 5233.964 & 3805.091 & 0.992 \\ [0.5ex]
$0.4$--$0.6$ & $10.0$--$10.5$ & \num{1313521} & 1.002 & 4156.797 & 3748.746 & 0.972 \\ [0.5ex]
$0.6$--$0.8$ & $9.0$--$9.5$   & \num{4717101} & 1.001 & 5701.216 & 5447.162 & 1.014 \\ [0.5ex]
$0.6$--$0.8$ & $9.5$--$10.0$  & \num{3779549} & 1.004 & 5291.588 & 5156.641 & 1.020 \\ [0.5ex]
$0.6$--$0.8$ & $10.0$--$10.5$ & \num{1898979} & 1.002 & 3872.312 & 5028.314 & 0.992 \\ [0.5ex]
$0.8$--$1.0$ & $9.0$--$9.5$   & \num{4362066} & 1.001 & 5299.914 & 5526.251 & 1.009 \\ [0.5ex]
$0.8$--$1.0$ & $9.5$--$10.0$  & \num{4416628} & 1.001 & 4867.143 & 5254.788 & 0.981 \\ [0.5ex]
$0.8$--$1.0$ & $10.0$--$10.5$ & \num{2067640} & 1.002 & 4025.050 & 4910.138 & 0.987 \\ [0.5ex]
$1.0$--$1.2$ & $9.0$--$9.5$   & \num{1875729} & 1.001 & 5368.448 & 5253.693 & 1.027 \\ [0.5ex]
$1.0$--$1.2$ & $9.5$--$10.0$  & \num{2961898} & 1.001 & 5288.332 & 5132.424 & 1.031 \\ [0.5ex]
$1.0$--$1.2$ & $10.0$--$10.5$ & \num{1480278} & 1.002 & 1965.738 & 2535.839 & 1.032 \\ [0.5ex]
\hline
\end{tabular}
\tablefoot{Convergence diagnostic statistics from NUTS sampling with \texttt{PyMC} for each mass-redshift bin from the posterior means. Statistics here include the minimum bulk effective sample size $\mathrm{ESS_{bulk,\,min}}$, the minimum tail effective sample size $\mathrm{ESS_{tail,\,min}}$, the maximum $\hat{R}_{\rm max}$, and the minimum BFMI, $\mathrm{BFMI_{min}}$, across all sampling chains.}
\par\bigskip
\caption{Per-bin inferred 3D shape parameters and class fractions.}
\label{tab:3D_shape_results}
\centering
\resizebox{\textwidth}{!}{
\footnotesize
\setlength{\tabcolsep}{2.5pt}
\begin{tabular}{ccccccccccccc}
\hline
\noalign{\vskip 2.5pt}
$z$ & $\Mstarwun$ & $N_{\rm galaxies}$ & $\mu_E$ & $\mu_T$ & $\mu_{\logten(A)}$ & $\sigma_E$ & $\sigma_T$ & $\sigma_{\logten(A)}$ & $\rho_{E,\logten(A)}$ & $f_{\rm prolate}$ & $f_{\rm oblate}$ & $f_{\rm spheroid}$ \\ [0.75ex]
\hline
\noalign{\vskip 2pt}
$0.0$--$0.2$ & $9.0$--$9.5$   & \num{306909}  & $0.809\pm0.002$ & $0.275\pm0.004$ & $0.584\pm0.002$ & $0.177\pm0.001$ & $0.196\pm0.004$ & $0.250\pm0.001$ & $0.854\pm0.002$ & $0.043\pm0.002$ & $0.763\pm0.003$ & $0.194\pm0.002$ \\ [0.75ex]
$0.0$--$0.2$ & $9.5$--$10.0$  & \num{109780}  & $0.806\pm0.001$ & $0.168\pm0.011$ & $0.652\pm0.001$ & $0.088\pm0.002$ & $0.180\pm0.009$ & $0.226\pm0.001$ & $0.875\pm0.003$ & $0.011\pm0.002$ & $0.965\pm0.003$ & $0.024\pm0.002$ \\ [0.75ex]
$0.0$--$0.2$& $10.0$--$10.5$  & \num{21000}   & $0.821\pm0.001$ & $0.008\pm0.008$ & $0.769\pm0.002$ & $0.059\pm0.002$ & $0.386\pm0.016$ & $0.169\pm0.002$ & $0.873\pm0.009$ & $0.113\pm0.012$ & $0.884\pm0.012$ & $0.003\pm0.001$ \\ [0.75ex]
$0.2$--$0.4$ & $9.0$--$9.5$   & \num{2694631} & $0.797\pm0.000$ & $0.277\pm0.002$ & $0.544\pm0.000$ & $0.130\pm0.000$ & $0.229\pm0.001$ & $0.236\pm0.000$ & $0.884\pm0.001$ & $0.080\pm0.001$ & $0.798\pm0.001$ & $0.122\pm0.000$ \\ [0.75ex]
$0.2$--$0.4$ & $9.5$--$10.0$  & \num{1462850} & $0.781\pm0.000$ & $0.108\pm0.006$ & $0.614\pm0.000$ & $0.117\pm0.001$ & $0.223\pm0.004$ & $0.221\pm0.000$ & $0.887\pm0.001$ & $0.019\pm0.001$ & $0.885\pm0.001$ & $0.096\pm0.001$ \\ [0.75ex]
$0.2$--$0.4$ & $10.0$--$10.5$ & \num{694841}  & $0.781\pm0.000$ & $0.001\pm0.001$ & $0.722\pm0.000$ & $0.086\pm0.000$ & $0.336\pm0.002$ & $0.187\pm0.000$ & $0.870\pm0.001$ & $0.073\pm0.002$ & $0.879\pm0.002$ & $0.048\pm0.001$ \\ [0.75ex]
$0.4$--$0.6$ & $9.0$--$9.5$   & \num{3983560} & $0.791\pm0.000$ & $0.341\pm0.001$ & $0.544\pm0.000$ & $0.093\pm0.000$ & $0.269\pm0.001$ & $0.208\pm0.000$ & $0.838\pm0.001$ & $0.177\pm0.001$ & $0.760\pm0.001$ & $0.063\pm0.000$ \\ [0.75ex]
$0.4$--$0.6$ & $9.5$--$10.0$  & \num{2386934} & $0.782\pm0.000$ & $0.000\pm0.000$ & $0.631\pm0.000$ & $0.086\pm0.000$ & $0.460\pm0.001$ & $0.205\pm0.000$ & $0.888\pm0.001$ & $0.166\pm0.001$ & $0.783\pm0.001$ & $0.050\pm0.000$ \\ [0.75ex]
$0.4$--$0.6$ & $10.0$--$10.5$ & \num{1313521} & $0.775\pm0.000$ & $0.000\pm0.000$ & $0.741\pm0.000$ & $0.094\pm0.000$ & $0.455\pm0.002$ & $0.194\pm0.000$ & $0.873\pm0.001$ & $0.161\pm0.001$ & $0.761\pm0.002$ & $0.078\pm0.001$ \\ [0.75ex]
$0.6$--$0.8$ & $9.0$--$9.5$   & \num{4717101} & $0.782\pm0.000$ & $0.445\pm0.001$ & $0.539\pm0.000$ & $0.078\pm0.000$ & $0.262\pm0.001$ & $0.189\pm0.000$ & $0.745\pm0.001$ & $0.281\pm0.001$ & $0.667\pm0.001$ & $0.052\pm0.000$ \\ [0.75ex]
$0.6$--$0.8$ & $9.5$--$10.0$  & \num{3779549} & $0.748\pm0.000$ & $0.286\pm0.002$ & $0.619\pm0.000$ & $0.104\pm0.000$ & $0.258\pm0.002$ & $0.205\pm0.000$ & $0.811\pm0.001$ & $0.119\pm0.001$ & $0.727\pm0.001$ & $0.154\pm0.001$ \\ [0.75ex]
$0.6$--$0.8$ & $10.0$--$10.5$ & \num{1898979} & $0.725\pm0.000$ & $0.219\pm0.002$ & $0.725\pm0.000$ & $0.142\pm0.000$ & $0.196\pm0.002$ & $0.208\pm0.000$ & $0.832\pm0.001$ & $0.024\pm0.001$ & $0.701\pm0.001$ & $0.275\pm0.001$ \\ [0.75ex]
$0.8$--$1.0$ & $9.0$--$9.5$   & \num{4362066} & $0.775\pm0.000$ & $0.505\pm0.000$ & $0.541\pm0.000$ & $0.067\pm0.000$ & $0.230\pm0.001$ & $0.172\pm0.000$ & $0.577\pm0.001$ & $0.349\pm0.001$ & $0.605\pm0.001$ & $0.047\pm0.000$ \\ [0.75ex]
$0.8$--$1.0$ & $9.5$--$10.0$  & \num{4416628} & $0.731\pm0.000$ & $0.350\pm0.001$ & $0.617\pm0.000$ & $0.102\pm0.000$ & $0.218\pm0.001$ & $0.194\pm0.000$ & $0.706\pm0.001$ & $0.122\pm0.001$ & $0.676\pm0.001$ & $0.202\pm0.001$ \\ [0.75ex]
$0.8$--$1.0$ & $10.0$--$10.5$ & \num{2067640} & $0.671\pm0.000$ & $0.258\pm0.002$ & $0.716\pm0.000$ & $0.160\pm0.000$ & $0.184\pm0.002$ & $0.207\pm0.000$ & $0.724\pm0.001$ & $0.022\pm0.001$ & $0.540\pm0.001$ & $0.438\pm0.001$ \\ [0.75ex]
$1.0$--$1.2$ & $9.0$--$9.5$   & \num{1875729} & $0.771\pm0.000$ & $0.547\pm0.001$ & $0.525\pm0.000$ & $0.059\pm0.000$ & $0.216\pm0.001$ & $0.161\pm0.000$ & $0.362\pm0.003$ & $0.419\pm0.001$ & $0.544\pm0.001$ & $0.038\pm0.000$ \\ [0.75ex]
$1.0$--$1.2$ & $9.5$--$10.0$  & \num{2961898} & $0.716\pm0.000$ & $0.392\pm0.001$ & $0.600\pm0.000$ & $0.096\pm0.000$ & $0.212\pm0.001$ & $0.185\pm0.000$ & $0.518\pm0.001$ & $0.151\pm0.001$ & $0.608\pm0.001$ & $0.241\pm0.001$ \\ [0.75ex]
$1.0$--$1.2$ & $10.0$--$10.5$ & \num{1480278} & $0.600\pm0.001$ & $0.205\pm0.009$ & $0.692\pm0.000$ & $0.166\pm0.000$ & $0.324\pm0.009$ & $0.200\pm0.000$ & $0.536\pm0.002$ & $0.067\pm0.002$ & $0.335\pm0.002$ & $0.599\pm0.001$ \\ [0.75ex]
\hline
\end{tabular}}
\tablefoot{Mean and standard deviations for the inferred seven 3D shape parameters and class fractions in each mass-redshift bin from random draws from the posterior. Uncertainties are statistical only.}
}]

\FloatBarrier
\twocolumn[{
\section{Visual examples of galaxies by class probability \label{app:Visualisation}}
\capfig
\centering
\includegraphics[angle=0,width=\hsize]{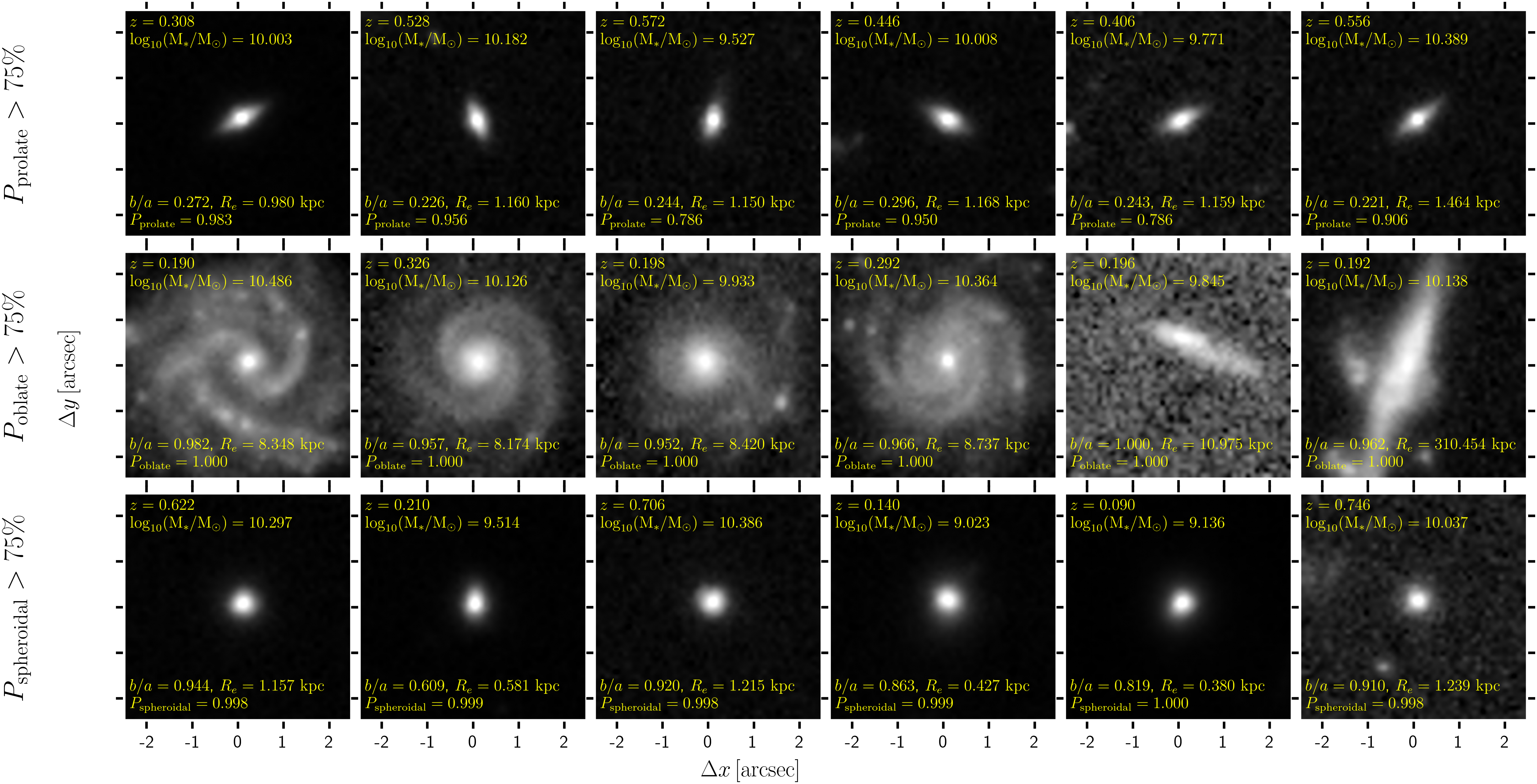}
\caption{Example $\ang{;;5} \times \ang{;;5}$ \Euclid VIS cutouts for prolate (top), oblate (middle), and spheroid (bottom) galaxies with $P_{\rm class}>75\%$. Each cutout shows the photometric redshift, stellar mass, Sérsic $b/a$ and $R_{\rm e}$, and class probability. Axes give the offset from the centre of the source in arcseconds.}
\label{fig:cutouts}

}]

Following section 3.4 of \citetalias{Pandya_2024} (also section 5.1 of \citetalias{Zhang_2019}), we used the posterior means of the model histograms to determine the relative contributions from each 3D ellipsoid class, assigning a class probability to each galaxy based only on observed $\logten(a)$ and $b/a$ values rather than population-level fractions. We then examined \Euclid VIS image cutouts\footnote{Cutouts were created using the \texttt{Cutana} software \citep{Cutana} on ESA Datalabs \citep{Datalabscite}.} for `high-probability' candidates (galaxies with individual class probabilities $\,>75\%$, an arbitrary threshold) across various mass-redshift bins. Figure \ref{fig:cutouts} shows several examples of these cutouts for the prolate (top row), oblate (middle row) and spheroid (bottom row) classes. Visual inspection reveals both consistencies and differences from figure 17 of \citetalias{Pandya_2024}. 

Within this sample, only 34 galaxies have $P_{\rm prolate}>75\%$. The cutouts in the top row of Fig.\,\ref{fig:cutouts} show the most common type of cutout seen for prolate candidates. Our high-probability prolate candidates appear more compact, with bright cores and only slightly extended features, making visual shape determination ambiguous. While these example candidates have the highest $P_{\rm prolate}$, they have an angular $R_{\rm e}$ only 1.1--1.4 times the VIS PSF FWHM and are therefore barely resolved, and most lie at $\Mstarwun>9.5$, outside the low-mass bins in which the prolate-to-oblate trend is found. The prolate candidates in the top two rows of figure 14 in \citetalias{Pandya_2024} exhibit clearer extended disc components in false-colour RGB images, though these are at higher redshifts than our sample.

Several galaxies with $P_{\rm oblate}>75\%$ are shown in the middle row of Fig.\,\ref{fig:cutouts}. The first four cutouts show oblate candidates that appear to be face-on spiral galaxies, consistent with cutouts from the middle two rows of figure 14 in \citetalias{Pandya_2024}. While most high-probability oblate candidates resemble these, the last two cutouts in the middle row show candidates that appear to be edge-on disc galaxies, in two different mass-redshift bins. As discussed in Sect.\,\ref{sc:Methodology}, prolate and oblate galaxies viewed edge-on are difficult to distinguish, since edge-on discs occupy the same low-$b/a$ region as intrinsically elongated systems. However, both edge-on examples here have very large $b/a$ values and observed sizes larger than the range spanned by the 3D ellipsoid model library ($\logten(a)\in[-1,1]$, corresponding to $a=0.1$--$10\,{\rm kpc}$). Thus, the classification is effectively an extrapolation, and the edge-on disc interpretation cannot be reliably distinguished from a genuinely elongated system for these objects on the basis of $b/a$ and $\logten(a)$ alone.

The final row of Fig.\,\ref{fig:cutouts} shows examples of high-probability spheroid candidates. These cutouts appear compact, round, and centrally concentrated, with smooth light profiles and no apparent disc components -- consistent with those in the bottom two rows of figure 14 in \citetalias{Pandya_2024}. We stress that these are illustrative examples, selected as those with the highest class probabilities computed from observed $b/a$ and $\logten(a)$ values alone. The same two projected parameters that drive our population-level inference cannot uniquely determine an individual galaxy's intrinsic 3D shape (Sect.\,\ref{ssc:Discussion-Limitations}), so they should not be interpreted as definitive morphological classifications. Combining \Euclid VIS and NIR imaging, or additional morphological measures, may refine per-object classifications. We defer such analyses to a future paper.

\twocolumn[{
\section{Per-bin observed, model, and residual histograms \label{app:Residuals}}
Figures \ref{fig:z0_0p2_logM9_9p5_residuals}--\ref{fig:z1_1p2_logM10_10p5_residuals} show, for each mass--redshift bin, the observed and posterior-mean model $b/a$--$\logten(a)$ histograms, together with the residuals $N_{\rm obs}-N_{\rm model}$ and the Pearson residuals $(N_{\rm obs}-N_{\rm model})/\sqrt{N_{\rm model}}$, computed over the cells used in the likelihood. The residuals indicate systematic differences between the data and the single trivariate Gaussian model that are not captured by the statistical uncertainties in Table\,\ref{tab:3D_shape_results}.
\par\medskip
\capfig
\centering
\includegraphics[angle=0,width=\hsize]{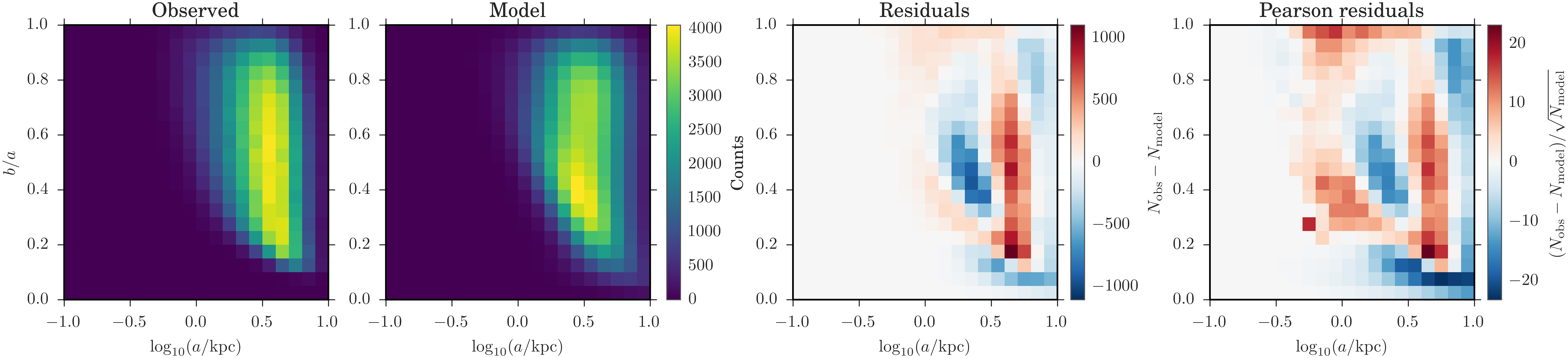} 
\caption{Observed (first panel) and posterior-mean model (second panel) $b/a$--$\logten(a)$ histograms for the $0.0 < z<0.2$, $9.0<\Mstarwun<9.5$ bin, with the residuals $N_{\rm obs}-N_{\rm model}$ (third panel) and the Pearson residuals $(N_{\rm obs}-N_{\rm model})/\sqrt{N_{\rm model}}$ (fourth panel). Colour bars give galaxy counts for the histograms, the difference in counts for the residuals, and the dimensionless Pearson residual for the last panel.}
\label{fig:z0_0p2_logM9_9p5_residuals}
\par\medskip
\centering
\includegraphics[angle=0,width=\hsize]{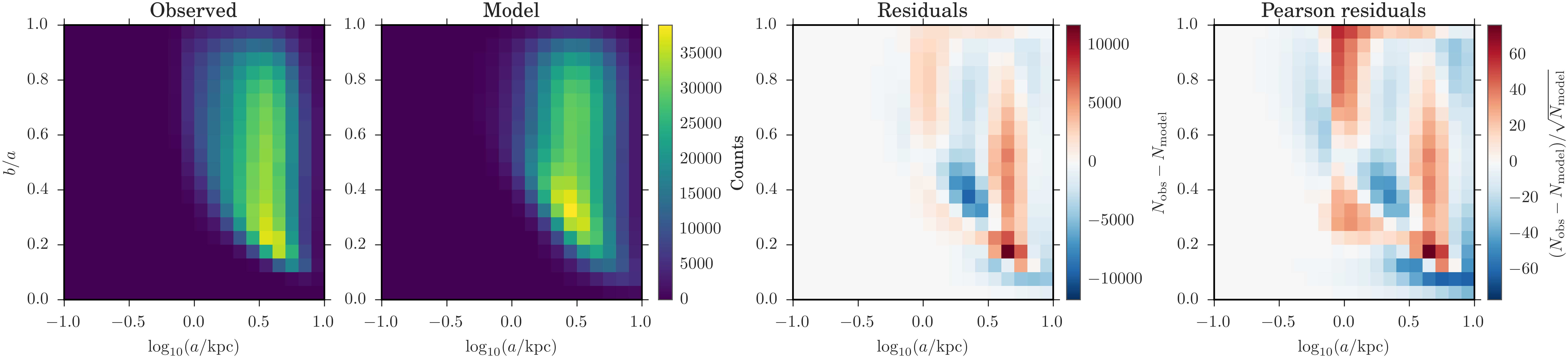} 
\caption{Same as Fig.\,\ref{fig:z0_0p2_logM9_9p5_residuals}, but for the $0.2<z<0.4$, $9.0<\Mstarwun<9.5$ bin.}
\label{fig:z0p2_0p4_logM9_9p5_residuals}
\par\medskip
\centering
\includegraphics[angle=0,width=\hsize]{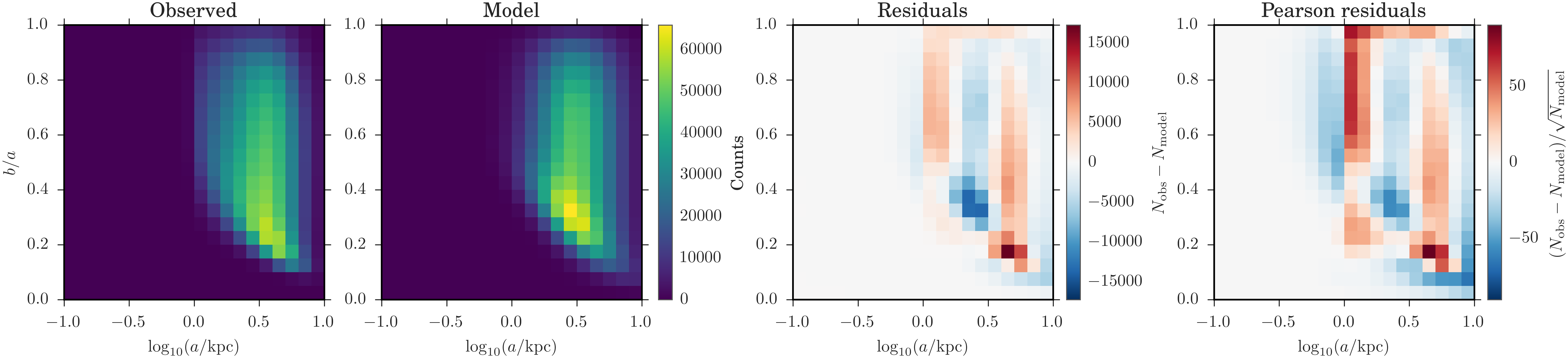} 
\caption{Same as Fig.\,\ref{fig:z0_0p2_logM9_9p5_residuals}, but for the $0.4<z<0.6$, $9.0<\Mstarwun<9.5$ bin.}
\label{fig:z0p4_0p6_logM9_9p5_residuals}
\par\medskip
\centering
\includegraphics[angle=0,width=\hsize]{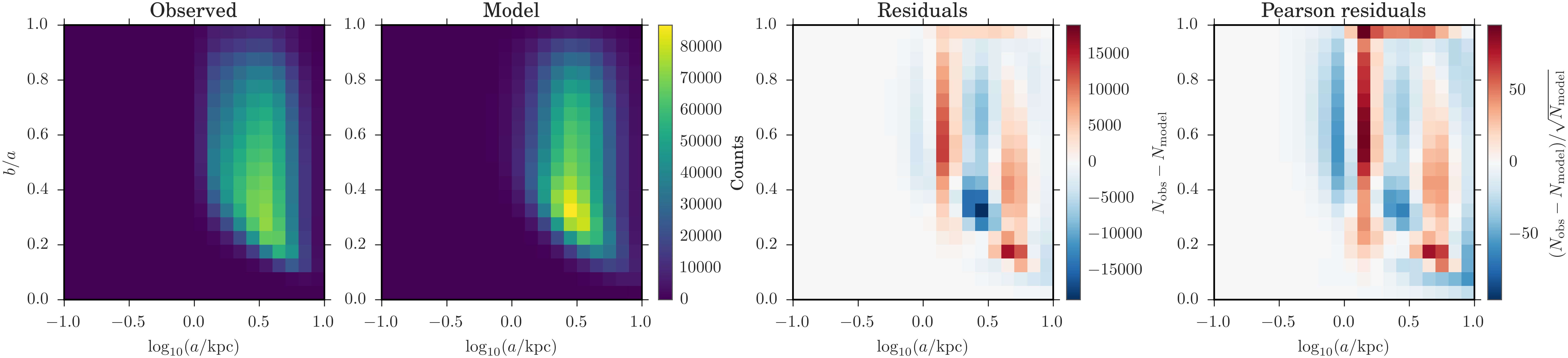} 
\caption{Same as Fig.\,\ref{fig:z0_0p2_logM9_9p5_residuals}, but for the $0.6<z<0.8$, $9.0<\Mstarwun<9.5$ bin.}
\label{fig:z0p6_0p8_logM9_9p5_residuals}
}]

\begin{figure*}[!t]
\centering
\includegraphics[angle=0,width=\hsize]{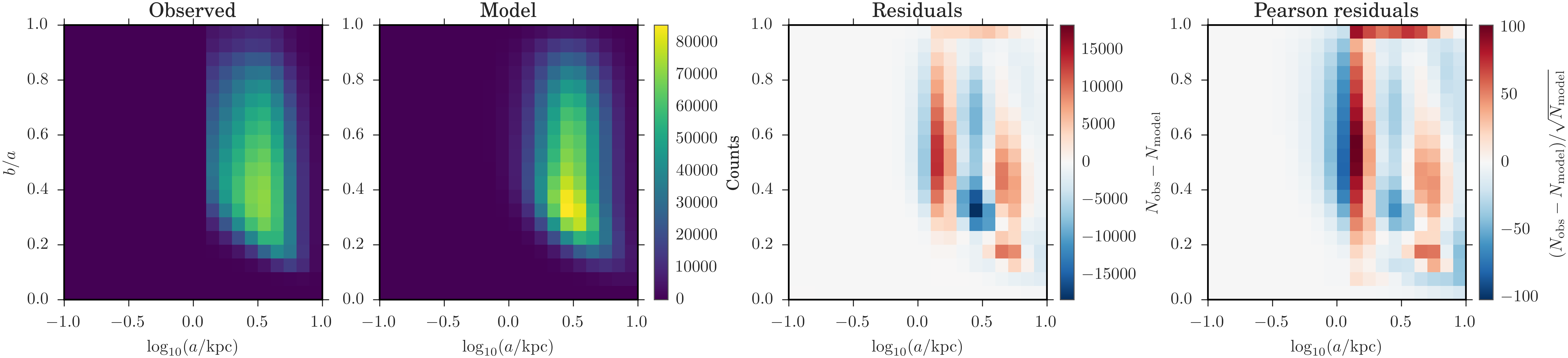} 
\caption{Same as Fig.\,\ref{fig:z0_0p2_logM9_9p5_residuals}, but for the $0.8<z<1.0$, $9.0<\Mstarwun<9.5$ bin.}
\label{fig:z0p8_1_logM9_9p5_residuals}
\end{figure*}

\begin{figure*}[!t]
\centering
\includegraphics[angle=0,width=\hsize]{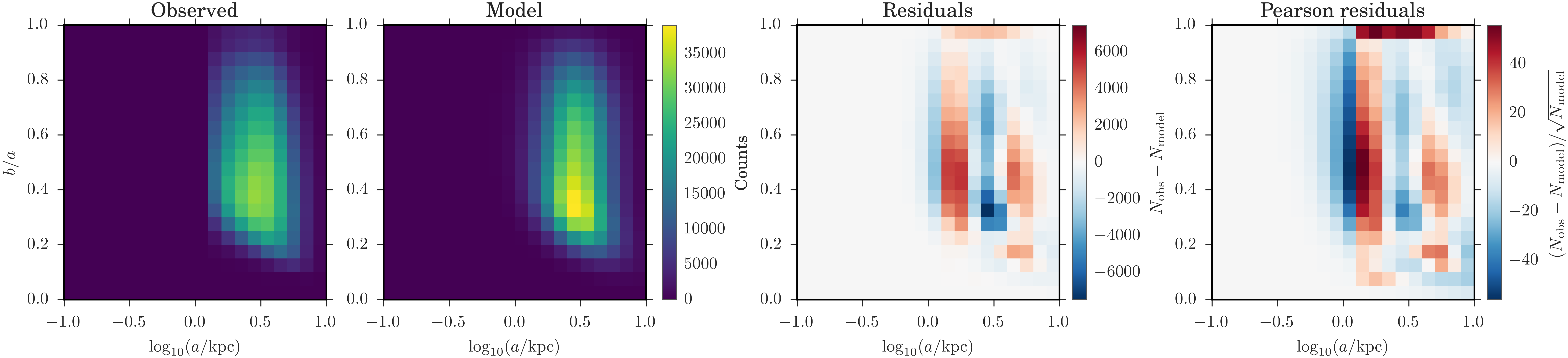} 
\caption{Same as Fig.\,\ref{fig:z0_0p2_logM9_9p5_residuals}, but for the $1.0<z<1.2$, $9.0<\Mstarwun<9.5$ bin.}
\label{fig:z1_1p2_logM9_9p5_residuals}
\end{figure*}

\begin{figure*}[!t]
\centering
\includegraphics[angle=0,width=\hsize]{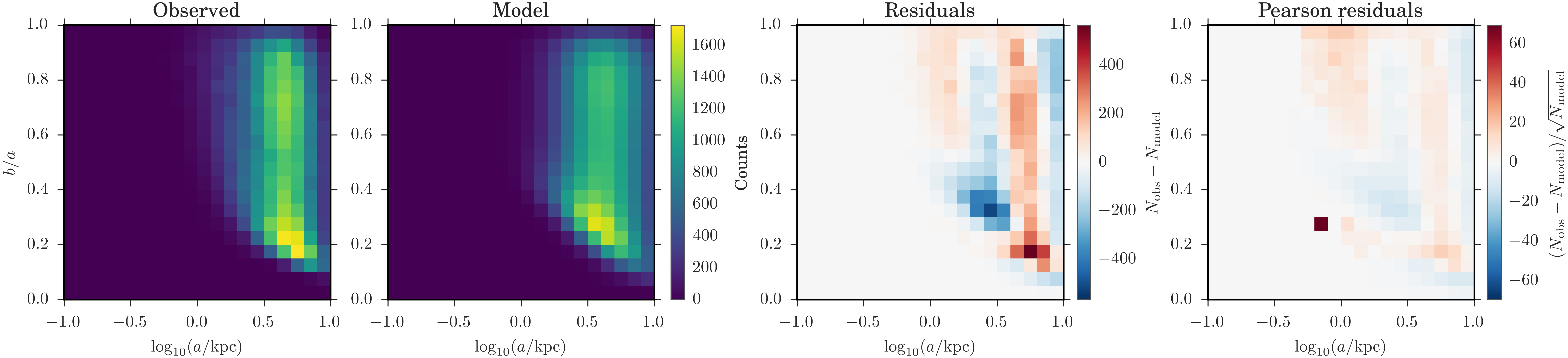} 
\caption{Same as Fig.\,\ref{fig:z0_0p2_logM9_9p5_residuals}, but for the $0.0\lesssim z<0.2$, $9.5<\Mstarwun<10.0$ bin.}
\label{fig:z0_0p2_logM9p5-10_residuals}
\end{figure*}

\begin{figure*}[!t]
\centering
\includegraphics[angle=0,width=\hsize]{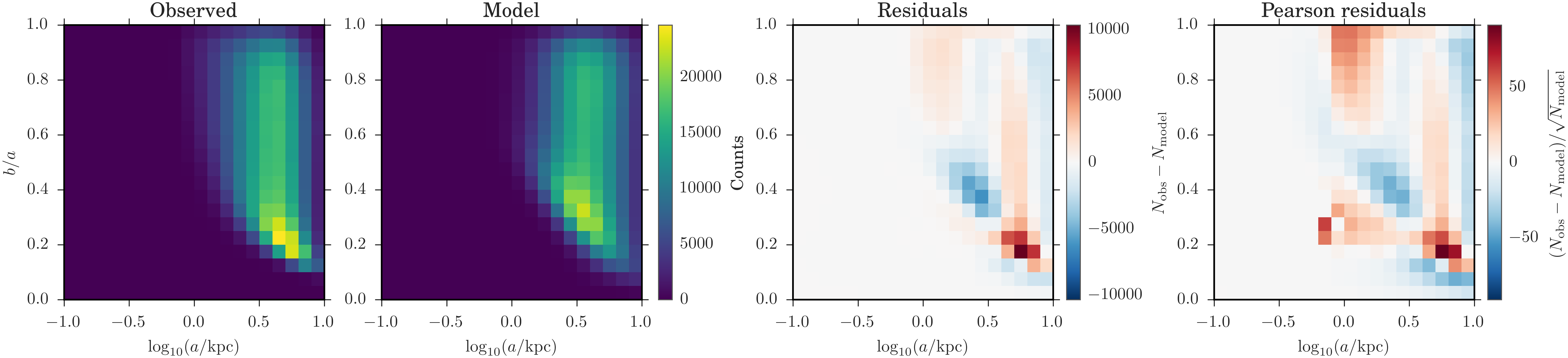} 
\caption{Same as Fig.\,\ref{fig:z0_0p2_logM9_9p5_residuals}, but for the $0.2<z<0.4$, $9.5<\Mstarwun<10.0$ bin.}
\label{fig:z0p2_0p4_logM9p5_10_residuals}
\end{figure*}

\begin{figure*}[!t]
\centering
\includegraphics[angle=0,width=\hsize]{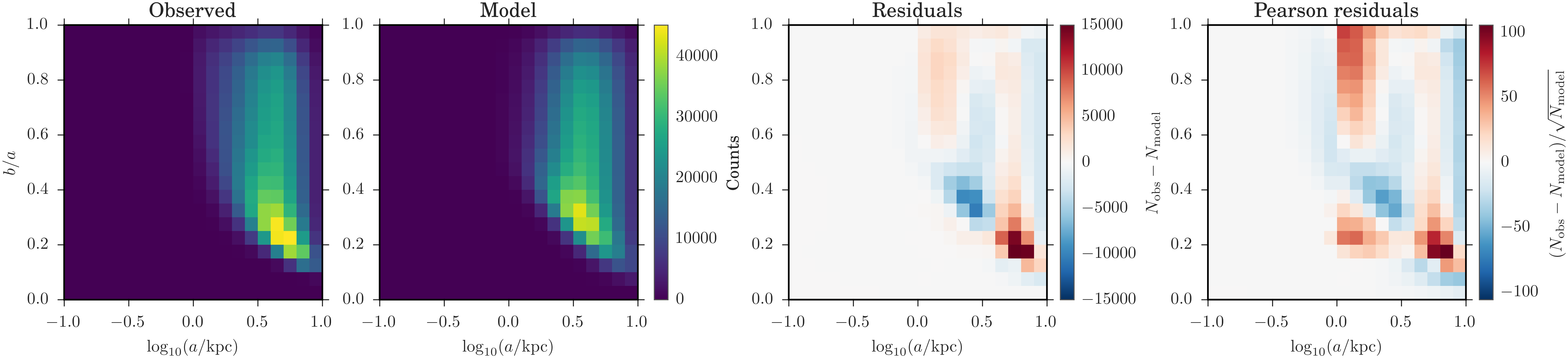} 
\caption{Same as Fig.\,\ref{fig:z0_0p2_logM9_9p5_residuals}, but for the $0.4<z<0.6$, $9.5<\Mstarwun<10.0$ bin.}
\label{fig:z0p4_0p6_logM9p5_10_residuals}
\end{figure*}

\begin{figure*}[!t]
\centering
\includegraphics[angle=0,width=\hsize]{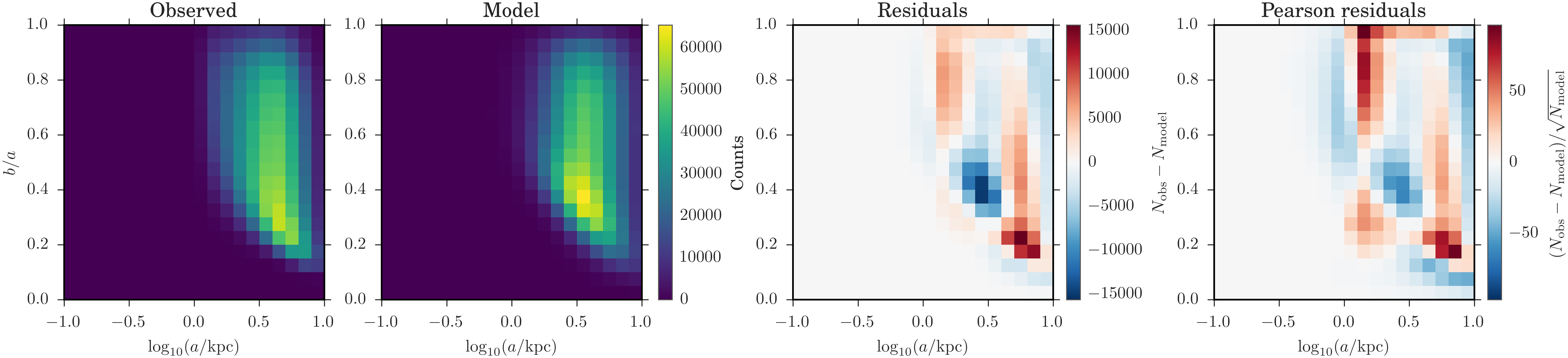} 
\caption{Same as Fig.\,\ref{fig:z0_0p2_logM9_9p5_residuals}, but for the $0.6<z<0.8$, $9.5<\Mstarwun<10.0$ bin.}
\label{fig:z0p6_0p8_logM9p5_10_residuals}
\end{figure*}

\begin{figure*}[!t]
\centering
\includegraphics[angle=0,width=\hsize]{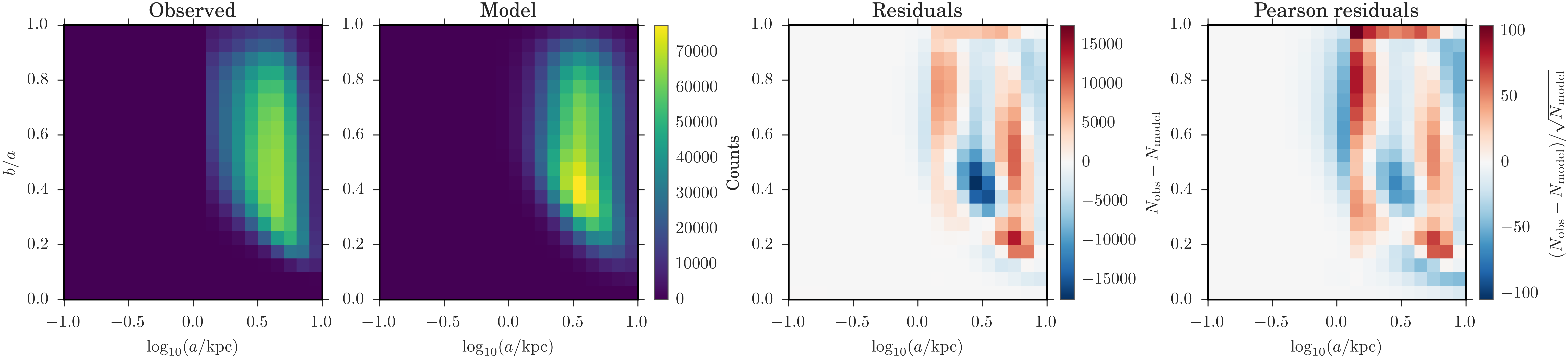} 
\caption{Same as Fig.\,\ref{fig:z0_0p2_logM9_9p5_residuals}, but for the $0.8<z<1.0$, $9.5<\Mstarwun<10.0$ bin.}
\label{fig:z0p8_1_logM9p5_10_residuals}
\end{figure*}

\begin{figure*}[!t]
\centering
\includegraphics[angle=0,width=\hsize]{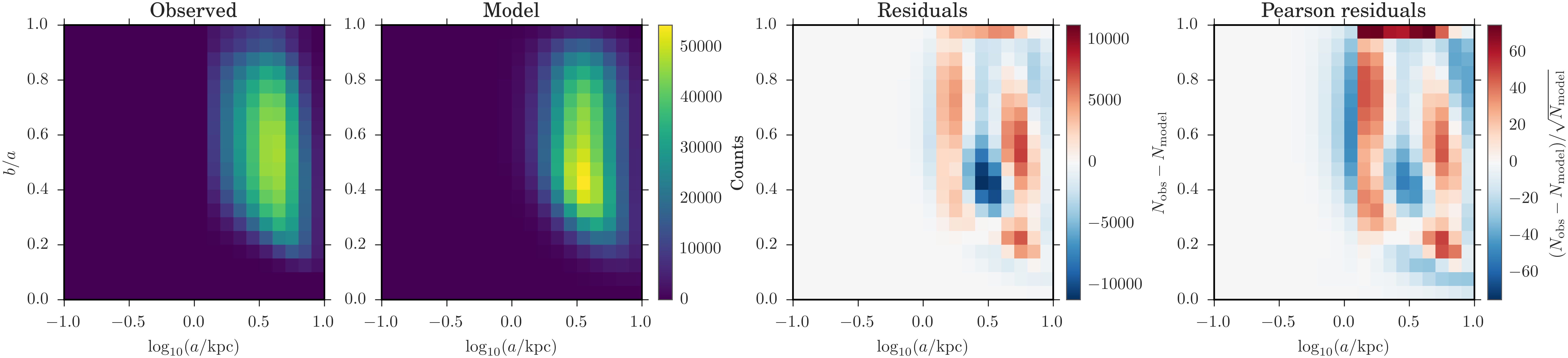} 
\caption{Same as Fig.\,\ref{fig:z0_0p2_logM9_9p5_residuals}, but for the $1.0<z<1.2$, $9.5<\Mstarwun<10.0$ bin.}
\label{fig:z1_1p2_logM9p5_10_residuals}
\end{figure*}

\begin{figure*}[!t]
\centering
\includegraphics[angle=0,width=\hsize]{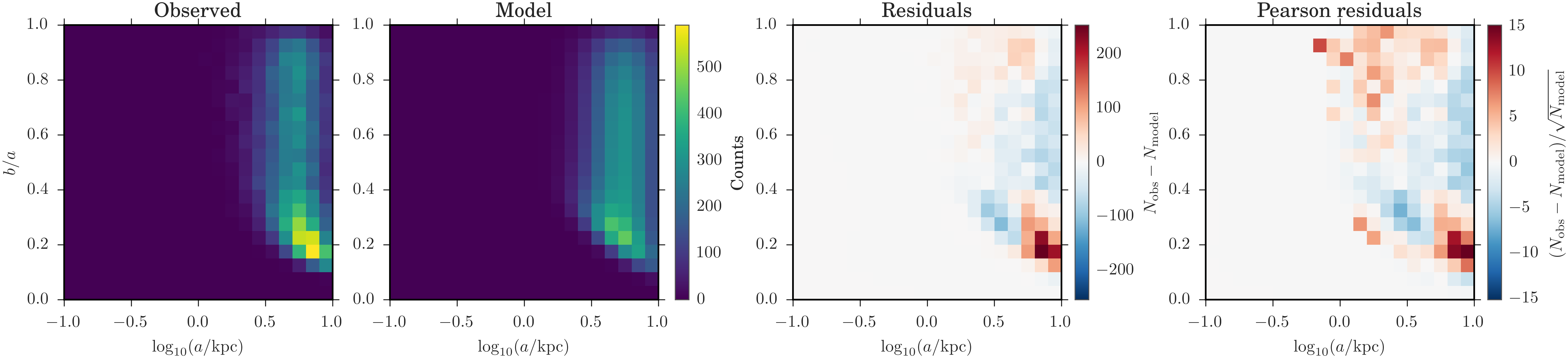} 
\caption{Same as Fig.\,\ref{fig:z0_0p2_logM9_9p5_residuals}, but for the $0.0\lesssim z<0.2$, $10.0<\Mstarwun<10.5$ bin.}
\label{fig:z0_0p2_logM10_10p5_residuals}
\end{figure*}

\begin{figure*}[!t]
\centering
\includegraphics[angle=0,width=\hsize]{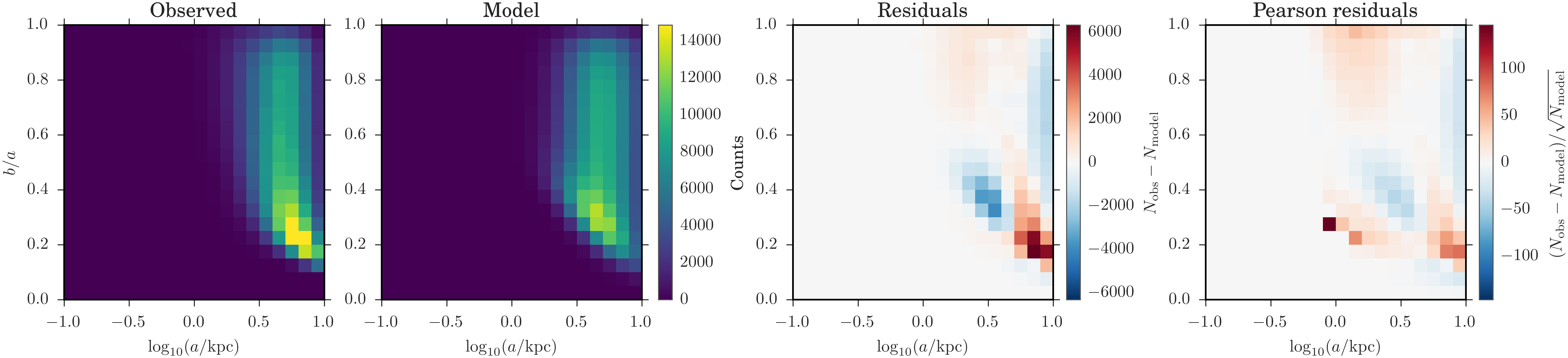} 
\caption{Same as Fig.\,\ref{fig:z0_0p2_logM9_9p5_residuals}, but for the $0.2<z<0.4$, $10.0<\Mstarwun<10.5$ bin.}
\label{fig:z0p2_0p4_logM10_10p5_residuals}
\end{figure*}

\begin{figure*}[!t]
\centering
\includegraphics[angle=0,width=\hsize]{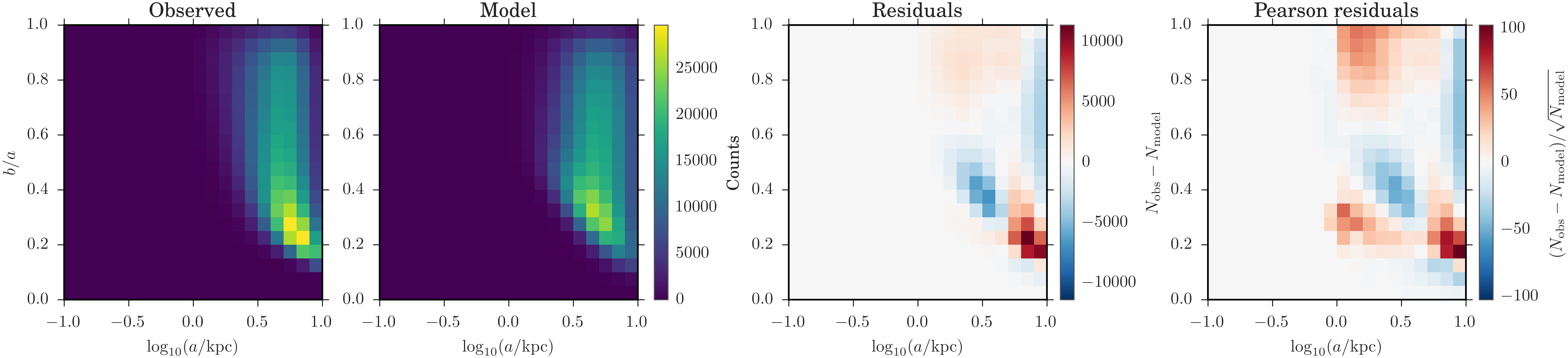} 
\caption{Same as Fig.\,\ref{fig:z0_0p2_logM9_9p5_residuals}, but for the $0.4<z<0.6$, $10.0<\Mstarwun<10.5$ bin.}
\label{fig:z0p4_0p6_logM10_10p5_residuals}
\end{figure*}

\begin{figure*}[!t]
\centering
\includegraphics[angle=0,width=\hsize]{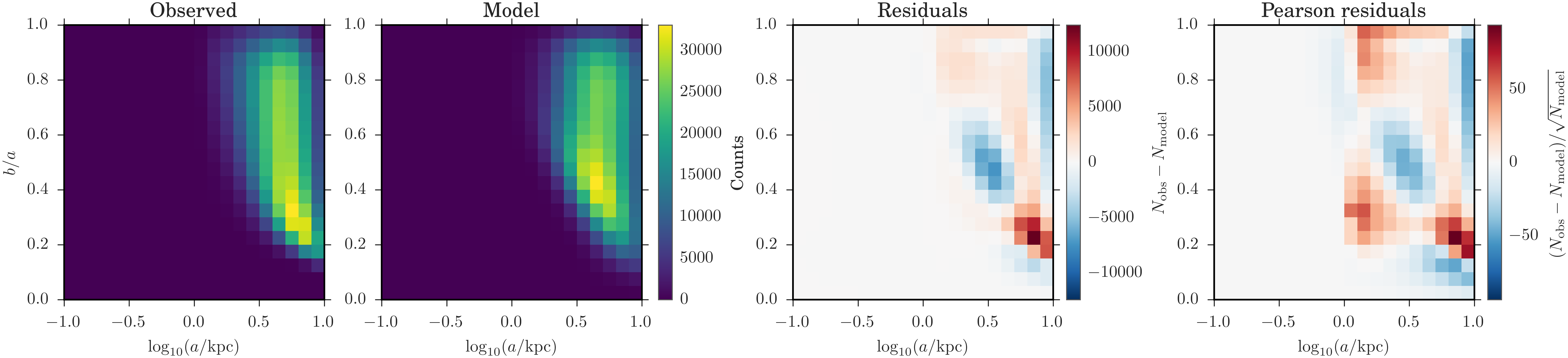} 
\caption{Same as Fig.\,\ref{fig:z0_0p2_logM9_9p5_residuals}, but for the $0.6<z<0.8$, $10.0<\Mstarwun<10.5$ bin.}
\label{fig:z0p6_0p8_logM10_10p5_residuals}
\end{figure*}

\begin{figure*}[!t]
\centering
\includegraphics[angle=0,width=\hsize]{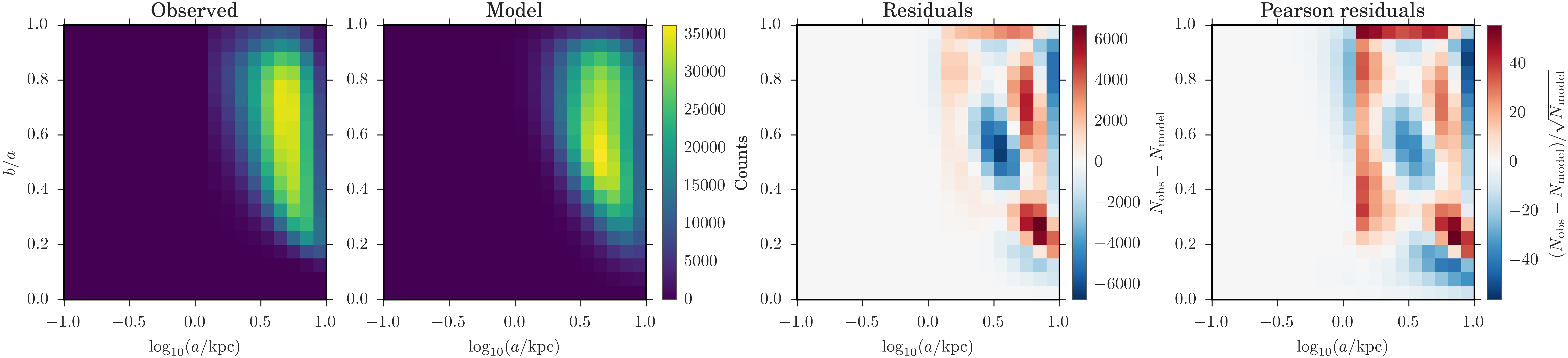} 
\caption{Same as Fig.\,\ref{fig:z0_0p2_logM9_9p5_residuals}, but for the $0.8<z<1.0$, $10.0<\Mstarwun<10.5$ bin.}
\label{fig:z0p8_1_logM10_10p5_residuals}
\end{figure*}

\begin{figure*}[!t]
\centering
\includegraphics[angle=0,width=\hsize]{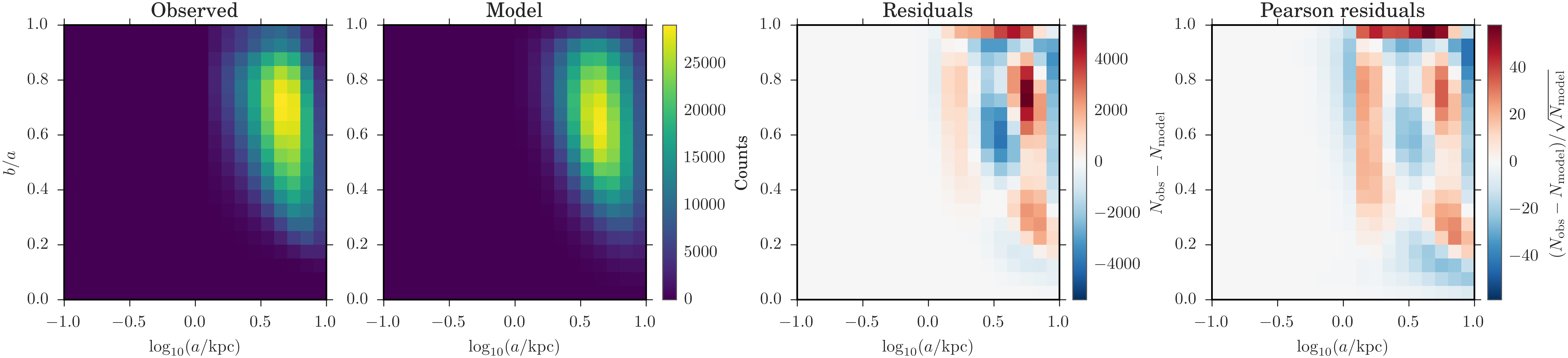} 
\caption{Same as Fig.\,\ref{fig:z0_0p2_logM9_9p5_residuals}, but for the $1.0<z<1.2$, $10.0<\Mstarwun<10.5$ bin.}
\label{fig:z1_1p2_logM10_10p5_residuals}
\end{figure*}

\FloatBarrier
\end{appendix}
\end{document}